\documentclass[12pt]{article}
\usepackage{booktabs}
\usepackage{amsfonts}
\usepackage{amsmath}
\usepackage{amssymb}
\usepackage{amsthm}
\usepackage{mathtools}
\usepackage{float}
\usepackage{rotating}
\usepackage{array}
\usepackage[dvipsnames]{xcolor}
\usepackage{fancyhdr}
\usepackage{graphicx}
\graphicspath{ {./images/} }
\usepackage{caption}
\usepackage{subcaption}
\usepackage{tabularx}
\usepackage{longtable}
\usepackage{color, colortbl}
\definecolor{Gray}{gray}{0.9}
\usepackage{hyperref}
\usepackage{threeparttable}
\usepackage{multicol}
\usepackage{multirow}
\usepackage[%
  backend=biber,
  style=apa,
  natbib=true,
]{biblatex}
\usepackage{enumitem} 
\usepackage[super]{nth} 
\usepackage{lscape}
\usepackage{setspace}
\usepackage{amssymb,amsmath}
\usepackage{amsfonts}
\usepackage{verbatim}
\usepackage{bbm}
\usepackage{multirow}
\usepackage[left=1in, right=1in, top=1in, bottom=1in]{geometry}
\usepackage{subcaption}
\usepackage{array}
\newcolumntype{H}{>{\setbox0=\hbox\bgroup}c<{\egroup}@{}}
\usepackage{hyperref}
\usepackage{color}
\hypersetup{citebordercolor= 1 1 1, linkbordercolor = 0 1 0}
\usepackage[export]{adjustbox}
\usepackage{bm}
\usepackage{textcomp}
\usepackage{epstopdf}
\usepackage{float}
\usepackage{placeins}
\usepackage[labelfont=bf, skip=0pt]{caption}
\usepackage{rotating}
\usepackage{subcaption}
\usepackage{tabularx}
\usepackage{threeparttable}
 \usepackage{tikz}
 \usetikzlibrary{decorations.pathreplacing}
 \tikzstyle{point}=[thick,fill=gray,gray]

\usepackage{pdflscape}

\usepackage{fontawesome5}

\newtheorem{assumption}{Assumption}

\usepackage{fancybox}
\usepackage[final]{pdfpages}

\title{Skill vs Education Types of Labour Mismatch and Their
Association with Earnings\thanks{I thank Arnab Bhattacharjee, Atanas Christev, Erkal Ersoy, Rachel Forshaw, David Jaeger, Lindsey Macmillan, Agnese Romiti, Rachel Scarfe, Mark Schaffer, Cristina Tealdi, and Gill Wyness for insightful comments. I also thank participants at the SGPE Conference 2023 and the Third Scotland and Northern England Conference in Applied Microeconomics for valuable feedback.}}

\author{Vsevolod Iakovlev\thanks{Heriot-Watt University, Edinburgh, United Kingdom. Email: \href{mailto:v.iakovlev@hw.ac.uk}{\texttt{v.iakovlev@hw.ac.uk}}}}

\date{\large \today}

\begin{document}

\maketitle

\begin{abstract}
        This paper analyses the distinction between educational and skill types of labour mismatch and their association with earnings. Drawing on cross-sectional data for 26 countries from the 1st Cycle of the \citet{OECD2012} Survey of Adult Skills (PIAAC), I examine educational and skill mismatch using a comprehensive set of education- and skill-based indicators, explore heterogeneity across worker characteristics, and investigate the sources of conflicting country-level correlations with earnings through an error components model. The results show that country-level unobserved heterogeneity induces endogeneity bias, with both its direction and magnitude varying across mismatch measures. Once unobserved heterogeneity is controlled for, over-education and over-skilling are associated with wage penalties, whereas under-education and under-skilling are linked to wage premiums. These findings highlight both conceptual and empirical distinctions between educational and skill mismatch and demonstrate the importance of indicator choice in the analysis.
    
    \vspace{0.5cm}
    
    \textsl{JEL Classification: D31, I24, J24} \\
    \hspace*{1.5em}\textsl{Keywords: Earnings, Education, Skill, Labour Mismatch}
\end{abstract}

\clearpage

\section{Introduction}\label{sec:1_intro}

Since \citeauthor{mincer1958investment}'s \textsl{\citetitle{mincer1958investment}} (\citeyear{mincer1958investment}), the relationship between education and earnings has been a central topic of interest for labour economists for over six decades. One line of research has focused on the concept of \textit{labour mismatch}, referring to the vertical discrepancy\footnote{Although the discrepancy across education fields, referred to as \textit{horizontal mismatch}, has been shown to be associated with wage penalties \citep{mcguinness2018skills}, this work focuses on \textit{vertical mismatch} across education and skill levels, due to its more prominent role in the debate on the skill-education distinction.} between the characteristics that define a worker’s competencies and the corresponding requirements of their job. A central challenge in analysing the link between mismatch and earnings lies in its measurement, particularly when studies seek to draw conclusions about workers’ \textit{skills}. In the absence of datasets containing direct skills measures, the traditional approach has been to use \textit{education} as a proxy. This practice has raised concerns among researchers that equating educational mismatch with skill mismatch may produce invalid empirical results \citep{allen2001educational}. As datasets with direct skill measures have become more widely available, this concern has gained prominence, leading to the development of alternative measures of skill mismatch \citep{allen2013skill, pellizzari2017new}. 

However, due to the latent nature of labour mismatch, the advantage of new skill-based measures over education-based ones is not self-evident. Moreover, labour market theories, such as human capital, job competition, and labour market friction models, provide theoretical justification for treating skill and educational mismatch as equivalent. This raises the question of whether adopting an alternative measurement framework yields different results for the analysis of wage penalties associated with labour mismatch. This paper contributes to the literature by comparing the outcomes of a comprehensive set of education- and skill-based measures across 26 countries and by examining the implications of selected indicators for estimating a labour mismatch specification of the \citet{mincer1974schooling} earnings function.

The paper begins by providing background to the research question and reviewing related literature in Section~\ref{sec:1_background}, followed by an outline of the educational and skill mismatch measurement frameworks in Section~\ref{sec:1_measures}. Section~\ref{sec:1_skill_vs_edu} examines key labour market theories’ interpretations of the relationship between skill and education, and their links with earnings. Following the description of the data in Section~\ref{sec:1_data}, Section~\ref{sec:1_param} presents a preliminary analysis of the measures’ output from several perspectives, including cross-country comparison, correlation analysis, contribution to out-of-sample prediction performance, and heterogeneity across multiple worker characteristics. Section~\ref{sec:1_method} outlines the econometric model, and Section~\ref{sec:1_analysis} discusses the estimation results. Finally, Section~\ref{sec:1_conclusion} concludes the analysis and identifies the directions for further research.

The main findings confirm the conclusion first made by \citet{allen2001educational} that education and skill mismatch should be distinguished both conceptually and empirically, as using one as a proxy for the other is unlikely to yield accurate results in analyses based on the Mincer earnings function. Specifically, the error component analysis shows that median earnings exhibit a positive relationship with average under-education at the country level, whereas their association with average under-skilling tends to be negative. Estimates for average over-matching are mixed and vary across measures. If country-level variation is not accounted for, unobserved heterogeneity may induce endogeneity bias, with its direction and magnitude depending on the mismatch indicator. Fixed-effects and random-effects estimates suggest that, once unobserved heterogeneity is controlled for, over-education and over-skilling are associated with wage penalties, while under-education and under-skilling are linked to wage premiums. Although the magnitude depends on the choice of measure, the findings are broadly consistent with the literature.\footnote{See, for example, \citet{perry2014can, mcguinness2018skills, cassidy2024increasing, jacobs2022wage, ragoobur2022education, wen2023educational, morsy2019youth, njifen2024education, cultrera2022educational, zhao2025wage}.} Substantial heterogeneity in educational and skill mismatch is also observed across gender and migration status, warranting separate investigation.

\section{Background and related literature}\label{sec:1_background}

In the late 1960s, economists turned their attention to education in an effort to explain income inequality. This focus followed earlier theories that linked the distribution of earnings to \textit{ability} and \textit{chance}. \citet{mincer1958investment} argues that neither of the two factors offers substantial insight into the causes of income inequality. The former is difficult to capture empirically. For instance, intelligence quotient is an inadequate measure, as its normal distribution contrasts sharply with the heavily skewed distribution of earnings. The latter proves equally unhelpful, since its stochastic nature precludes the development of meaningful economic intuition.

Building on the work of \citet{friedman1953choice}, which establishes a link between rational choice and the distribution of personal income, \citet{mincer1958investment} proposes a model that emphasises investment in training, while assuming identically distributed ability and chance. In this model, occupations require varying amounts of training during which individuals receive no income. Consequently, workers are entitled to compensation determined by the present value of their lifetime earnings at the time an occupation is chosen. Among its implications, \citeauthor{mincer1958investment} demonstrates a positive relationship between training and earnings. However, it is not until his \citeyear{mincer1974schooling} model that the classic Mincer equation appears:
\begin{equation}\label{eq:mincer}
    \ln{w} = a_0 + \delta s + \beta_0 x + \beta_1 x^2 + \varepsilon ,
\end{equation}
which establishes a relationship between the natural logarithm of earnings, $\ln{w}$, on the LHS, and schooling $s$, along with linear and quadratic experience terms, $x$ and $x^2$, on the RHS.\footnote{For the derivation of Equation~\ref{eq:mincer}, as well as discussion and empirical analysis of the two models’ implications, see \citet{heckman2003fifty}.} This section reviews the development of labour mismatch indicators and their application within the Mincer equation, followed by the recent advances in the literature.

\subsection{Early research on education mismatch}

The empirical validity of certain theoretical implications of the Mincer equation was at the centre of debate among labour economists for several decades. For example, using data from 1940-1950, \citet{heckman2003fifty} show that, consistent with Mincer's model, no amount of experience enables a worker with less schooling to earn more than one with greater schooling. However, they reject this hypothesis with 1960-1970 data and, using 1980-1990 data, find convergence in the log-earnings/experience profiles at some schooling levels. Likewise, \citet{rumberger1987impact} demonstrates that the apparently straightforward relationship between schooling and earnings becomes less clear when schooling does not align with job requirements. His work summarises the debate on the validity of the human capital model in the context of educational surplus. This view rests on the premise that firms maximise the use of workers' skills by adapting to changes in prices and production technology through the substitution of inputs. In this framework, earnings are linked to workers' productivity, which is argued to be connected to schooling via skills. Consequently, a surplus of schooling should still exert a positive effect on earnings, since firms employ and fully utilise workers as long as their marginal product exceeds the wage. 

This is not the case, however, in the job competition model developed by \citet{thurow1975generating}, in which firms hire workers based on the estimated costs of training, predicted from observable characteristics such as education. Since firms focus on solving the cost-minimisation problem, they may choose not to utilise a schooling surplus. As a result, the return to schooling surplus for workers can potentially be zero or even negative. \citet{rumberger1987impact} addressed this issue by modifying Equation~\ref{eq:mincer} to what later became commonly known as the \textit{over-, required-, and under-education}\footnote{Note that original \citeauthor{rumberger1987impact}'s (\citeyear{rumberger1987impact}) specification does not feature under-education.} (\textit{ORU}) Mincer equation:
\begin{equation}\label{eq:oru}
    \ln{w} = a_0 + \delta_o s_o + \delta_r s_r + \delta_u s_u + \boldsymbol{\beta}\mathbf{X} + \varepsilon,
\end{equation}
where over- and under-education are defined by the difference between the individual level of attained schooling, $s$, and the required level of schooling, $s_r$, as 
\begin{equation*}
s_o = \begin{cases}
     s - s_r & \text{if } s > s_r \\
     0 & \text{if } s \leq s_r
\end{cases} \text{ and } s_u = \begin{cases}
     s_r - s & \text{if } s < s_r \\
     0 & \text{if } s \geq s_r
\end{cases},
\end{equation*}
respectively. Accordingly, \citeauthor{rumberger1987impact} argues that, if the human capital theory is incorrect, we should expect $\widehat{\delta}_o$ in (\ref{eq:oru}) to be non-positive and $\widehat{\delta}_r$ to exceed $\widehat{\delta}$ in (\ref{eq:mincer}). This specification typically places a lesser focus on experience, which is now incorporated into the vector of covariates $\mathbf{X}$, along with other personal characteristics.

The ORU specification has been widely employed in the literature to estimate the association between educational mismatch and earnings, yielding generally consistent results. \citet{hartog2000over} reviews 45 sets of results from five countries, covering different periods between 1969 and 1992, and using three main measurement frameworks: job analysis, realised matches, and direct self-assessment. The reviewed studies largely agree that (i) $\widehat{\delta}_r > \widehat{\delta}_o$, (ii) $\widehat{\delta}_o > 0$ but $\widehat{\delta}_o < \widehat{\delta}_r$, and (iii) $\widehat{\delta}_u < 0$. \citet{hartog2000over} finds that (ii) is the only result that holds universally, whereas (i) and (iii) exhibit exceptions for certain years and countries. Nonetheless, all three patterns are observed across the different measurement frameworks. This is a significant finding, underscoring the robustness of these conclusions and the apparent insensitivity of the ORU specification to the measurement method.

However, \citeauthor{hartog2000over}'s review omits the results of models using alternative inputs. \citet{verdugo1989impact} were the first to approach educational mismatch not in terms of the gap in years of schooling but as a classification problem. They modify Equation~\ref{eq:oru} by replacing $s_o$ and $s_u$ with binary variables for over- and under-education:  
\begin{equation}\label{eq:oru_class}
    \ln{w} = a_0 + \delta_o \, overed + \delta_u \, undered + \boldsymbol{\beta}\mathbf{X} + \varepsilon,
\end{equation}
and report results that contradict patterns (ii) and (iii).\footnote{\citet{verdugo1992surplus} also propose a version of the model with continuous variables for over- and under-education.} Their work sparked a controversy over the validity of this approach. The results were challenged by \citet{cohn1992impact}, who suspected model misspecification, and by \citet{gill1992surplus}, who questioned the empirical strategy. \citeauthor{verdugo1989impact} responded in their \citeyear{verdugo1992surplus} reply, rejecting the idea that their findings were driven by comparisons between low- and high-paid jobs, emphasising that their analysis covered multiple occupational groups with similar earnings. Furthermore, they defend the use of binary over- and under-education variables alongside years of schooling, arguing that highly educated workers are not necessarily productive in their jobs, thereby undermining the assumption that total returns to education exceed occupation-specific returns. However, the controversy persisted. \citet{hartog2000over} lists other studies using \citeauthor{verdugo1989impact}'s specification that find similar results but still exclude it as a model producing outlier results. Nevertheless, because respondents can be classified into one of the three categories using any mismatch measure, Equation~\ref{eq:oru_class} is preferred for analyses involving multiple measurement frameworks, ensuring comparability.

\subsection{The development of skill mismatch measures}

Another measurement issue of particular interest to this paper arises from the distinction between education and skill. \citet{allen2001educational} analyse the application of assignment theory to educational mismatch. While the specific assumptions vary across models, common features include an explicitly formulated assignment problem linking workers’ personal characteristics to their earnings \citep{sattinger1993assignment}. \citeauthor{allen2001educational} argue that assignment theory implies educational mismatch to be both a necessary and sufficient condition for skill mismatch, and vice versa.\footnote{It is unclear which model \citet{allen2001educational} refer to; however, their description of assignment theory closely resembles the differential rents model developed by \citet{ricardo1951principles}.} Their empirical findings suggest that skill mismatch among Dutch university graduates does not account for a substantial share of the wage penalties associated with educational mismatch, thereby contradicting the predictions of assignment theory. Although their results indicate that over-skilling exerts its own negative effect on wages, its magnitude appears to be small.

Their work has initiated a new branch of literature dedicated to the distinction between skill and educational mismatch, which they establish by analysing workers’ responses to two survey questions designed to capture skill underutilisation or deficit.\footnote{This measure is often referred to as direct self-assessment (DSA); see Section~\ref{sec:1_param} for details.} A similar approach is adopted by \citet{di2006education}. Although their results contradict those of \citet{allen2001educational} in the context of on-the-job search, they also find no reduction in the wage penalties associated with over-education among Italian university graduates when accounting for over-skilling. \citeauthor{di2006education} interpret these findings as inconsistent with assignment theory and suggest that the discrepancy between skill and education could instead be explained by simple variations in ability, which are only weakly related to earnings. This conclusion, however, does not align with the findings of \citet{green2010overqualification}. Building on earlier work by \citet{green2007there}, the authors draw on data from several UK surveys to distinguish between the effects of ``formal'' over-qualification, defined as cases in which workers manage to utilise their skills despite having a higher level of education than required, and ``real'' over-qualification, which combines over-education with skill underutilisation. Their analysis reveals a steeper increase in pay penalties associated with real rather than formal over-qualification, supporting the predictions of assignment theory. Thus, while there is little agreement on the importance of differentiating between skills and education, the degree to which these factors explain wage penalties arising from labour mismatch remains uncertain.

\citet{leuven2011overeducation} highlight several identification problems that may account for the inconsistency of empirical results, including unobserved heterogeneity arising from reliance solely on educational mismatch and measurement error commonly associated with self-reported measures of skill mismatch. In principle, both issues could be addressed through instrumental variable methods, the most widely used solution to unobserved heterogeneity. However, this approach has met with limited success due to the lack of credible instruments. Furthermore, \citet{leuven2011overeducation} argue that existing measures of skill mismatch lack a sound theoretical foundation, relying instead on data-driven approaches.\footnote{Some studies form a separate branch of the literature by using surveys compatible with databases containing skill requirements to construct measures of skill mismatch. For instance, \citet{lindenlaub2017sorting}, \citet{lise2020multidimensional}, and \citet{guvenen2020multidimensional} employ O*NET descriptors aggregated through Principal Component Analysis (PCA).} This suggests that acquiring a reliable measure of skills alone would not resolve the problem. In particular, comparing an individual’s skills to the occupation-specific average using a statistically determined cutoff implicitly assumes that all workers fully utilise their skill endowment, regardless of match quality.

To address this issue, \citet{pellizzari2017new} propose a measure for skill mismatch grounded in formal theory, outlined in Section~\ref{sec:1_param}. The explicit economic model differentiates their framework from alternative measures that rely on direct skill data, such as that of \citet{desjardins2011analysis}, who compare a worker’s position in the skill distribution with their position in the distribution of corresponding task engagement scores. Empirically, the model can be summarised as a comparison between a worker’s skills and occupation-specific critical values, defined by the 5th and 95th percentiles of the skill distribution among well-matched workers, identified using self-reported measures. An earlier version of this approach, introduced by \citet{oecd2013first}, was criticised by \citet{allen2013skill}, who argued that linking skill requirements to occupational groups reduces heterogeneity on the demand side while leaving supply-side heterogeneity unchanged. This, in turn, may induce a strong covariance between workers’ skills and the corresponding requirements, producing controversial outcomes. As an alternative, \citet{allen2013skill} develop their own measure, conceptually similar to that of \citet{desjardins2011analysis}, but based on the difference between standardised values for skills and relevant task engagement.

\subsection{Contemporary empirical practices}

Recent applied studies employ both types of skill mismatch measures. For instance, applying the model of \citet{pellizzari2017new} to the first round of PIAAC data, \citet{pivovarova2022immigrants} find that immigrants in the United States are more likely to be over-matched, particularly during their early years in the country, which results in lower wages and standards of living. Alternatively, using the measure proposed by \citet{allen2013skill} together with repeated cross-sections from the 1994 IALS, 2003 ALL, and 2012 PIAAC surveys, \citet{shin2021trends} report that skill mismatch in the United States is largely driven by job-related variables rather than personal characteristics such as gender and migration status. Although the latter study draws on a broader set of datasets, the divergence in findings may be attributable to the alternative measures of skill mismatch, thereby demonstrating the importance of methodological choices even within a single type of labour mismatch.

The discrepancy between mismatch measures, which partly motivates this work, has prompted several methodological studies comparing their construction and outcomes across different contexts. \citet{flisi2014occupational} provide a comprehensive review of 20 specifications of three education- and four skill-based measures applied to PIAAC data for 17 European countries. Their analysis focuses on a narrower set of five specifications, selected using Principal Component Analysis (PCA) and combined into a single measure, to explore their socio-economic determinants, including country of residence, education level, gender, age, and migration status. They find that women are more likely to be over-skilled and simultaneously mismatched in both skills and education, but spot no gender differences in over-education. Older workers are reported to face a higher likelihood of both over-education and over-skilling compared with middle-aged workers, while younger workers display a greater risk of being over-skilled only. Finally, their most intuitive result suggests that higher-educated workers are more prone to all types of over-matching, including simultaneous mismatch.

I build on the agenda set by \citet{flisi2014occupational} but take a different direction in both context and methodology when comparing skill- and education-based measures of labour mismatch. Specifically, the outputs of three education-based and three skill-based measures, each with multiple specifications, are compared across a range of socio-economic characteristics. Unlike \citet{flisi2014occupational}, however, the present analysis does not combine the various frameworks into a single composite measure; instead, each measure is examined separately. This approach preserves the distinctive features of the individual frameworks, which are unique in their definitions of labour mismatch and may capture different aspects of matching outcomes. In addition, the specifications are selected not only for their predictive power but also for their structural distinctiveness, rather than solely for their ability to explain variation in the data.

Another key difference lies in the context. Rather than estimating average predicted probabilities of mismatch, this work examines the explanatory power of mismatch measures within the Mincer earnings function. A related attempt was made by \citet{desjardins2011analysis}, who focus exclusively on skill mismatch. Their measure is based on a precursor measure to \citet{allen2013skill}, initially developed by \citet{krahn1998literacy}. They report that being over-skilled (high skill scores and low skill engagement) in literacy is associated with an average wage penalty of $-4\%$, whereas under-skilled workers (low skill scores and high skill engagement) tend to earn wages that are, on average, $21\%$ higher. However, because in both cases mismatched workers are compared only with the low-skilled subset of the well-matched workers, rather than with the entire pool of well-skilled workers, these estimates provide only limited insight into the association between skill mismatch and earnings.

This paper improves upon their methodology by employing the more recently developed frameworks of \citet{pellizzari2017new} and \citet{allen2013skill}, which represent the two principal approaches to measuring skill mismatch in the literature. Moreover, rather than computing country-specific estimates as in \citet{flisi2014occupational} or including a set of corresponding dummies in the statistical model as in \citet{desjardins2011analysis}, this study exploits cross-country variation by implementing a market-level fixed effects model, with markets defined as country-specific industries. This approach not only accounts for unobserved heterogeneity but also captures market-level variation, which, as the paper demonstrates, constitutes an essential component of the association between labour mismatch and earnings.

\section{Labour mismatch measurement frameworks}\label{sec:1_measures}

As noted in Section~\ref{sec:1_background}, much of the disagreement regarding the effect of mismatch on earnings stems from differences in measurement methodologies. The literature proposes a wide range of mismatch measurement frameworks, which can be classified according to various characteristics.

Firstly, given the focus of this paper, it is reasonable to classify the measures according to the nature of the input variable --- either skill or education. Education-based measures have certain advantages over skill-based measures. The former often rely on data standardised by the International Standard Classification of Education (ISCED) and the International Standard Classification of Occupations (ISCO), and are intuitively clear; for example, comparing an individual’s years of schooling with the number of years their employer required for the job provides an immediate understanding of the discrepancy. The latter, on the other hand, are dataset-specific and require a theoretical framework for economic interpretation. Moreover, unlike education, which can be interpreted as an investment in human capital or as a signal, skill assessment data does not necessarily reflect the level of skill that workers apply in their occupations, thereby weakening its connection with earnings.

Another common distinction is between objective and subjective measures. Subjective measures rely on the genuineness and precision of respondents, making them vulnerable to measurement error. Examples include \textit{direct and indirect self-assessment} (DSA and ISA). DSA is obtained by asking respondents to evaluate their level of skill or education in relation to the day-to-day demands of their job. \citet{verhaest2006impact} differentiate these measures based on whether the questions concern the interviewee’s education or their skill deployment. Traditionally, reviews of mismatch measures, such as that by \citet{flisi2017measuring}, classify DSA as an educational mismatch measure. However, considering its similarity to some self-reported skill mismatch measures, such as the one used by \citet{allen2001educational}, and its focus on job performance rather than formal requirements, both versions of DSA can be considered skill mismatch measures. 

ISA, by contrast, is based on responses regarding workers’ attained and required levels of education. \citet{verhaest2006impact} further differentiate ISA measures based on whether the questions refer to the education required to perform the job or the education required to secure the job. It can be argued that the former pertains to skill, while the latter relates to education. However, both questions address the extent to which respondents’ education equips them with the skills needed to perform their jobs, with the second question additionally contextualising this within a specific firm’s recruitment process. Although the two may have different implications in some labour market theories (e.g., human capital and signalling), they both refer to education in its formal sense and are therefore measures of educational mismatch.

Two widely used objective educational mismatch measures that do not rely on respondents’ self-assessment are \textit{job analysis} (JA) and \textit{realised matches} (RM), first employed by \citet{rumberger1987impact} and \citet{verdugo1989impact}, respectively. JA derives the mismatch level by comparing workers’ attained education with the required education for their occupational group. The latter is typically defined by an occupation classification system, such as that of \citet{international2012international}. This measure is straightforward and regarded by some as ``conceptually superior'' \citep{hartog2000over}. Nevertheless, in practice, it can be challenging to identify an appropriate classification for the dataset in question. Moreover, generalised and infrequently updated classifications, such as ISCO, may fail to capture the specific features of particular labour markets or periods.

Unlike JA, RM defines the required education based on the observed distribution. In this framework, a worker is considered over-educated (under-educated) if their attained education is more than one standard deviation above (below) the modal or mean education for their occupational group. A key criticism of RM is the arbitrariness of the classification thresholds. While one standard deviation is commonly used, some studies adopt alternative cutoffs, such as 0.5 standard deviations, which \citet{tsay2005influence} justify due to underestimation concerns. Another critical parameter in RM is the central tendency measure. Some studies, including the original by \citet{verdugo1989impact}, prefer the mean. However, since educational distributions often exhibit multiple peaks, the mode may be a more suitable choice. Additionally, its lower sensitivity to outliers suggests it may yield more accurate results than the mean \citep{kiker1997overeducation}.

Before moving on to objective measures of skill mismatch, it is essential to acknowledge the conceptual challenges associated with designing such measures. As mentioned, skill measures often lack a clear economic interpretation. DSA circumvents this issue by estimating skill mismatch without requiring a preliminary measurement of the skill itself. However, this approach only permits a subjective measure.

An objective measure requires a theoretical framework linking skill measurement to the concept of skill mismatch. One such framework is the model developed by \textit{\citeauthor{pellizzari2017new}} (PF) (\citeyear{pellizzari2017new}), which defines the required skill thresholds in relation to the optimal skill deployment choices of well-matched workers. Using the original notation, suppose each worker $i$ is characterised by a skill endowment $\eta_i$ and a level of skill $s_i$ deployed at job $j$, such that the following utility function is maximised:
\begin{equation}\label{eq:utility}
    U_{ij} = w_{ij} - \mathbf{1}[y_{ij} < 0] F - c_i(s_i),
\end{equation}
where $w_{ij} = \gamma_i y_{ij}$ denotes the wage, proportional to the output of the match $y_{ij}$ with $\gamma_i \geq 0$; $F \geq 0$ is a cost incurred when producing negative output; and $c_i(s_i) = \mathbf{1}[s_i > \eta_i] \delta s_i $ is the cost of deploying skill $s_i$, equal to $\delta s_i$ with $\delta \geq 0$ if the deployed skill exceeds the endowment $\eta_i$, and zero otherwise. The output $y_{ij}$ is a function of deployed skill characterised by a locally constant marginal product that decreases beyond a threshold, a fixed operational cost $k_j $, and a return to deployed skill $\beta_j$:
\begin{equation}\label{eq:output}
    y_{ij} = \begin{cases}
    \beta_j s_i - k_j & \text{if } s_i\leq max_j \\
    \beta_j max_j - k_j & \text{if } s_i > max_j,
    \end{cases}
\end{equation}
where $min_j$ and $max_j$ are critical points in the skills distribution defined by the values of $s$ that result in zero and maximum output, respectively. \citeauthor{pellizzari2017new}'s model then uses these two critical values to define skill mismatch. A worker is described as over-skilled (under-skilled) if their skill endowment is above (below) $max_j$ ($min_j$), and well-skilled if $min_j \leq \eta_i \leq max_j$. It can be shown that the optimal skill deployment values are $s_{\text{well-skilled}}^* = \eta_i$, $s_{\text{under-skilled}}^* = min_j$, and $s_{\text{over-skilled}}^* \in [max_j, \eta_i]$.

Empirical identification of $min_j$ and $max_j$ is based on respondents’ answers to two interview questions designed to reveal (i) their ability to perform a more demanding job with their current skill endowment, and (ii) their need for additional training to perform the current job. Workers who respond negatively to both questions are assumed to be well-skilled. Since the optimal skill deployment level for well-skilled workers corresponds to their endowment, $min_j$ and $max_j$ can be derived from the distribution of $\eta_i$ within each occupational group. Over- and under-skilled workers are then classified according to the derived values of $min_j$ and $max_j$. Although \citeauthor{pellizzari2017new}’s framework relies on DSA to identify well-matched workers, it classifies the remainder in a distribution-driven manner, thus addressing the skill deployment issue mentioned at the beginning of this section. Most importantly, it links skill assessment data to earnings.
  
Although \citet{pellizzari2017new} developed the economic model above, the measure itself was first introduced in a report by \citet{oecd2013first}. Shortly thereafter, \citet{allen2013skill} criticised the framework, arguing that by computing the classification thresholds within occupation groups, PF reduces heterogeneity on the demand side without making a corresponding adjustment on the supply side. Their concern is that the PF approach overlooks the potentially strong correlation between skill scores and requirements within occupation groups, leading to paradoxical results.

To address this issue, they propose an alternative indicator that compares a worker’s skill score with their skill engagement (skill use), which is referred to as the \textit{Allen-Levels-van-der-Velden} (ALV) measure. Specifically, their measure involves computing z-scores for both the skill and engagement variables. When multiple variables describing skill engagement are available, their average is used instead. The skill engagement z-scores are then subtracted from those of the skill variable. A worker is classified as over-skilled (under-utilised) if the difference between the z-scores exceeds $1.5$, and under-skilled (over-utilised) if the value is below $-1.5$.

A similarity can be spotted between this measure and the one developed by \citet{krahn1998literacy}, which classifies workers into four groups defined by their position in the distributions of skill scores and skill engagement: high-skilled match (high score and high engagement), low-skilled match (low score and low engagement), high-skilled mismatch (high score and low engagement), and low-skilled mismatch (low score and high engagement).\footnote{Due to the greater statistical robustness of ALV, the alternative proposed by \citet{krahn1998literacy} is not included in the analysis.} This suggests that skill engagement-based measures constitute another established class of skill mismatch measures, alongside skill requirement-based measures \citep{oecd2013first, pellizzari2017new} and self-reported measures \citep{allen2001educational, di2006education, green2010overqualification}.
   
The final feature that distinguishes mismatch measures, though by no means the least important, is the unit of measurement. In the context of educational mismatch, this distinction dates back to the \citeyear{verdugo1992surplus} debate between \citeauthor{verdugo1992surplus}, who advocated measuring mismatch using binary variables for over- and under-education, and \citeauthor{cohn1992impact} and \citeauthor{gill1992surplus}, who defended the traditional measurement based on years of \mbox{over-,} \mbox{under-,} and required schooling. These two approaches have been reported to yield results leading to opposite conclusions \citep{hartog2000over}. In the context of skill mismatch, this distinction remains relevant. While the DSA measure is necessarily binary, the PF and ALV measures can be adapted to accommodate a continuous scale. Although binary measures may suffer from reduced variation, they facilitate comparability across measures, which is essential for this study.

In summary, the literature presents a variety of labour mismatch measurement frameworks that can be differentiated in several ways. Furthermore, measures such as RM involve a choice of parameters that affect their resulting classification. This raises two key questions: which measures should be used for the analysis of earnings, and how this choice influences the results.
\section{Skill vs education in labour market theories}\label{sec:1_skill_vs_edu}

It could be argued that the primary reason for using educational mismatch as a proxy for skill mismatch is purely practical --- the limited availability of direct skill data. However, employing this approximation in empirical analysis still requires theoretical justification, often drawn from labour market theories that interpret the relationship between education and skill, as well as their connection to earnings. Therefore, the validity of this empirical approximation has implications for the validity of these theories. It is thus helpful to outline which frameworks are supported and which are challenged by the results of this work.

\citet{leuven2011overeducation} provide a brief overview of the role of educational mismatch in six different labour market theories. To offer a summary of various interpretations of the relationship between skill and education in the context of earnings, this section discusses the same set of theories, supplemented by assignment theory, which is represented by three different models, as outlined by \citet{sattinger1993assignment}. The objective of this section is to identify theories that distinguish between skill and education so minimally that assuming equivalence between the two concepts is appropriate for applied research.

\textit{Human capital theory} \citep{becker1964human} considers education as an investment, repaid by an increase in a worker’s productivity that leads to higher earnings. Productivity and earnings have a direct link that originates from neoclassical economic theory, which suggests that profit-maximising firms hire workers as long as their marginal product exceeds the costs of employing them. Thus, earnings are determined by workers’ productivity. Moreover, it is assumed that firms can adapt to changes in prices and production technologies by substituting inputs; thus, the positive relationship between productivity and earnings remains robust even in the presence of over-education. However, the link between education and productivity is indirect and operates through skill.\footnote{\citet{rumberger1987impact} provides an overview of explanations for the education–skill–productivity relationship.} 

In the human capital view, education is only one factor contributing to a worker’s skill. Nonetheless, it assumes the positive relationship between the two is strong enough to generate returns that partially explain income inequality. This is where the human capital model brings education and skill close to equivalence. In practice, it enables the analysis of the Mincer function without explicitly accounting for skill, which effectively imposes this equivalence.\footnote{Although the Mincer function usually includes terms for experience, it acts as another factor influencing skill alongside education rather than as a direct measure of skill.}

In contrast to human capital theory, which associates marginal products with workers, the \textit{job competition model} \citep{thurow1975generating} links marginal products to jobs \citep{rumberger1987impact}. In this model, firms fill positions while aiming to minimise the costs of training. These costs are determined by the gap between the job requirements and the workers’ skill levels. Since skills are assumed to be unobservable, firms use education as a proxy to predict training costs and inform hiring decisions. Like human capital theory, the job competition model acknowledges that skill and education are distinct concepts but assumes one strongly predicts the other. This leads to similar implications for the formulation of statistical models using the Mincer equation.

The relationship between education and skill in \textit{assignment theory} is less straightforward. \citet{sattinger1993assignment} identifies three main types of assignment models, each adopting slightly different approaches to these two concepts. The model that arguably contrasts most with human capital and job competition theories is the \textit{linear programming optimal assignment model} \citep{koopmans1957assignment}. This model assumes no hierarchical relationship between jobs and workers. Although both workers and ``machines'' differ in certain characteristics, there is no direct link between these and earnings. Instead, earnings are determined by the outputs associated with alternative assignments. It is consistent with \citeauthor{koopmans1957assignment}’s model to view skill and education as two of a worker’s characteristics, the relationship between which is ambiguous. However, neither is directly linked to earnings in the Mincerian sense, as the effect of varying skill and education levels on the outcome of the assignment problem is not immediately apparent. 

In contrast, the \textit{differential rents model} \citep{ricardo1951principles} allows for a hierarchical assignment, whereby higher-skilled workers tend to be allocated to more demanding jobs. In \citeauthor{sattinger1993assignment}’s formulation, each worker is characterised solely by their skill, though it is suggested that other attributes such as education or ability could be used instead. This does not imply equivalence between these concepts, but rather that the framework can potentially be extended to encompass a range of characteristics that influence the productivity of the worker-job match.

Lastly, \citeauthor{roy1951some}’s (\citeyear{roy1951some}) \textit{sectoral model} combines features of the two assignment models discussed above. Formally, it can be expressed as a special case of the linear programming optimal assignment problem and, like the differential rents model, is solved through decisions made by profit- and earnings-maximising firms and workers. Unlike the previous models, \citeauthor{roy1951some}’s framework is formulated in terms of occupations rather than individual jobs (e.g., ``catching rabbits'' or ``fishing for trout'', as in the original formulation). Earnings are therefore determined by how many projects each worker completes (rabbits caught or trout fished), given their choice of occupation. Nonetheless, the role of skill or a skill measure is similar. Rather than entering the earnings function directly, skill represents a natural or acquired inclination of a worker towards a particular occupation, influencing the assignment problem. Although assignment theory does not allocate distinct roles to education and skill (unlike human capital or job competition models, where one predicts the other), it also does not explicitly assume equivalence between the two concepts, as noted by \citet{allen2001educational}. Instead, it focuses on the solution of the assignment problem as the primary determinant of the earnings distribution.

\citeauthor{spence1973job}'s (\citeyear{spence1973job}) \textit{signalling model} provides education and skill with the most distinct roles so far. Similar to human capital theory, education is considered an investment. However, due to asymmetric information, the return does not stem from a direct effect on skill or productivity, but rather from education serving as a costly signal of a worker’s innate skill. This rests on the premise that a naturally high-skilled worker can attain a certain level of education more easily than a naturally low-skilled worker. Consequently, there exists a threshold education level that would be unprofitable for a low-skilled worker to pursue, since the premium earned by being perceived as high-skilled falls short of the costs of obtaining that education. 

Intuitively, \citeauthor{spence1973job}’s model implies a positive relationship between skill and education because higher skill reduces the cost of attaining higher education levels. However, this relationship is far from equivalent compared to the human capital and job competition models, since workers may sometimes be incentivised to misrepresent their skills through educational choices. Moreover, the signalling model offers an intriguing implication for the discussion of educational mismatch. The extent of observed educational mismatch in an economy depends on how the required education is defined. Job requirements, as determined by firms, are likely to be distribution-based and reflect the signalling value of education, given the skill distribution among workers. If education serves purely as a signalling mechanism and two equally skilled workers with different education levels perform the job equally well, then the education level actually required to perform the job is zero, implying that any positive investment in education results in over-education.
 
Another theory contributing to the educational mismatch literature is \citeauthor{sicherman1990theory}’s (\citeyear{sicherman1990theory}) \textit{model of career mobility}, which focuses on workers’ career paths. In this view, workers aim to maximise their expected lifetime earnings by allocating finite time across various jobs. Earnings are defined as a function of human capital, which in turn increases with education and innate skill. This creates two potential sources of returns to education: a direct return through increased human capital and an indirect return through improved career paths. A notable implication of over-education in this model is that a worker may choose a job with lower requirements than their attained education if it offers a higher probability of promotion. The career mobility model suggests a relationship between education and skill that incorporates elements from the human capital, signalling, and assignment theories. Specifically, education is considered an investment in human capital, but skill is exogenously fixed, and no direct linkage between the two is assumed. Arguably, \citeauthor{sicherman1990theory} position education and skill as far from equivalent as possible.

Some economists explain the phenomenon of over-education through the macroeconomic concept of \textit{labour market frictions}. These typically refer to imperfect information and insurance markets, heterogeneity, slow mobility, labour market capacity, and related factors, and are often modelled using a matching function with the number of unemployed workers and vacancies as inputs \citep{pissarides2000equilibrium}. Such models often define earnings through a wage bargaining equation based on the Nash bargaining approach, which does not explicitly incorporate either skill or education. \citet{leuven2011overeducation} reviews several studies that include the concept of skill in search and matching models. These models generally feature two equilibria: one in which no worker chooses a job with requirements below their attained education, and another in which some workers do, a choice argued to depend on factors such as the gap between productivity levels and the skill distribution across workers \citep{albrecht2002matching}, productivity and quit rates \citep{gautier2002unemployment}, and the opportunity to pursue higher-return jobs \citep{dolado2009job}. However, none of these studies distinguishes between skill and education, instead using the two concepts interchangeably, thereby assuming their equivalence.

Finally, it is worth noting the role of \textit{preferences} in the educational mismatch literature. Intuitively, workers who derive greater utility from the process of acquiring education are more likely to become over-educated, and vice versa. This introduces an additional source of return to education through a utility function, complicating the relationship between earnings and educational decisions. \citet{oosterbeek2000schooling} conduct an empirical analysis using a Cobb-Douglas utility function with the net present value of lifetime earnings and schooling as inputs. They find that the marginal rate of return to schooling is higher than that to earnings, suggesting the standard Mincerian model underestimates returns to schooling. Additionally, they find that social background and ``innate ability'' are positively related to preferences for schooling and negatively associated with the marginal rate of return. The relationship between skill and education in the context of preferences and earnings resembles that in the career mobility model. The difference lies in the sources of returns to education: in \citeauthor{sicherman1990theory}'s (\citeyear{sicherman1990theory}) model, the indirect effect of education on returns comes from career path improvement, whereas the preferences theory suggests returns also arise from positive utility gained during the schooling process, which is unrelated to earnings.

In summary, the literature offers a broad spectrum of interpretations regarding the relationship between skill and education in the context of earnings. Theories such as human capital, job competition, labour market frictions, and assignment models facilitate the use of educational data for the analysis of skill mismatch. However, it is more challenging to adopt a similar empirical framework when considering signalling, career mobility, preferences, and assignment theories.
\section{Survey of Adult Skills (PIAAC)}\label{sec:1_data}

The data for the analysis come from the First Cycle of the \citeauthor{OECD2012} Survey of Adult Skills (PIAAC), conducted between 2011 and 2012 in 35 countries, of which 26 are included. A distinctive feature of the PIAAC dataset is its direct assessment of three core abilities: evaluating and engaging with written texts, interpreting and communicating numerical information, and performing practical tasks using digital communication tools. These are measured on a 500-point scale and referred to as \textit{literacy}, \textit{numeracy}, and \textit{problem-solving}, respectively. In addition, the survey collects extensive information on demographic characteristics, education and training, social and linguistic background, employment status and income, as well as the use of cognitive, interactional, social, physical, and learning skills. This section begins with a brief note on the motivation for using the PIAAC dataset, followed by a description of the data cleaning process, variable construction, and summary statistics.

The PIAAC dataset has recently become a common choice for analysing labour mismatch in relation to productivity, earnings, horizontal mismatch, job tasks, cognitive skills, university enrollment rates, and other topics. For example, \citet{mcgowan2015labour} find that over-skilling and under-qualification are the main drivers of the negative relationship between workers’ productivity and labour mismatch. In their follow-up work (\citeyear{mcgowan2017skills}), using the second round of the survey, they show that better allocation of skills is associated with higher productivity. \citet{nieto2017overeducation} demonstrate that the well-documented wage penalty associated with over-education is partly explained by the lower skill levels of over-educated workers compared to their equally educated peers in more demanding jobs. \citet{montt2017field} evaluates both vertical and horizontal mismatch, finding that working outside one’s field does not lead to a wage penalty unless the worker is also over-educated. \citet{pouliakas2015heterogeneity} use skill mismatch to estimate cognitive skill demand and analyse its relationship with task complexity, finding the two to be significantly related. Finally, \citet{castro2022magnitude} examines the effect of a shock in tertiary education participation on the extent of over-education and over-skilling. They find that the impact on skill mismatch is similar to that on education mismatch, although this result appears to hold only for LAC countries, where over-skilling estimates are lower than in OECD countries. In summary, PIAAC data are widely used to investigate various aspects of labour mismatch, making it a strong choice for methodological analysis of measurement frameworks.

\begin{figure}[h!]
    \caption{Hourly earnings including bonuses (USD PPP) and its natural logarithm}
    \label{fig:hist_earn}
    \centering
    {\includegraphics[scale=1, trim=0cm 0cm 0cm 0cm, clip]{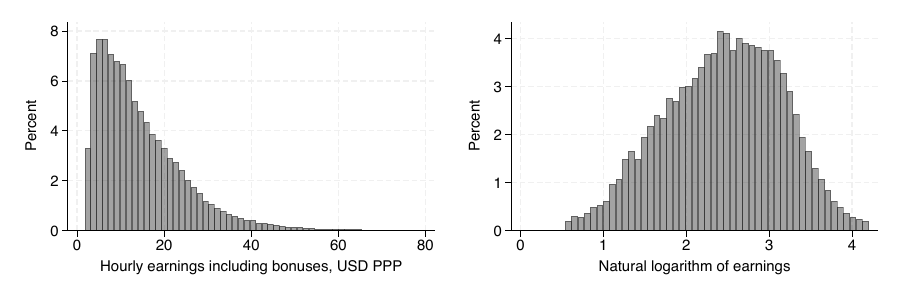}}
    \begin{tablenotes}
		\footnotesize
    		\item \textit{Notes:} The distribution of hourly earnings is trimmed at the 1st and 99th percentiles to avoid outliers.
    \end{tablenotes}
\end{figure}

The variables essential to the analysis include the country of residence, employment status, hourly earnings including bonuses (PPP-adjusted USD), and the current job’s occupation group at the 1-digit level of the International Standard Classification of Occupations 2008. Observations with missing values for any of these variables are excluded from the dataset. In addition, respondents who are unemployed or outside the labour force are omitted from the analysis. Other variables used are non-essential and may contain missing values.
\begin{table}[h!]
	\centering
    \footnotesize
    \caption{Summary statistics: Earnings and education}
    \label{tab:sumstat_earn_ed}
    \begin{tabular}{l | rrrrrr}
\toprule
{} & Earnings     & Highest     & Attained  & Required  & Attained   & Required \\  
   &    & qual.     & ISCO SL   & ISCO SL   & years of edu.      & year of edu.    \\
\midrule
N &  72,563 &  68,888 &     68,888 &     72,563 &  71,854 &  70,996 \\
Mean  &     14.60 &      8.02 &         2.67 &         2.57 &     13.14 &     12.55 \\
Std   &     10.20 &      3.84 &         0.90 &         0.98 &      2.97 &      3.34 \\
Min   &      1.71 &      1 &         1 &         1 &      3 &      3 \\
Q1   &      6.97 &      5 &         2 &         2 &     11 &     11 \\
Q2   &     11.96 &      6 &         2 &         2 &     13 &     12 \\
Q3   &     19.63 &     12 &         4 &         4 &     15 &     15 \\
Max   &     66.77 &     16 &         4 &         4 &     23 &     23 \\
\bottomrule	
\end{tabular}

    \begin{tablenotes}
		\footnotesize
    		\item \textit{Notes:} Earnings --- hourly earnings including bonuses (PPP-adjusted USD). Qualification (qual.) --- ISCED 1997 level.
    \end{tablenotes}
\end{table}
Table~\ref{tab:sumstat_earn_ed} presents summary statistics for the earnings and education variables. As expected, the distribution of hourly earnings is notably right-skewed (see Figure~\ref{fig:hist_earn}). The highest qualification is a categorical variable on a 1 to 16 scale, where 1 corresponds to no qualification or below International Standard Classification of Education 1997 (ISCED) level 1, and 14 corresponds to ISCED level 6 (doctoral degree).\footnote{Values of 15 and 16 are not ordered and correspond to a foreign qualification and an unidentified higher education degree, respectively.} The attained ISCO skill level (SL)\footnote{ISCO skill level is derived from an education variable and is not a direct measure of skill.} has four categories and is computed from the highest qualification using a mapping provided by the International Labour Organisation (\citeyear{international2012international}). Specifically, ISCED level 1 is equivalent to attained SL 1; ISCED levels 2 through 4 correspond to attained SL 2, ISCED level 5b maps to SL 3, and ISCED levels 5a and 6 correspond to SL 4.

The required ISCO SL is derived by mapping respondents’ ISCO occupation groups onto the same 1 to 4 scale \citep{international2012international}: managers in ISCO group 1 are split into the high- and low-skilled categories with required SL 4 and 3, respectively; professionals in ISCO group 2 have a required SL of 4; technicians and associate professionals in ISCO group 3 have a required SL of 3; workers in ISCO groups 4 to 8 are required to have SL of 2; and the elementary occupations in ISCO group 9 have a required SL of 1. Finally, the last two variables in Table~\ref{tab:sumstat_earn_ed} refer to years spent in full-time education and years of education required for a respondent’s current job, respectively. The distributions of attained and required education are similar across both sets of variables, with mean values slightly higher for obtained education.

\begin{figure}[h!]
    \caption{Attained qualification and ISCO skill level}
    \label{fig:hist_qual}
    \centering
    {\includegraphics[scale=1, trim=0cm 0cm 0cm 0cm, clip]{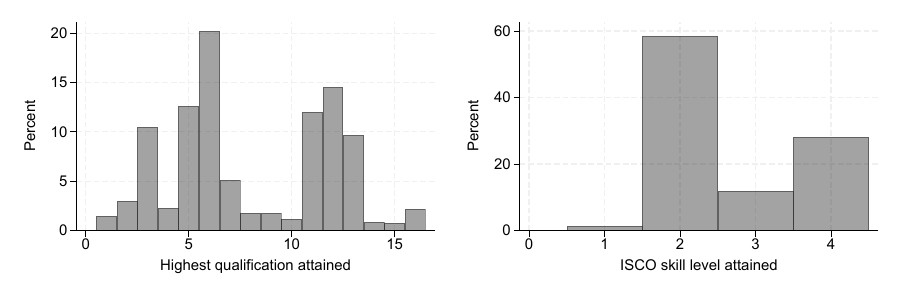}}
\end{figure}

One potential criticism concerns the reduction of the highest qualification variable from 16 categories to four ISCO skill levels. This may raise concerns that the ISCO skill level is a poor proxy for the distribution of qualification levels. However, Figure~\ref{fig:hist_qual} shows that most observations are concentrated around several main qualification values: 3 (ISCED level 2), 5 to 7 (ISCED level 3), and 11 to 13 (ISCED level 5), corresponding to high school dropouts, high school graduates, and university graduates, respectively. According to the ISCO mapping \citep{international2012international}, respondents with the highest qualification of 1 are assigned skill level 1; those with qualifications between 2 and 10 are assigned skill level 2; a qualification value of 11 corresponds to skill level 3; and values from 14 to 16 are mapped to skill level 4. Given this mapping and the distributions in Figure~\ref{fig:hist_qual}, the ISCO skill level can be considered an appropriate approximation for the highest qualification.

Table~\ref{tab:sumstat_skills} displays summary statistics for the variables used to measure skill mismatch. The first two are binary variables indicating whether respondents feel capable of handling more demanding duties and whether they perceive a need for additional training to perform their current duties effectively. The results suggest that 84\% report being under-challenged, while 36\% report needing further training. Although these indicators are not direct measures of skill, they are used to compute the DSA and Pellizzari-Fichen frameworks. 

\begin{table}[h!]
	\centering
    \footnotesize
    \caption{Summary statistics: Skills}
    \label{tab:sumstat_skills}
    \begin{tabular}{l | rrrrr}
\toprule
{} &   Not challenged &  Need training &       Literacy &       Numeracy &       Problem-solving \\
\midrule
N &  72,563 &   72,563 &  72,561 &  72,561 &  51,224 \\
Mean  &      0.84 &       0.36 &    271.88 &    268.40 &    279.61 \\
Std   &      0.37 &       0.48 &     46.01 &     50.03 &     41.79 \\
Min   &      0 &       0 &     66.39 &     24.85 &      8.04 \\
Q1   &      1 &       0 &    243.75 &    238.32 &    252.10 \\
Q2   &      1 &       0 &    276.45 &    272.73 &    282.43 \\
Q3   &      1 &       1 &    304.57 &    303.51 &    309.25 \\
Max   &      1 &       1 &    410.65 &    430.98 &    480.01 \\
\bottomrule
\end{tabular}

    \begin{tablenotes}
		\footnotesize
    		\item \textit{Notes:} Not challenged --- answered positively to ``Do you feel that you have the skills to cope with more demanding duties than those you are required to perform in your current job?'' Need training --- answered positively to ``Do you feel that you need further training in order to cope well with your present duties?''
    \end{tablenotes}
\end{table}

\begin{figure}[h!]
    \caption{Skill scores}
    \label{fig:hist_skills}
    \centering
    {\includegraphics[scale=0.9, trim=0cm 0cm 0cm 0cm, clip]{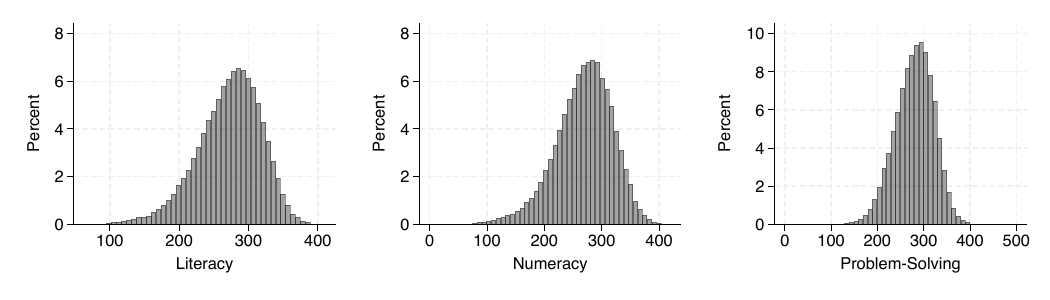}}
\end{figure}

The following three variables report the results of the PIAAC assessments of literacy, numeracy, and problem-solving skills, each scored on a 500-point scale. These variables are calculated as the average of the ten plausible values provided in the dataset for each score. Table~\ref{tab:sumstat_skills} and Figure~\ref{fig:hist_skills} show that the distributions of all three skills are broadly similar, approximating a Gaussian curve with slight negative skewness. It should be noted, however, that problem-solving data are available for only 71\% of respondents. Of the 20,817 missing observations, 36\% are due to data unavailability in France, Italy, and Spain.\footnote{See Table~\ref{tab:cntry_count}.} Table~\ref{tab:psl_missing} shows that, relative to workers with observed problem-solving data, the remaining 13,254 observations with missing values are characterised by 44\% lower median earnings, two ISCED categories lower educational attainment, 10\% lower literacy, 15\% lower numeracy, and an average age that is 8 years higher.\footnote{See Appendix~\ref{app:1_summary_stats}.} This pattern suggests non-random missingness and potential selection on unobserved characteristics, which may lead to biased estimates. To address this concern, the sensitivity of the mismatch coefficients to selection on unobserved variables is examined using the methodology proposed by \citet{oster2019unobservable}. Specifically, the degree of selection on unobserved covariates required to drive the mismatch coefficients to zero, denoted by $\delta$, is estimated under the assumption of proportional selection on observed and unobserved controls.\footnote{The structural controls of the Mincer equation (education, literacy, numeracy, problem-solving, tenure, and tenure-squared) are included in all regressions; hence, selection on them does not contribute to the estimation of $\delta$. Following \citet{oster2019unobservable}, $R_{\text{max}}$ is set to $1.3 \times R^2_\text{POLS}$.} Selection on unobservables is considered unlikely to overturn the results if $\hat{\delta} > 1$.

Table~\ref{tab:occ_count} presents the frequency of occupation groups and their occupation-specific averages for earnings, highest qualification attained, and skill variables. The occupation groups are primarily based on 1-digit ISCO codes, with the managers group split into high- and low-skilled subgroups using 2-digit codes, reflecting their distinct ISCO skill level requirements. The armed forces occupations are omitted due to their small representation in the data. Furthermore, because some labour mismatch measures are distribution-based, respondents in country-specific occupation groups with fewer than 30 observations are excluded from the analysis. 
\begin{table}[h!]
    \centering
    \footnotesize
    \caption{Frequencies and averages: ISCO occupation groups}
    \label{tab:occ_count}
    \begin{tabular}{l | rrrrrrrrr}
\toprule
Occupation &  N &  Frac. &  Median   & Median    & Mean &     Mean &     Mean & Mean & Mean \\ 
group &   &           & earn.  & qual.     & lit.  & num.  & pr.slv. & gender & age \\
\midrule
HS managers                              &      3,592 &      0.05 &  22.43 &    12.0 &  294.71 &  298.06 &  294.73 &      0.40 &  43.78 \\
Professionals                                      &     14,882 &      0.21 &  18.22 &    12.0 &  294.43 &  293.58 &  292.51 &      0.62 &  40.60 \\
Techc-s \& assoc.            &     10,691 &      0.15 &  14.99 &    11.0 &  283.73 &  282.65 &  286.77 &      0.53 &  40.26 \\
LS managers                                &       554 &      0.01 &  14.17 &     7.0 &  280.62 &  277.81 &  287.16 &      0.50 &  39.67 \\
Clerical support                           &      7,778 &      0.11 &  12.56 &     6.0 &  278.51 &  273.06 &  282.62 &      0.69 &  39.18 \\
Craft \& related                   &      7,727 &      0.11 &  10.06 &     5.0 &  255.74 &  254.85 &  264.90 &      0.13 &  38.38 \\
Operat. \& assem.        &      5,915 &      0.08 &   9.32 &     5.0 &  253.75 &  251.10 &  259.49 &      0.19 &  40.46 \\
Service and sales                          &     13,697 &      0.19 &   9.24 &     6.0 &  264.17 &  256.28 &  271.97 &      0.69 &  36.70 \\
Element. occup.                             &      7,141 &      0.10 &   7.86 &     5.0 &  239.44 &  230.84 &  257.78 &      0.55 &  39.89 \\
Agric. \& fishery &       586 &      0.01 &   7.08 &     3.0 &  218.15 &  210.36 &  254.02 &      0.25 &  38.83 \\
\bottomrule
\end{tabular}

    \begin{tablenotes}
		\footnotesize
    		\item \textit{Notes:} Rows are sorted by median hourly earnings, including bonuses (PPP-adjusted USD). Qualification (qual.) -- ISCED 1997 level.
    \end{tablenotes}
\end{table}
The number of observations in each occupation group ranges from $554$ to $14,882$. This coverage could be improved by using 2-digit or 3-digit codes more extensively; however, such refinement would likely result in substantial data loss, as smaller occupation subgroups have higher rates of missing values. The table indicates that occupation groups with lower median earnings tend to have lower median qualification levels and lower mean skill scores.

Finally, Table~\ref{tab:cntry_count} mirrors the structure of Table~\ref{tab:occ_count} but reports statistics by country. The variance in country-specific sample sizes is smaller than that observed across occupations. However, Canada, Peru, Hungary, Singapore, Germany, Turkey, Austria, the United States, and Sweden were excluded because their earnings data are available only in deciles. In addition, Italy, France, and Spain lack data on problem-solving, while Estonia does not report data on the highest qualification attained. The table indicates that, in contrast to occupations, country-specific averages for qualification and skill scores do not exhibit a clear association with median earnings.

\begin{table}[h!]
    \centering
    \footnotesize
    \caption{Frequencies and averages: Countries}
    \label{tab:cntry_count}
    \begin{tabular}{l | rrrrrrrrr}
\toprule
\multirow{2}{*}{Country} &  N &  Frac. &  Median   & Median    & Mean &     Mean &     Mean & Mean & Mean \\ 
&   &           & earn.  & qual.     & lit.  & num.  & pr.slv. & gender & age \\
\midrule
Denmark            &       4,426 &       0.06 &  23.07 &     7.0 &  274.42 &  283.81 &  283.13 &      0.51 &  43.29 \\
Norway             &       3,114 &       0.04 &  22.78 &     8.0 &  285.89 &  287.77 &  291.09 &      0.51 &  39.52 \\
Netherlands        &       3,125 &       0.04 &  19.63 &     6.0 &  292.23 &  288.20 &  291.16 &      0.51 &  39.90 \\
Belgium            &       2,688 &       0.04 &  19.53 &     7.0 &  281.33 &  286.04 &  283.07 &      0.50 &  40.17 \\
Ireland            &       2,724 &       0.04 &  19.17 &     8.0 &  276.65 &  266.96 &  281 &      0.57 &  39.15 \\
Finland            &       3,115 &       0.04 &  17.13 &    10.0 &  298.13 &  292.67 &  293.64 &      0.52 &  41.30 \\
United Kingdom     &       4,680 &       0.06 &  14.78 &     6.0 &  280.46 &  271.16 &  283.47 &      0.59 &  39.73 \\
New Zealand        &       3,521 &       0.05 &  14.52 &     8.0 &  283.93 &  273.46 &  289.44 &      0.56 &    NaN \\
Italy              &       1,762 &       0.02 &  14.16 &     6.0 &  259.81 &  260.60 &     NaN &      0.49 &  41.46 \\
France             &       3,572 &       0.05 &  13.92 &     6.0 &  268.32 &  263.42 &     NaN &      0.50 &  41.08 \\
Korea              &       3,000 &       0.04 &  12.64 &     6.0 &  274.85 &  267.31 &  285.04 &      0.46 &  39.68 \\
Japan              &       3,186 &       0.04 &  12.44 &    11.0 &  300.68 &  293.02 &  297.96 &      0.49 &  41.52 \\
Spain              &       2,430 &       0.03 &  12.38 &     6.0 &  259.57 &  255.52 &     NaN &      0.49 &  39.82 \\
Israel             &       2,507 &       0.03 &   9.49 &     6.0 &  257.17 &  254.49 &  271.41 &      0.51 &  37.11 \\
Greece             &       1,174 &       0.02 &   8.48 &     6.0 &  257.60 &  259.97 &  259.28 &      0.53 &  39.24 \\
Slovenia           &       2,168 &       0.03 &   8.31 &     6.0 &  261.26 &  264.55 &  266.51 &      0.50 &  41.24 \\
Czech Republic     &       2,566 &       0.04 &   8 &     6.0 &  279.93 &  281.20 &  285.80 &      0.52 &  38.63 \\
Estonia            &       3,662 &       0.05 &   7.36 &     NaN &  277.64 &  274.34 &  274.46 &      0.58 &  41.03 \\
Slovak Republic    &       2,419 &       0.03 &   6.88 &     6.0 &  279.28 &  284.11 &  280.73 &      0.51 &  40.48 \\
Poland             &       3,851 &       0.05 &   6.67 &     6.0 &  276.55 &  268.23 &  278.20 &      0.44 &  31.08 \\
Chile              &       2,309 &       0.03 &   6.63 &     6.0 &  225.65 &  211.16 &  252.83 &      0.50 &  38.65 \\
Lithuania          &       2,680 &       0.04 &   6.01 &     9.0 &  271.36 &  273.80 &  260.40 &      0.61 &  42.58 \\
Ecuador            &       1,654 &       0.02 &   4.80 &     6.0 &  195.59 &  188.42 &  227.87 &      0.41 &  35.78 \\
Russian Federation &       1,468 &       0.02 &   4.79 &    12.0 &  285 &  279.78 &  287.85 &      0.63 &  36.31 \\
Kazakhstan         &       2,534 &       0.03 &   4.66 &     9.0 &  255.27 &  251.43 &  266.32 &      0.57 &  38.50 \\
Mexico             &       2,228 &       0.03 &   4.22 &     3.0 &  226.17 &  216.68 &  257.25 &      0.39 &  35.87 \\
\bottomrule
\end{tabular}

    \begin{tablenotes}
		\footnotesize
    		\item \textit{Notes:} Rows are sorted by median hourly earnings, including bonuses (PPP-adjusted USD). Qualification (qual.) -- ISCED 1997 level. Problem
    \end{tablenotes}
\end{table}
\section{Measurement frameworks output and selection}\label{sec:1_param}

Section~\ref{sec:1_measures} reviews a variety of skill and educational mismatch measurement frameworks. Although only six primary measures are considered --- \textit{job analysis} (JA), \textit{realised matches} (RM), \textit{indirect} (ISA) and \textit{direct self-assessments} (DSA), \textit{Pellizzari-Fichen} (PF), and \textit{Allen-Levels-van-der-Velden} (ALV) --- some frameworks require the researcher to select parameter values that determine the resulting mismatch classification (e.g., measure of central tendency and number of standard deviations for RM). Including the whole variety of specifications in the econometric analysis is impractical for this study. Therefore, a preliminary investigation is required to select the relevant measures for analysing the Mincer function. This is done in three steps.

First, country-specific labour shares of \mbox{under-,} \mbox{well-,} and over-matched workers are mapped onto a colour spectrum, where each data point’s position corresponds to its place in the distribution, and countries are ordered by their median earnings. Second, correlation matrices are presented in a similar manner. Finally, log-earnings are regressed on a set of controls and all mismatch measure specifications using the \textit{least absolute shrinkage and selection operator} (Lasso). The graphical analysis aims to reveal country-level patterns between earnings and mismatch levels, and to identify the degree of similarity among classification results. Meanwhile, Lasso is employed to select the specifications most helpful in predicting earnings by minimising the \textit{mean squared prediction error} (MSPE) and eliminating the rest. The final part of this section examines heterogeneity in the outputs of the selected mismatch measures across various worker characteristics, guiding further research.

\subsection{Mismatch shares cross-country comparison}\label{sec:1_shares}

\begin{figure}[!htbp]
	\centering
	\caption{Country-specific mismatch shares: Selected measures}
	\label{fig:sel_hm}
	{\includegraphics[scale=0.4, trim=0cm 0cm 0cm 0cm, clip]{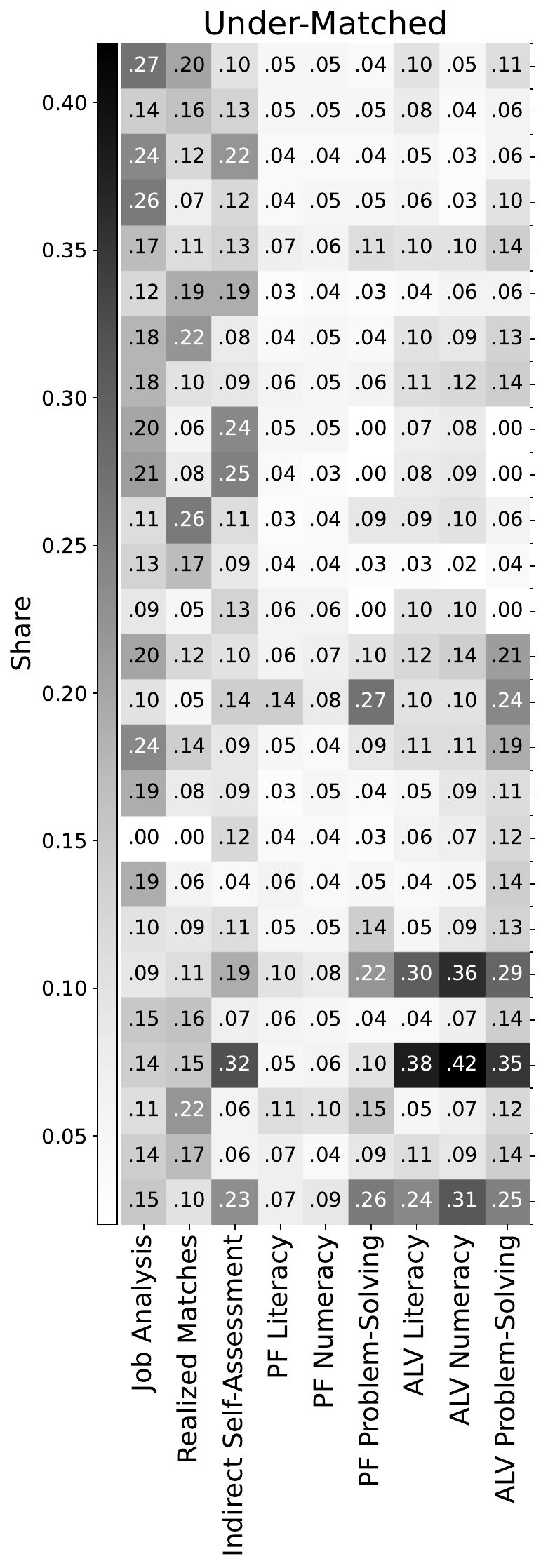}}
	{\includegraphics[scale=0.4, trim=0cm 0cm 0cm 0cm, clip]{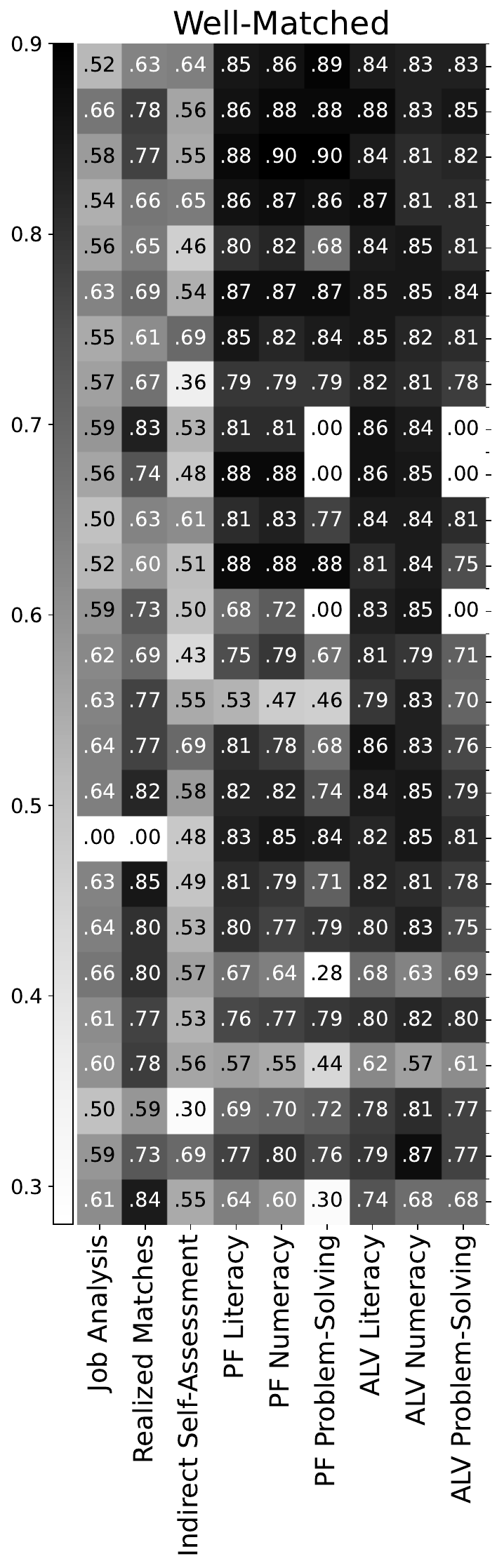}}
	{\includegraphics[scale=0.4, trim=0cm 0cm 0cm 0cm, clip]{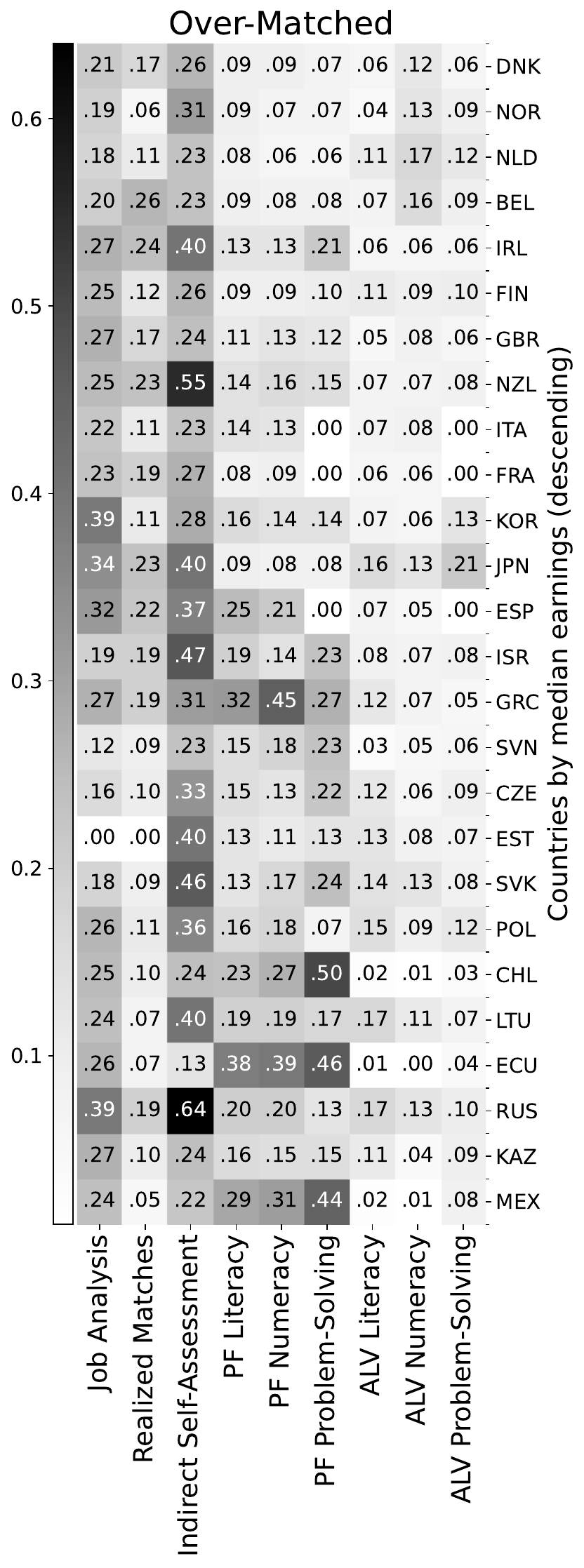}}
	\begin{tablenotes}
		\footnotesize
    		\item \textit{Notes:} Rows are sorted by the country-specific median hourly earnings, including bonuses (PPP corrected USD).
    		\item Selected parameters:
    		\item \begin{itemize}
    			\item[] Realised matches (RM): mode $\pm 1$ standard deviation threshold \\
    			ISA Indirect self-assessment (ISA): 1 year gap \\
    			Pellizzari-Fichen (PF): 5th and 95th percentile critical values \\
    			Allen-Levels-van-der-Velden (ALV): 1.5 points difference in the z-scores
    		\end{itemize}
	\end{tablenotes}
\end{figure}

Let us begin the discussion with a brief overview of the mismatch measures selected for further analysis in this section. Figure~\ref{fig:sel_hm} presents country-specific shares of under-, well- and over-matched workers, computed using three educational mismatch measures --- JA, RM, and ISA --- and two skill mismatch measures --- PF and ALV --- applied to the literacy, numeracy, and problem-solving scores. The countries are sorted in descending order by their median earnings. The key takeaway from this figure is that educational and skill mismatch may exhibit opposing relationships with earnings. Specifically, the shares of well-matched workers, according to JA, are generally larger in countries with lower median earnings. A similar pattern is observed in the shares of well-matched workers computed using RM. ISA does not display this pattern: countries in the top quartile of the earnings distribution have high well-matched shares, although those with relatively low well-matched shares are still clustered in the third quartile. On the other hand, the PF measure applied to each of the three PIAAC skill measures produces shares that decrease as countries’ median earnings decrease. A similar trend is observed in the shares produced by ALV in the lower quartile of the median earnings distribution. These patterns are reflected to varying extents in the shares of under- and over-matched workers.\footnote{Among other countries, Chile, Ecuador, and Mexico may be of particular interest to researchers due to their relatively high levels of ALV-computed under-skilling and PF-computed over-skilling.} Although the statistical significance of these associations is uncertain, the fact that educational and skill mismatch is often described in the literature as equivalent makes this a sufficient starting point for the investigation.

One potential economic interpretation of the observed patterns is consistent with the signalling view of education \citep{spence1973job} expanded to accommodate cases in which education is attained by a substantial share of low-skilled workers. Specifically, because advanced degrees are more commonly attained in higher-income countries\footnote{The country-level correlation between median earnings and years of education and between median earnings and ISCED level is 0.31 and 0.13, respectively.}, they may be less useful to employers for differentiating between low- and high-skilled workers. The necessity to observe skills directly, rather than relying on educational credentials, provides employers with an incentive to invest in their own screening mechanisms, leading to greater educational mismatch and lower skill mismatch. By contrast, when education is relatively rare, it may be perceived as a strong signal of skill, thereby contributing to lower educational mismatch and greater skill mismatch. However, the plausibility of this explanation is challenging to test empirically and is therefore left for future research.

The remainder of this section reviews each of the six considered measures and their alternative parameter specifications. Note that JA, which compares workers’ ISCO skill levels derived from their highest qualification with the ISCO skill level requirements for the corresponding occupational groups defined by the International Labour Office (\citeyear{international2012international}), is the only indicator without alternative specifications. The other frameworks include alternative parameter settings that require careful examination.

Figure~\ref{fig:rm_hm} maps the shares produced by RM using information on workers’ highest attained qualifications (ISCED) converted into ISCO skill levels and current occupational groups.\footnote{See Appendix~\ref{sec:A_heatmaps}.} Section~\ref{sec:1_measures} notes that the literature features multiple specifications of RM, which differ in the measure of central tendency and the number of standard deviations used to determine classification thresholds. The figure shows that for mean-based RM, a switch from the commonly used one standard deviation cutoff to either 0.5 or 1.5 standard deviations results in considerable differences between the shares; for example, the UK's over-education varies between 5\% and 29\%. In contrast, mode-based RM is less sensitive to the threshold value, despite the coefficient on standard deviation varying over a wider range of 0.1 to 2, compared with 0.5 to 1.5 in mean-based RM. The corresponding values for the UK range between 10\% and 22\%. This is due to the mode’s lower sensitivity to outliers and the nature of the ISCO skill levels metric, which has only four categories. Consequently, the effect of changing the threshold value on mode-derived mismatch shares is less smooth. The resulting shares for mean-based RM show no clear association with median earnings. However, for mode-based RM, the majority of countries in the upper half of the median earnings distribution (Denmark, Belgium, Ireland, etc.) feature relatively low shares of well-educated workers, primarily reflected in over-education and, to a lesser extent, under-education shares.

Figure~\ref{fig:isa_hm} presents mismatch shares computed using ISA.\footnote{See Appendix~\ref{sec:A_heatmaps}.} This measure compares the number of years of formal education required for a worker’s current job with their highest level of education attained, imputed into years of education. The resulting mismatch classification is determined by the gap between the years of attained and required education that allows a worker to be labelled as ``well-educated''. The results are computed for gaps of 1 to 5 years. As expected, the shares of well-matched workers are higher when a wider gap is used; e.g., 69\% of UK workers are classified as well-educated with a gap of 4 years, and 91\% with a gap of 1 year. Although clustering of countries with lower mismatch shares in the third quartile of the median earnings distribution is not immediately apparent in Figure~\ref{fig:sel_hm}, comparing shares across alternative gap sizes makes it more visible.

The mismatch shares reported in Figure~\ref{fig:dsa_hm} are computed using DSA. This measure classifies workers based on their responses to two questions: (i) ``Do you feel that you have the skills to cope with more demanding duties than those you are required to perform in your current job?'' and (ii) ``Do you feel that you need further training in order to cope well with your present duties?''. A respondent is considered over-skilled if they answer affirmatively to (i) and negatively to (ii), under-skilled --- negatively to (i) and affirmatively to (ii), and well-skilled if they answer negatively to both questions. The column labelled \textit{Double-Positive (DP) error} in Figure~\ref{fig:dsa_hm} contains the share of respondents who answered affirmatively to both questions. A cautious interpretation of these cases is to treat them as measurement error and exclude them from the analysis. However, as shown in the figure, such responses account for more than a quarter of the sample in most countries. Consequently, in some cases, a researcher may prefer to classify these respondents as well-skilled, arguing that they are likely uncertain about their skill potential and therefore are either well-matched or borderline cases. This interpretation can be considered a separate skill mismatch measure, referred to as \textit{``relaxed'' direct self-assessment}. As implied above, this approach rests on the following assumption:

\begin{assumption}[Relaxed DSA Homogeneity]
\label{asm:relaxed_dsa}
	The distribution of matching outcomes is identical for respondents with affirmative answers to both (i) and (ii) and respondents with negative answers to both (i) and (ii).
\end{assumption}
The plausibility of this assumption rests on the similarity between the well-skilled respondents, who answer both (i) and (ii) negatively, and the DP ones, who answer affirmatively. As shown in Table~\ref{tab:dsa_count}, while well-skilled respondents have median earnings of \$13.52 per hour, DP respondents earn a lower median wage of \$10.82 per hour. In addition, the well-skilled and DP respondents display a qualification gap, which mirrors that between the over- and under-skilled groups (ISCED levels 6 and 8, respectively). Finally, workers classified as well-matched under DSA are the oldest of the four groups, with a mean age of 44.35, whereas those with double-positive responses are the youngest, with a mean age of 37.88. Taken together, these findings weaken the argument that DP respondents are comparable to the well-matched group.

One reason why relaxed DSA may be both interesting and problematic in the context of PIAAC data is the potential association between DP errors and country-specific median earnings. Figure~\ref{fig:dsa_hm} suggests that the proportion of DP respondents increases with higher median earnings. Notably, unlike the results of education-based measures, the shares of well-matched individuals computed using regular DSA show a positive relationship with median earnings. However, when DP observations are integrated into the well-matched group in relaxed DSA, the large number of DP errors reverses this association, resulting in a negative relationship between well-matched shares and earnings. Although this may be considered an additional finding relevant for users of PIAAC data, it has limited economic applicability because the theoretical basis for the association is unclear. Moreover, it indicates a potential violation of Assumption~\ref{asm:relaxed_dsa}, which would render relaxed DSA unsuitable for the purposes of this study.

The first 18 columns in Figures~\ref{fig:pf_well_hm}, \ref{fig:pf_under_hm}, and \ref{fig:pf_over_hm} present the shares of well-, under- and over-skilled workers, respectively, computed using the Pellizzari-Fichen framework. The measure is applied separately to each of the three skill scores. Additionally, the shares are computed using both regular and relaxed DSA: \textit{regular PF} employs classification thresholds based on the skill distributions of respondents who answered negatively to both questions (i) and (ii), whereas \textit{relaxed PF} utilises respondents who gave the same answer to both questions, whether ``yes'' or ``no''. Finally, the original \citet{pellizzari2017new} classification is based on the 5th and 95th percentiles of the well-matched workers’ skill score distribution. This specification is referred to as the 10\% version, and its results are compared to those of the 5\% and 20\% versions of the PF framework, corresponding to the 2.5th and 75.5th percentiles and the 10th and 90th percentiles, respectively.

Let us first consider the output classification of regular PF. The effect of alternative classification thresholds is similar to that observed in the cases of ISA and RM: ``stricter'' specifications produce higher under- and over-matched shares and, consequently, lower well-matched shares. However, unlike ISA and RM, the PF measure yields mismatch shares that are negatively associated with country-specific median earnings. Figure~\ref{fig:pf_well_hm} shows that countries with higher median earnings tend to have a larger share of well-skilled workers. The results are broadly consistent across the three skill domains, except in a few countries, such as Chile and Mexico, where the problem-solving well-matched shares differ considerably from the literacy and numeracy ones. As shown in Figures~\ref{fig:pf_under_hm} and \ref{fig:pf_over_hm}, this pattern is mainly driven by over-skilling and, to a lesser extent, under-skilling: the share of over-skilled workers is negatively associated with country-specific median earnings.

It is worth noting, however, that this pattern is not observed in the shares produced by the relaxed versions of PF. On the contrary, the relaxed shares exhibit noticeably lower variance across countries compared with the results of all other labour mismatch measures. This is because DP errors appear to be negatively correlated with median earnings (see Figure~\ref{fig:dsa_hm}). Recall the distinction between regular and relaxed PF: in the relaxed version, the pool of well-skilled workers --- whose skill distribution is used to compute the classification thresholds --- is expanded to include respondents who gave double-positive answers. This expansion increases the variance of the skill distributions, which in turn leads to wider thresholds and, consequently, lower shares of under- and over-skilled individuals and higher shares of well-skilled individuals. Since countries with lower median earnings tend to have a larger share of DP respondents, their PF shares are affected to a greater extent. The reason behind the negative correlation between the share of DP respondents and median earnings, however, remains unclear.

The final set of specifications presented in Figures~\ref{fig:pf_well_hm} to \ref{fig:pf_over_hm} is produced using ALV. As with the previous measures, specifications that allow for larger differences in z-scores classify fewer workers as mismatched. The shares of the well-matched are generally similar across countries, except for those in the lowest quartile of the median earnings distribution, where the shares display a pattern similar to PF. In the case of under-skilling, this pattern is less pronounced, with higher values concentrated around the middle of the median earnings distribution. Notably, Chile, Ecuador, and Mexico exhibit particularly high shares of under-skilled workers. Interestingly, these countries also record the highest shares of DP errors (see Figure~\ref{fig:dsa_hm}). Since DSA and ALV are not based on common variables, this may indicate a broader measurement error issue. Consequently, the shares of over-skilled workers are relatively small in these three outlier countries, while in the remaining countries, the pattern is the reverse of that observed for under-skilling, with higher values concentrated in the tails of the distribution.

\subsection{Correlation analysis}\label{sec:1_correlation}

Figure~\ref{fig:corr_hm} presents correlation matrices for a subset of the labour mismatch measures discussed above. For RM, the mode $\pm 1$ standard deviation is employed as a classification threshold, since it is robust to outliers and widely used in the literature. For ISA, the 1-year gap is adopted, while for PF, the 5th and 95th percentiles are used as critical values, and for ALV, a 1.5-point z-score allowance is applied. Each of these specifications reflects common practice in the literature. As noted above, alternative thresholds primarily affect the ``strictness'' of the mismatch measure and the representativeness of the resulting shares. Nonetheless, it remains essential to compare the performance of regular and relaxed measures, given the association between DP errors and country-specific median earnings.

The figure suggests three main clusters in the correlation matrix: education-based measures, skill-based measures, and DSAs. Among them, JA's classification aligns most closely with RM, particularly in the case of under-education, where the correlation reaches 70\%. JA also shows a moderate correlation with ISA, reaching 25\% for over-matching, whereas the associations with the other measures are weaker or conflicting. RM exhibits a similar level of association with ISA, with a correlation of 21\% for over-matching, but weaker associations with the remaining measures. ISA, in turn, appears to have only mild to negligible associations with both the skill-based measures and DSA.

Within the skill-based cluster, each of the three regular PF measures exhibits its highest correlation with the corresponding relaxed specification: 70\%, 74\%, and 62\% for the under-matched classifications in literacy, numeracy, and problem-solving, respectively. PF mismatch in literacy and numeracy also display a considerably stronger association with each other than with problem-solving, with correlations of 58\% compared to 41\% and 38\% for the well-matched classifications. By contrast, ALV shows much weaker correlations, both with the PF measures (22\%, 16\%, and 19\% for literacy, numeracy, and problem-solving, respectively) and across its own skill domains (29\% between literacy and numeracy, 22\% between literacy and problem-solving, and 11\% between numeracy and problem-solving). Lastly, both versions of the DSA classification exhibit very low correlations, with all reported values ranging between $-4\%$ and $6\%$. This result is unsurprising, as DSA typically categorises the majority of respondents as over-matched, whereas the other measures tend to classify most respondents as well-matched.

The absence of correlation between the classification output produced by either regular or relaxed DSA and that of other measures makes it difficult to place DSA within the categories of either education-based or skill-based mismatch. This poses a challenge for interpreting econometric results derived from DSA in comparison with those based on alternative measures. Furthermore, the large share of DP errors raises the risk of selection bias. A potential solution is to adopt the relaxed version of DSA, in which DP errors are reclassified as well-matched, provided that the assumption of relaxed DSA homogeneity holds. However, as described in the previous subsection, Table~\ref{tab:dsa_count} indicates that Assumption~\ref{asm:relaxed_dsa} may be difficult to defend. On the other hand, the averages displayed in the table also suggest that omitting the DP errors from the analysis may reduce the representation of younger, better-educated, but lower-earning workers. Moreover, since the distribution of workers across the DSA categories does not appear to have an intuitive association with age, education, or earnings, the direction of the potential selection bias is ambiguous. Overall, although regular and relaxed DSA may serve as useful statistical predictors, they provide little basis for drawing inferences about labour mismatch.

Although DSA’s classifications of over- and under-matching may be considered unreliable, its definition of well-matched workers can still serve as input for other measures. Both the regular and relaxed PF frameworks generate alternative classifications, yet their correlation is sufficiently strong to treat them as belonging to the same class of skill mismatch measures. Nevertheless, the economic interpretation of the differences between the regular and relaxed specifications hinges on how DP respondents are understood, which requires separate investigation. For this reason, the relaxed PF measures are not pursued further in the present analysis. This leaves a final selection of six labour mismatch measures, as outlined at the beginning of the section.

\subsection{Out-of-sample prediction performance}\label{sec:1_oos}

This section compares labour mismatch measures based on their out-of-sample (OOS) prediction performance. To this end, each measure specification is used as input to Lasso (\cite{frank1993statistical}; \cite{tibshirani1996regression}), a regularised regression framework. Following the formulation in \citet{ahrens2020lassopack}, the Lasso estimator minimises the mean squared error subject to both overall and predictor-specific penalties on the absolute values of the coefficient estimates ($l_1$ regularisation), denoted by $\lambda$ and $\psi$, respectively:
\begin{equation}
	\boldsymbol{\hat{\beta}}_{lasso}(\lambda) = \text{arg} \min \frac{1}{n}\sum_{i=1}^n(y_i-\mathbf{x}_i'\boldsymbol{\beta})^2+\frac{\lambda}{n}\sum_{j=1}^p\psi_j |\beta_j|.
\end{equation}

Lasso may exclude some of the $p$ covariates by setting their coefficients to zero, thereby serving as a model-selection tool. The selection is determined by $\lambda$, which is tuned to optimise out-of-sample (OOS) prediction performance via 10-fold cross-validation (CV) \citep{geisser1975predictive}. Specifically, the data is split into 10 folds, each of which serves as the validation set once, while the model is estimated on the remaining nine folds. 

\begin{figure}[h!]
    \caption{Mean-squared prediction error as a function of $\ln(\lambda)$}
    \label{fig:cvlasso}
    \centering
    \includegraphics[scale=0.7, trim=0cm 0cm 0cm 0cm, clip]{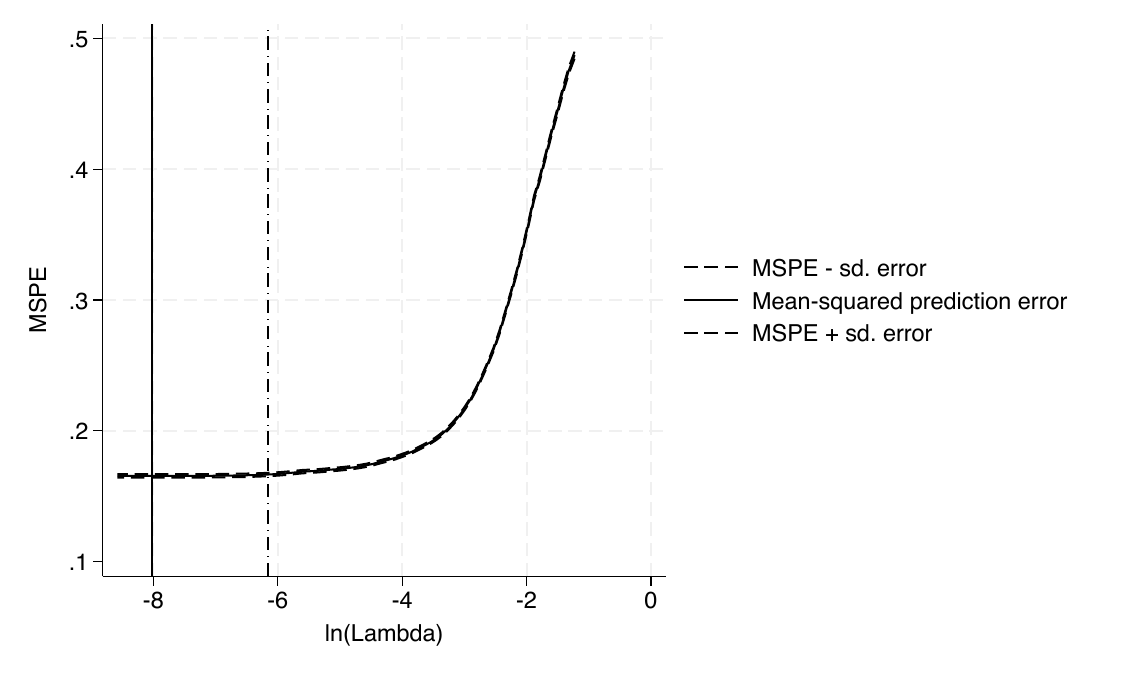}
    \begin{tablenotes}
		\footnotesize
    		\item \textit{Notes:} The solid and dashed vertical lines correspond to $\lambda_{lopt}$ and $\lambda_{lse}$, respectively.
    \end{tablenotes}
\end{figure}

After predictions are obtained for all validation sets, the mean squared prediction error (MSPE) is calculated by comparing predicted and actual values. The value of $\lambda$ that minimises the MSPE, denoted $\lambda_{lopt}$ \textit{(lambda-optimum)}, is selected for analysis (see Figure~\ref{fig:cvlasso}). Alternatively, a more parsimonious specification can be chosen by selecting the largest $\lambda$ within one \textit{standard error (SE)} of the one minimising the MSPE, $\lambda_{lse}$ \textit{(lambda-SE)}. Finally, because several mismatch measures are highly correlated, the model is estimated using adaptive Lasso \citep{zou2006adaptive}, which applies predictor-specific penalty loadings $\psi_j = 1/|\hat{\beta}_{0,j}|^{\theta}$ and requires less restrictive assumptions about the correlation between the predictors.

Lasso is applied to the following regression equation:
\begin{equation}
	\ln{w_i} = \alpha + \boldsymbol{\delta} \mathbf{M}_i + \boldsymbol{\beta} \mathbf{X}_i + \varepsilon_i,
\end{equation}
where $\ln w_i$ denotes the natural log of individual earnings, $\mathbf{M}_i$ is the vector of mismatch measures, each defined under a set of alternative parameter specifications\footnote{\textit{JA}; \textit{RM} (mean-based with $0.5$, $1$, or $1.5$ SD thresholds; mode-based with $0.1$, $1$, or $2$ SD thresholds); \textit{DSA} (regular or relaxed); \textit{ISA} (1 to 5 year gaps); \textit{PF} (regular or relaxed; literacy, numeracy, or problem-solving based, with $0.025$, $0.05$, or $0.1$ quantile thresholds); \textit{ALV} (literacy, numeracy, or problem-solving based, with $1$, $1.5$, or $2$ z-score gaps).} and $\mathbf{X}_i$ is the vector of control variables.\footnote{Controls include personal characteristics (\textit{female, age, age$^2$, tenure}), migration variables (\textit{migrated after age 16, years in country}), country and industry fixed effects, and the variables used in constructing the mismatch measures (\textit{education, ISCO required, years of schooling, years to get job, not challenged, need training, literacy, numeracy, problem-solving, literacy use, numeracy use, problem-solving use}).} The base category for each mismatch measure is the well-matched group. Hence, the vector $\boldsymbol{\delta}$ contains the coefficients for the under- and over-matching. This exercise aims to identify the non-zero elements of $\hat{\boldsymbol{\delta}}$, i.e., the mismatch specifications that contribute to optimal out-of-sample prediction performance.

The model training results are displayed in Table~\ref{tab:1_lasso_models}. The most ``generous'' specification, despite the highest overall penalty parameter $\lambda_{lopt} \approx 0.007$, is the Adaptive Lasso, followed by the standard Lasso with $\lambda_{lopt} \approx 0.0003$, and the most parsimonious model with $\lambda_{lse} \approx 0.003$. Although each model employs a different value of $\lambda$, the resulting MSPE remains close to $0.16$, with only slight variation. This suggests that covariates excluded by the more selective models contribute little to the OOS predictive performance of the Mincer earnings function.

\begin{table}[h!]
    \centering
    \footnotesize
    \caption{Lasso models}\label{tab:1_lasso_models}
    \begin{tabular}{l|c c c}
    \toprule
                            & Lasso with $\lambda_{lopt}$ & Lasso with $\lambda_{lse}$ & Adaptive Lasso with $\lambda_{lopt}$ 	\\
    \midrule
         Lambda             & 0.00027	&  0.00255  & 0.00645	\\
         MSPE               & 0.16530   &  0.16696	& 0.16537 	\\
         Standard deviation & 0.00167   &  0.00161  & 0.00167 	\\
    \hline \\[-1.8ex]
    	Selected controls   & 122 		& 79 		& 142 \\
         Observations		& 42,922	    & 42,922.   & 42,922.   \\
    \bottomrule
    \end{tabular}
    \begin{tablenotes}
    \footnotesize
    	\item \textit{Notes:} The lists of selected controls and post-estimation results are available in Table~\ref{tab:cvlasso}, Appendix~\ref{sec:A_cvlasso}. The list of 144 initial regressors includes \textit{JA}; \textit{RM} (mean-based with $0.5$, $1$, or $1.5$ SD thresholds; mode-based with $0.1$, $1$, or $2$ SD thresholds); \textit{DSA} (regular or relaxed); \textit{ISA} (1 to 5 year gaps); \textit{PF} (regular or relaxed; literacy, numeracy, or problem-solving based, with $0.025$, $0.05$, or $0.1$ quantile thresholds); \textit{ALV} (literacy, numeracy, or problem-solving based, with $1$, $1.5$, or $2$ z-score gaps); personal characteristics (\textit{female, age, age$^2$, tenure}); migration variables (\textit{migrated after age 16, years in country}); country and industry fixed effects, and the variables used in constructing the mismatch measures (\textit{education, ISCO required, years of schooling, years to get job, not challenged, need training, literacy, numeracy, problem-solving, literacy use, numeracy use, problem-solving use}).
    \end{tablenotes}
\end{table}

Table~\ref{tab:cvlasso} presents the post-estimation results for the three Lasso specifications.\footnote{See Appendix~\ref{sec:A_cvlasso}.} The Adaptive Lasso selects the most encompassing model. In addition to the majority of the controls, it includes JA, both DSA specifications, ISA with different gaps, a range of mean- and mode-based RM, as well as numerous PF and ALV specifications. In total, 142 out of 144 variables are selected. The standard Lasso with $\lambda_{lopt}$ yields a more conservative model: while most controls are retained, all DSA specifications are excluded, and most other measures lose variants that differ only by the strictness of classification thresholds. Finally, the most parsimonious model is obtained with the Lasso using $\lambda_{lse}$. This excludes some controls, leaving only one or two specifications per measure and mismatch type (i.e., under- and over-matching). Lasso slightly favours the stricter versions of ISA, mode-based RM over mean-based RM, regular over relaxed PF, and it assigns only a few non-zero coefficients to ALV. 

Overall, the results suggest that the labour mismatch measure makes a meaningful contribution to the OOS prediction performance of the Mincer earnings function. Despite including all measure components among the controls, even the most parsimonious Lasso does not entirely exclude any measures except DSA. This indicates that there are no strong data-driven reasons to revise the selection made in this section thus far.

\subsection{Heterogeneity}\label{sec:1_heterogeneity}

The final part of this section presents the graphical heterogeneity analysis of earnings and labour mismatch across key worker characteristics: gender, age, migration status, education, and skills.\footnote{See Appendix~\ref{sec:A_heterogeneity}.} The analysis compares the distributions of mismatch shares among workers within groups defined by these characteristics. Because country-level shares would not provide a sufficient number of groups, the calculations are conducted for 346 markets, where a market is defined as a country-specific industry with at least 30 workers. The mismatch shares are based on the selected measures: JA, mode-based RM with $\pm1$ standard deviation thresholds, 1-year gap ISA, and the $0.05$ quantile PF and 1 z-score AVL applied to literacy, numeracy, and problem-solving scores.

To begin with, Figures~\ref{fig:wage_gaps_1} and \ref{fig:wage_gaps_2} present the distributions of log earnings across various worker categories. The top panel of Figure~\ref{fig:wage_gaps_1} indicates a negative wage gap between male and female workers. The middle panel illustrates positive wage gaps between older workers (over 45 years) and middle-aged workers (30 to 44 years), as well as between middle-aged and younger workers (under 30 years). The bottom panel suggests a positive wage gap between migrant and local workers. Figure~\ref{fig:wage_gaps_2} provides a similar breakdown by workers' education, measured using the four ISCO skill levels and PIAAC skill scores categorised into quartiles. The top panel shows positive wage gaps between each consecutive ISCO skill level, indicating that more educated workers tend to earn higher wages. A similar pattern is observed in the three panels corresponding to literacy, numeracy, and problem-solving. The only exception occurs for workers in the fourth quartile of the problem-solving distribution, who do not appear to earn more than those in the third quartile.

Figures~\ref{fig:hst_u_by_gender} and~\ref{fig:hst_o_by_gender} present the shares of under- and over-matched workers by gender. The top three panels of Figure~\ref{fig:hst_u_by_gender} show negative gender mismatch gaps, indicating that women are less likely to be under-educated than men. In contrast, the PF- and ALV-based plots display little to no differences in under-matching between male and female workers. The over-matching distributions in Figure~\ref{fig:hst_o_by_gender} reveal that women are more likely to be over-educated. The ALV-based results resemble those of the education-based measures, although with smaller magnitudes. By contrast, the PF-based shares display clear negative gender mismatch gaps, indicating that women are less likely to be over-skilled. Overall, these findings suggest that gender is an important determinant of labour mismatch, and that the observed associations vary depending on the chosen measure of mismatch.

Another commonly used control that shows a clear association with mismatch outcomes is age. The top three panels of Figure~\ref{fig:hst_u_by_age} indicate that older workers are more likely to be under-educated than both middle-aged and younger workers. A similar pattern is observed in the under-skilling shares computed using PF and ALV, with the exception of numeracy-based ALV, where the differences in the distributions are less pronounced. These results are mirrored in the shares of over-matched workers presented in Figure~\ref{fig:hst_o_by_age}. Older workers are less likely to be over-educated and over-skilled, which is supported by all measures except RM. One possible explanation is that younger workers face greater difficulty in securing jobs that match their level of skill and education due to limited experience. This highlights the distinction between skill and experience, underscoring the importance of controlling for both.

Let us now consider the differences in mismatch distributions by migration status. The under-education shares in Figure~\ref{fig:hst_u_by_mig} show a negative migration mismatch gap, indicating that migrant workers are less likely to be under-educated. In contrast, both sets of skill-based measures display positive mismatch gaps. This conflicting pattern is mirrored in the over-matching shares in Figure~\ref{fig:hst_o_by_mig}, which show that migrant workers are more likely to be over-educated but less likely to be over-skilled. This conclusion is consistently supported across all measures. The higher incidence of over-education among migrant workers may be explained by language and cultural barriers that prevent them from accessing jobs aligned with their qualifications. Since respondents are classified as migrants only if they entered a country after the age of 16, the higher level of under-skilling may reflect differences in school systems between the country of birth and the country of residence. However, this does not fully explain the consistent divergence observed between the education- and skill-based measures.

Finally, Figures~\ref{fig:hst_w_by_skill} to~\ref{fig:hst_o_by_sl} show limited variation in educational mismatch across the quartiles of the skill distributions and skill mismatch across ISCO skill levels.

\section{The market-level error components model}\label{sec:1_method}

The statistical analysis is based on \citeauthor{verdugo1989impact}'s (\citeyear{verdugo1989impact}) version of the Mincer earnings function, as defined in Equation~\ref{eq:oru_class}. To address the concern regarding unobserved heterogeneity raised in Section~\ref{sec:1_param}, the equation is modified using an \textit{error components model}. 

Suppose $N = \sum_{j=1}^M n_j$ workers $i$ are employed in $M$ labour markets $j$, where each $j$ corresponds to a country-specific industry according to the International Standard Industrial Classification of All Economic Activities Revision 4 (ISIC). Then, Equation~\ref{eq:oru_class} can be expressed as:  
\begin{equation}\label{eq:oru_vv_panel}
	\ln{w_{ij}} = a + \beta_o\, over_{ij} + \beta_u\, under_{ij} + \boldsymbol{\gamma}\mathbf{X}_{ij} + \mu_j + \eta_{ij},
\end{equation}
where $\ln{w_{ij}}$ is the natural logarithm of earnings, $over_{ij}$ is a binary variable equal to $1$ if worker $i$ is over-matched and $0$ otherwise, $under_{ij}$ is defined similarly, $\mathbf{X}_{ij}$ is a vector of controls, $\mu_j$ captures market-level unobserved heterogeneity, and $\eta_{ij}$ is an idiosyncratic error term such that $E[over_{ij}\eta_{ij}] = E[under_{ij}\eta_{ij}] = 0$.

Since $\mu_j$ may be correlated with the regressors of interest, estimating (\ref{eq:oru_vv_panel}) using \textit{pooled ordinary least squares} (POLS) allows the $\hat{\beta}$s to be influenced by unobserved heterogeneity, potentially resulting in biased and inconsistent estimates. Nevertheless, $\hat{\beta}_o^{POLS}$ and $\hat{\beta}_u^{POLS}$ remain informative, as they capture variation both within and between markets, and therefore constitute the first specification for the analysis.

To obtain more accurate estimates, the model must correct for $\mu_j$. Following \citet{mundlak1978pooling}, let us specify the relationship between unobserved heterogeneity and the regressors as
\begin{equation}\label{eq:mundlak_uh}
	\mu_j = \phi_o\, \overline{over}_{j} + \phi_u\, \overline{under}_{j} + \boldsymbol{\omega}\overline{\mathbf{X}}_j + \nu_j,
\end{equation}
where $\bar{\cdot}_j$ denotes the average across individuals $i$ within market $j$. Substituting this expression for $\mu_j$ into (\ref{eq:oru_vv_panel}) yields
\begin{align}\label{eq:oru_mundlak}
\begin{split}
	\ln{w_{ij}} = \alpha +& \beta_o\, over_{ij} + \beta_u\, under_{ij} + \boldsymbol{\gamma}\mathbf{X}_{ij} \\ 
	+& \phi_o\, \overline{over}_{j} + \phi_u\, \overline{under}_{j} + \boldsymbol{\omega}\overline{\mathbf{X}}_j + \nu_j + \eta_{ij}.
\end{split}
\end{align}

It can be shown that applying \textit{generalised least squares} (GLS) to estimate the $\beta$s in Equation~\ref{eq:oru_mundlak} is equivalent to the \textit{fixed effects} (FE) estimation of Equation~\ref{eq:oru_vv_panel}, i.e. $\hat{\beta}^{FE} = \hat{\beta}^{GLS}$ \citep{mundlak1978pooling}. These estimates are important for two reasons. First, $\hat{\beta}_o^{FE}$ and $\hat{\beta}_u^{FE}$ capture the individual-level association between labour mismatch and earnings. Second, comparing the FE and POLS estimates provides insight into the direction and magnitude of the bias induced by unobserved market-specific factors. Consequently, the FE estimation constitutes the second specification.

To investigate the variation in unobserved factors further, it is helpful to consider the \textit{between estimates} (BE) of the $\beta$s. These can be obtained by summing the GLS coefficient estimates for the variable of interest and its corresponding market-level average in Equation~\ref{eq:oru_mundlak}, i.e. $\hat{\beta}^{BE} = \hat{\beta}^{GLS} + \hat{\phi}^{GLS}$ \citep{mundlak1978pooling}. It is important to note that the interpretation of the BE differs from those produced by POLS and FE. Since $over_{ij}$ and $under_{ij}$ are binary variables, $\overline{over_{j}}$ and $\overline{under_{j}}$ represent the shares of workers for whom the respective variables take the value of one. Hence, $\hat{\beta}_o^{BE}$ ($\hat{\beta}_u^{BE}$) corresponds to the average change in earnings associated with moving from a market with no over-matched (under-matched) workers to a fully over-matched (under-matched) one, while the constant $\alpha^{BE}$ represents the average log-earnings in a fully well-matched market.\footnote{An arguably more helpful approach is to divide the BE by 10 and interpret them as the increase in earnings associated with a 10\% rise in average over- or under-matching.} Although the $\beta^{BE}$ cannot be directly compared with $\beta^{FE}$ and $\beta^{POLS}$, they are essential for understanding the market-level association between mismatch and earnings, and therefore complement the FE estimates in the second specification.

Finally, the analysis employs \textit{random effects} (RE) as an alternative approach to combining the within and between sources of variation. The RE estimates are obtained by applying GLS to the quasi-demeaned version of Equation~\ref{eq:oru_vv_panel}: 
\begin{multline}\label{eq:oru_mundlak_re}
	\ln{w}_{ij} - \theta \overline{\ln{w}}_{j} = \alpha(1-\theta) + \beta_o (over_{ij} - \theta\overline{over}_{j}) + \beta_u (under_{ij} - \theta\overline{under}_{j}) \\
	+ \boldsymbol{\gamma}(\mathbf{X}_{ij} - \theta\overline{\mathbf{X}}_{j}) + \mu_j(1-\theta) + \eta_{ij} - \theta\overline{\eta}_{j}.
\end{multline} 
Unlike POLS, which gives equal weight to the fixed effects and between estimates, the RE estimator produces a matrix-weighted average of the $\hat{\beta}^{FE}$ and $\hat{\beta}^{BE}$ estimates, with the inverse of their respective variances as weights \citep{baltagi2008econometric}. Since the RE approach may yield efficiency gains relative to POLS, the $\hat{\beta}^{RE}$ estimates are presented as the third specification. The RE estimation relies on orthogonality of the mismatch variables with respect to $\mu_j$, i.e., $E[over_{ij}\mu_{j}] = E[under_{ij}\mu_{j}] = 0$. This assumption is plausible only if the controls are sufficiently correlated with $\mu_j$ to account for the relevant unobserved heterogeneity. The RE estimates are therefore interpreted as complementary to, rather than substitutes for, the FE results. Furthermore, the plausibility of the RE assumption is evaluated by testing whether the coefficients on the averages of the mismatch variables in the \citeauthor{mundlak1978pooling} model, $\phi_o$ and $\phi_u$, are jointly zero.

It should also be noted that $\hat{\beta}_o$ and $\hat{\beta}_u$ may be biased in the presence of other sources of endogeneity, such as omitted variables or measurement error. For instance, if workers' education and skill variables do not fully explain their productivity, highly productive workers may end up in more demanding jobs, thereby biasing the under-matching wage premiums upward. Similarly, workers who are less productive than their education and skills suggest may appear over-matched and push the corresponding wage penalties further away from zero. Moreover, missing values in problem-solving and double-positive DSA errors may induce additional endogeneity in $\hat{\beta}_o$ and $\hat{\beta}_u$, as discussed in Sections~\ref{sec:1_data} and \ref{sec:1_param}.
\section{Results}\label{sec:1_analysis}

This section presents the results from estimating \citeauthor{verdugo1989impact}’s (\citeyear{verdugo1989impact}) adaptation of the Mincer earnings function, which are reported in Tables \ref{tab:ja} to \ref{tab:alvp}.\footnote{See Appendix \ref{sec:A_error_comp}.} Each of the nine tables corresponds to one of the selected mismatch measures.\footnote{The selected measures are: \textit{job analysis} (JA); \textit{realised matches} with a mode $\pm 1$ standard deviation threshold (RM); \textit{indirect self-assessments} with a one-year gap (ISA); \textit{Pellizzari-Fichen} with the 5th and 95th percentile critical values (PF); and \textit{Allen-Levels-van-der-Velden} with a $1.5$ difference in the z-score (ALV).} The three estimators described in Section \ref{sec:1_method} --- POLS, \citet{mundlak1978pooling} FE, and RE --- employ the same set of controls: gender, age, age-squared, tenure, tenure-squared, immigration status, health status and its interaction with age, number of children and its interaction with gender, years of attained education, and the natural logarithms of literacy, numeracy, and problem-solving scores. Additionally, the Mundlak FE model includes market-specific averages of both the regressors of interest and the covariates, enabling the calculation of BE.\footnote{See Section~\ref{sec:1_method} for more details.}

The first column of Table~\ref{tab:1_sum_res_over} summarises the POLS results for over-matching. The coefficient estimates for JA and RM are relatively similar, indicating that being over-educated is associated with respective wage penalties of $-10.3\%$ and $-6.3\%$, with confidence intervals ranging from $-14\%$ to $-1\%$. The coefficient for ISA is larger, suggesting a $-18.8\%$ reduction in earnings, with a confidence interval of $[-23\%, -15\%]$. The wage penalties linked to over-skilling in literacy, numeracy, and problem-solving, as measured by PF, are concentrated around $-30.3\%$, with confidence intervals spanning $-37\%$ to $-23\%$. The corresponding ALV estimates display greater variation: over-skilling in literacy is associated with the largest wage penalty of $-38.9\%$, $[-43\%, -34\%]$, whereas over-skilling in numeracy and problem-solving is linked to smaller reductions in earnings of $-8.2\%$ and $-18.8\%$, with confidence intervals of $[-13\%, -3\%]$ and $[-23\%, -17\%]$, respectively. Turning to under-matching POLS estimates presented in the first column of Table~\ref{tab:1_sum_res_under}, the education-based measures yield similar estimates, indicating a $10.5\%$ wage premium for JA and RM, and a slightly higher premium of $12.7\%$ for ISA. The corresponding confidence intervals range from $6\%$ to $16\%$. By contrast, the results for skill-based measures are more distinct. Over-skilling in literacy and numeracy, as defined by PF, is associated with substantial wage premiums of $24\%$ and $25.4\%$, respectively, with confidence intervals ranging from $19\%$ to $32\%$. The problem-solving estimate is smaller, at $13.2\%$ with a confidence interval of $[7\%, 19\%]$. The ALV estimate for literacy is similar to that of PF, at $22.6\%$, $[19\%, 26\%]$, whereas the estimates for numeracy and problem-solving are considerably lower, at $5.7\%$ and $8.7\%$, with confidence intervals between $1\%$ and $12\%$. Finally, it should be noted that the estimates of \citeauthor{oster2019unobservable}'s $\delta$ (\citeyear{oster2019unobservable}) exceed the rule-of-thumb benchmark of 1 for all mismatch measures, indicating that selection on unobservables would need to be stronger than selection on observables to fully explain the estimated wage penalties.

\begin{table}[h!]
    \centering
    \footnotesize
    \caption{Over-education and over-skilling wage penalties}\label{tab:1_sum_res_over}
    \begin{tabular}{c | c| c c c c}
    \toprule
    & & POLS        & FE       & BE        & RE  \\
    \midrule
        \multirow{6}{*}{\parbox{3cm}{\centering Education-based measures}}   & JA     &          -0.103$^{***}$&          -0.126$^{***}$&          -0.185         &          -0.126$^{***}$\\
                                                                             &        &   [-0.14,-0.07]         &   [-0.14,-0.11]         &    [-0.56,0.19]         &   [-0.14,-0.11]         \\
                                                                             & RM     &          -0.063$^{*}$  &          -0.164$^{***}$&           0.306         &          -0.162$^{***}$\\
                                                                             &        &   [-0.12,-0.01]         &   [-0.18,-0.15]         &    [-0.15,0.77]         &   [-0.18,-0.14]         \\
                                                                             & ISA    &          -0.188$^{***}$&          -0.130$^{***}$&          -0.783$^{***}$&          -0.131$^{***}$\\
                                                                             &        &   [-0.23,-0.15]         &   [-0.14,-0.12]         &   [-1.08,-0.49]         &   [-0.14,-0.12]         \\
        \hline \\[-1.8ex]
        \multirow{12}{*}{\parbox{3cm}{\centering Skill-based measures}}      & PF-L   &          -0.289$^{***}$&          -0.051$^{***}$&           0.089         &          -0.058$^{***}$\\
                                                                             &        &   [-0.33,-0.25]         &   [-0.08,-0.02]         &    [-0.31,0.48]         &   [-0.08,-0.03]         \\
                                                                             & PF-N   &          -0.329$^{***}$&          -0.079$^{***}$&          -0.051         &          -0.086$^{***}$\\
                                                                             &        &   [-0.37,-0.29]         &   [-0.10,-0.06]         &    [-0.40,0.30]         &   [-0.11,-0.06]         \\
                                                                             & PF-P   &          -0.283$^{***}$&          -0.057$^{***}$&           0.047         &          -0.064$^{***}$\\
                                                                             &        &   [-0.34,-0.23]         &   [-0.08,-0.03]         &    [-0.27,0.36]         &   [-0.09,-0.04]         \\
                                                                             & ALV-L  &          -0.389$^{***}$&          -0.161$^{***}$&          -1.731$^{***}$&          -0.171$^{***}$\\
                                                                             &        &   [-0.43,-0.34]         &   [-0.18,-0.14]         &   [-2.23,-1.23]         &   [-0.19,-0.15]         \\
                                                                             & ALV-N  &          -0.082$^{**}$ &          -0.075$^{***}$&           0.675         &          -0.077$^{***}$\\
                                                                             &        &   [-0.13,-0.03]         &   [-0.09,-0.06]         &    [-0.05,1.40]         &   [-0.09,-0.06]         \\
                                                                             & ALV-P  &          -0.198$^{***}$&          -0.092$^{***}$&          -1.356$^{***}$&          -0.096$^{***}$\\
                                                                             &        &   [-0.23,-0.17]         &   [-0.11,-0.08]         &   [-2.12,-0.59]         &   [-0.11,-0.08]         \\
    \bottomrule
    \end{tabular}
    \begin{tablenotes}
\footnotesize
\item \textit{Note:} * \( p < 0.05 \), ** \( p < 0.01 \), *** \( p < 0.001 \). 95\% confidence intervals in brackets. Measures of mismatch: JA is job analysis with the education requirement defined using the ISCO \citep{international2012international}; RM is realised matches with the occupation-specific mode $\pm1$ standard deviation classification thresholds; ISA is indirect self-assessment allowing for a one-year gap; PF is Pellizzari-Fichen with the critical values set at the 5th and 95th percentiles; ALV is Allen-Levels-Van-der-Velden with the allowed z-score difference of 1.5. PF and ALV are calculated using literacy (L), numeracy (N) and problem-solving (P). Controls include gender, age, age-squared, tenure, tenure-squared, immigration status, health status and its interaction with age, number of children and its interaction with gender, years of attained education, and the natural logarithms of literacy, numeracy, and problem-solving scores. The estimates are computed using PIAAC's final full sample weight (SPFWT0). Full results for the nine measures are reported in Tables~\ref{tab:ja} to \ref{tab:alvp}, respectively.
\end{tablenotes}
\end{table}

\begin{table}[h!]
    \centering
    \footnotesize
    \caption{Under-education and under-skilling wage penalties}\label{tab:1_sum_res_under}
    \begin{tabular}{c | c| c c c c}
    \toprule
    & & POLS        & FE       & BE        & RE  \\
    \midrule
        \multirow{6}{*}{\parbox{3cm}{\centering Education-based measures}}   & JA     &           0.106$^{***}$&           0.074$^{***}$&           0.413         &           0.076$^{***}$\\
                                                                             &        &     [0.07,0.15]         &     [0.06,0.09]         &    [-0.04,0.86]         &     [0.06,0.10]         \\
                                                                             & RM     &           0.103$^{***}$&           0.071$^{***}$&           0.773$^{***}$&           0.073$^{***}$\\
                                                                             &        &     [0.06,0.15]         &     [0.05,0.09]         &     [0.32,1.22]         &     [0.05,0.09]         \\
                                                                             & ISA    &           0.127$^{***}$&           0.098$^{***}$&           1.587$^{***}$&           0.098$^{***}$\\
                                                                             &        &     [0.09,0.16]         &     [0.08,0.12]         &     [1.09,2.09]         &     [0.08,0.12]         \\
        \hline \\[-1.8ex]
        \multirow{12}{*}{\parbox{3cm}{\centering Skill-based measures}}      & PF-L   &           0.240$^{***}$&           0.140$^{***}$&          -1.031$^{*}$  &           0.144$^{***}$\\
                                                                             &        &     [0.19,0.29]         &     [0.11,0.17]         &   [-2.04,-0.02]         &     [0.12,0.17]         \\
                                                                             & PF-N   &           0.254$^{***}$&           0.146$^{***}$&          -0.250         &           0.150$^{***}$\\
                                                                             &        &     [0.19,0.32]         &     [0.12,0.17]         &    [-1.36,0.86]         &     [0.12,0.18]         \\
                                                                             & PF-P   &           0.132$^{***}$&           0.090$^{***}$&          -0.570$^{*}$  &           0.091$^{***}$\\
                                                                             &        &     [0.07,0.19]         &     [0.07,0.11]         &   [-1.03,-0.11]         &     [0.07,0.11]         \\
                                                                             & ALV-L  &           0.226$^{***}$&           0.137$^{***}$&           1.938$^{***}$&           0.141$^{***}$\\
                                                                             &        &     [0.19,0.26]         &     [0.12,0.15]         &     [1.43,2.45]         &     [0.12,0.16]         \\
                                                                             & ALV-N  &           0.057$^{*}$  &           0.100$^{***}$&           1.317$^{***}$&           0.101$^{***}$\\
                                                                             &        &     [0.01,0.10]         &     [0.08,0.12]         &     [0.75,1.88]         &     [0.08,0.12]         \\
                                                                             & ALV-P  &           0.087$^{***}$&           0.068$^{***}$&          -0.587         &           0.069$^{***}$\\
                                                                             &        &     [0.06,0.12]         &     [0.05,0.08]         &    [-1.46,0.29]         &     [0.06,0.08]         \\
    \bottomrule
    \end{tabular}
    \begin{tablenotes}
\footnotesize
\item \textit{Note:} * \( p < 0.05 \), ** \( p < 0.01 \), *** \( p < 0.001 \). 95\% confidence intervals in brackets. Measures of mismatch: JA is job analysis with the education requirement defined using the ISCO \citep{international2012international}; RM is realised matches with the occupation-specific mode $\pm1$ standard deviation classification thresholds; ISA is indirect self-assessment allowing for a one-year gap; PF is Pellizzari-Fichen with the critical values set at the 5th and 95th percentiles; ALV is Allen-Levels-Van-der-Velden with the allowed z-score difference of 1.5. PF and ALV are calculated using literacy (L), numeracy (N), and problem-solving (P). Controls include gender, age, age-squared, tenure, tenure-squared, immigration status, health status and its interaction with age, number of children and its interaction with gender, years of attained education, and the natural logarithms of literacy, numeracy, and problem-solving scores. The estimates are computed using PIAAC's final full sample weight (SPFWT0). Full results for the nine measures are reported in Tables~\ref{tab:ja} to \ref{tab:alvp}, respectively.
\end{tablenotes}
\end{table}

The second column of Tables~\ref{tab:1_sum_res_over} and \ref{tab:1_sum_res_under} presents the results of the FE estimation using \citet{mundlak1978pooling} formulation. When unobserved heterogeneity is accounted for, the under-matching coefficients for most mismatch measures decline in magnitude but retain their signs and statistical significance. Exceptions to this pattern include the over-education coefficients for JA and RM, as well as the under-skilling coefficient for ALV-Numeracy, which increase in magnitude by approximately $2\%$, $10\%$, and $4\%$, respectively. The reductions in magnitude vary across measures, ranging from $1\%$ to $25\%$, indicating differences in the extent of heterogeneity bias affecting the POLS coefficients. It is also worth noting that the FE estimates appear more precise than those from POLS, producing narrower confidence intervals.

The RE results, reported in the last column of Tables~\ref{tab:1_sum_res_over} and \ref{tab:1_sum_res_under}, show little to no difference from the FE estimates. However, a joint test of the coefficients on the market-level mean values of the mismatch variables rejects the null of joint insignificance at the 5\% level for all mismatch measures except JA and PF-Numeracy. This indicates that market-level unobserved heterogeneity is correlated with most mismatch variables, implying that the estimates from the FE model should be preferred, although the similarity between FE and RE results suggests that the bias is likely to be small. The finalised individual-level estimates for JA, RM, and ISA indicate that over-education is associated with wage penalties of $-12.6\%$, $-16.4\%$, and $-13\%$, respectively, while under-education is linked to the corresponding wage premiums of $7.4\%$, $7.1\%$, and $9.8\%$. The JA and ISA results are close to the average of 74 estimates reviewed by \citet{mcguinness2018skills}, who report an over-education wage penalty of $-13.6\%$, while the RM estimate is slightly larger in magnitude. For under-education, all three estimates are consistent with the range reported by \citet{njifen2024education}, who find wage premiums between $2.9\%$ and $9.5\%$. According to PF, the wage differentials associated with over- and under-skilling average $-6.2\%$ and $12.5\%$ across skills, respectively. The corresponding average estimates computed with ALV are $-10.9\%$ and $10.2\%$. For over-skilling, the average PF estimate is about $-1.3\%$ greater in magnitude than the $-7.5\%$ wage penalty reported across 11 studies by \citet{mcguinness2018skills}, whereas the $8\%$ under-skilling wage premium documented by \citet{perry2014can} is close to the ALV average.

Focusing solely on market-level variation, the BE are calculated as the sum of the mismatch coefficients and the coefficients on the corresponding market-level means. As displayed in the third column of Tables~\ref{tab:1_sum_res_over} and \ref{tab:1_sum_res_under}, the BE for ISA indicates that a $10\%$ increase in educational mismatch at the market level is associated with changes in average earnings of $-7.83\%$ for over-education. The corresponding estimates for RM and ISA under-education suggest associated premiums of $7.73\%$ and $15.87\%$, respectively. The results based on JA provide no evidence of an association between either under- or over-education and earnings at the market level. The coefficients based on the PF measure are largely insignificant, with the exception of under-skilling in literacy and problem-solving, which are associated with market-level wage penalties of $-10.31\%$ and $-5.7\%$, respectively. The ALV results suggest that a $10\%$ increase in literacy and problem-solving over-skilling is linked to reductions in earnings of $-17.31\%$ and $-13.56\%$, respectively. The corresponding ALV coefficients for under-skilling in literacy and numeracy indicate market-level wage premiums of $19.38\%$ and $13.17\%$, while the confidence interval for problem-solving is too wide for the estimate to be considered statistically significant.

Overall, the results are consistent with the literature, which reports individual-level wage penalties associated with over-education and over-skilling, as well as wage premiums linked to under-education and under-skilling \citep{perry2014can, mcguinness2018skills, cassidy2024increasing, jacobs2022wage, ragoobur2022education, wen2023educational, morsy2019youth, njifen2024education, cultrera2022educational, zhao2025wage}. Furthermore, the magnitude of the RE coefficients is comparable to that reported in other studies. The FE estimates suggest that unobserved heterogeneity increases the variance of the estimates and tends to drive them away from zero; however, the magnitude of this effect varies across measures, and in some cases, the bias appears to push the coefficients towards zero. Finally, the BE indicate that the direction of market-level associations can vary across measures, even within the same type of mismatch.

Additional inferences can be drawn from the RE estimates for the coefficients on the control variables reported in Appendix~\ref{sec:A_error_comp}. The coefficient for gender is significantly negative for all mismatch measures, at around $-12\%$, broadly consistent with the EU gender wage gap reported by \citet{commission2022gender}. The coefficients for age and tenure suggest that each additional year is associated with approximately $3.3\%$ and $1\%$ higher earnings, respectively, while their squared polynomials are significantly negative but extremely small in magnitude, indicating minimal non-linearity. Migrating after the age of 16 is associated with earnings that are between $-6.3\%$ and $-4.4\%$ lower compared with local workers. An extra year of education corresponds to an increase in earnings of between $5\%$ and $7\%$. Finally, the skill variables suggest that a $10\%$ increase in literacy, numeracy, and problem-solving scores is associated with earnings gains of $0.9\%$-$3.4\%$, $2.6\%$-$4.6\%$, and $1.2\%$-$3\%$, respectively, across the models. Overall, there is no evidence of substantial differences in the behaviour of the control variables across models using different measures of labour mismatch.
\section{Conclusion and research perspectives}\label{sec:1_conclusion}

This study investigates the implications of using skill-based and education-based measures of labour mismatch for estimating \citeauthor{verdugo1989impact}’s (\citeyear{verdugo1989impact}) version of the Mincer earnings function. The analysis examines the relationship between mismatch and earnings at both the individual and market levels. To this end, an error-components model is applied to cross-sectional data from 26 countries, drawn from the OECD Survey of Adult Skills (PIAAC).

The results suggest wage penalties linked to over-education and over-skilling, whereas under-education and under-skilling are associated with wage premiums. This pattern is consistent with the findings of \citet{verdugo1989impact} and with the broader empirical literature \citep{mcguinness2018skills}. However, the coefficient estimates vary considerably across indicators, demonstrating that both the choice of input variable and the measure of labour mismatch are critical for the analysis. Furthermore, fixed-effects estimation reveals that the Pooled OLS coefficients on mismatch variables are biased by market-level unobserved heterogeneity to different degrees and in some cases in opposite directions (e.g. JA- and RM-based vs. ISA-based over-education).

Given the above, two main conclusions can be drawn: (i) education and skill mismatch should be distinguished both conceptually and empirically, as using one as a proxy for the other is unlikely to yield accurate results in the analysis of the Mincer earnings function; and (ii) the coefficient estimates for under- and over-matching are sensitive to the choice of mismatch measure. The first point implies that researchers should exercise caution when drawing economic inferences from analyses of mismatch in qualifications, training, or other education-related characteristics. The second point suggests that the outcomes of mismatch analysis are often specific to the chosen measure, and their interpretation depends on the underlying features of the workers and jobs that the measure captures. Although the fact that different mismatch measures may yield distinct results might appear inconvenient, it also provides an opportunity to extract insights into labour markets from multiple perspectives. If adopted, this approach could support more effective policymaking, as interventions directed at skills (e.g., assisting career transitions) may prove effective according to skill-based indicators, even if they are unlikely to improve the outcomes of education-based measures, and vice versa. However, since the specific worker and job characteristics driving these divergences across measures remain unclear, the procedure for using them to draw robust economic inferences is not yet well defined.

Finally, the heterogeneity analysis highlights two crucial patterns. First, the graphical analysis shows that women are more likely to be over-educated, whereas men are more likely to be under-educated and over-skilled. Second, immigrant workers are less likely to be under-educated and over-skilled, yet more likely to be over-educated and under-skilled. A more rigorous investigation is required to uncover the sources of these disparities across gender and immigration status. Developing such analyses, alongside strategies for drawing further insights from the diversity of mismatch measures, is left for future research.

\clearpage
\printbibliography

\clearpage
\appendix
\section{Additional summary statistics}\label{app:1_summary_stats}

\begin{table}[htbp!]\centering
	\centering
    \footnotesize
    \caption{Frequencies and averages: Problem-solving missing values}
    \label{tab:psl_missing}
    \begin{tabular}{l|rrrrrrrrr}
\toprule
{} &  N &  Frac. &  Median   & Median    & Mean &     Mean &     Mean & Mean & Mean \\ 
&   &           & earn.  & qual.     & lit.  & num.  & pr.slv. & gender & age \\
\midrule
Observed data        &    50,121 &    0.71 &  13.10 & 7.00 & 279.21 & 278.31 & 279.70 & 1.53 & 35.37 \\
France, Spain, Italy &    7,563  &    0.11 &  13.60 & 6.00 & 263.79 & 260.42 & NaN & 1.49 & 40.78 \\
Other missing data  &    13,254 &    0.19 &  7.36  & 5.00 & 249.60 & 236.11 & NaN & 1.49 & 43.63 \\
\bottomrule
\end{tabular}

\end{table}

\begin{table}[htbp!]\centering
	\centering
    \footnotesize
    \caption{Frequencies and averages: JA}
    \label{tab:ja_count}
    \begin{tabular}{l|rrrrrrrrr}
\toprule
{} &  N &  Frac. &  Median   & Median    & Mean &     Mean &     Mean & Mean & Mean \\ 
&   &           & earn.  & qual.     & lit.  & num.  & pr.slv. & gender & age \\
\midrule
Over-matched  &   16,890 &    0.23 &  11.20 &    11.0 &  271.17 &  265.72 &  282.32 &      0.56 &  38.15 \\
Under-matched &   11,635 &    0.16 &  15.65 &     6.0 &  274.75 &  273.75 &  280.82 &      0.51 &  42.17 \\
Well-matched  &   40,363 &    0.56 &  12.09 &     6.0 &  270.83 &  267.44 &  278.63 &      0.49 &  39.10 \\
NaN   &    3,675 &    0.05 &   7.36 &     NaN &  277.60 &  274.28 &  274.48 &      0.58 &  41.02 \\
\bottomrule
\end{tabular}

\end{table}

\begin{table}[htbp!]\centering
	\centering
    \footnotesize
    \caption{Frequencies and averages: RM}
    \label{tab:rm_mode_1_count}
    \begin{tabular}{l|rrrrrrrrr}
\toprule
{} &  N &  Frac. &  Median   & Median    & Mean &     Mean &     Mean & Mean & Mean \\ 
&   &           & earn.  & qual.     & lit.  & num.  & pr.slv. & gender & age \\
\midrule
Over-matched  &           9,988 &           0.14 &  13.68 &    12.0 &  285.30 &  282.29 &  288.30 &      0.57 &  38.02 \\
Under-matched &           9,328 &           0.13 &  15.36 &     7.0 &  273 &  270.14 &  281.06 &      0.53 &  42.37 \\
Well-matched  &          49,572 &           0.68 &  11.67 &     6.0 &  268.54 &  264.83 &  277.76 &      0.50 &  39.08 \\
NaN   &           3,675 &           0.05 &   7.36 &     NaN &  277.60 &  274.28 &  274.48 &      0.58 &  41.02 \\
\bottomrule
\end{tabular}

\end{table}

\begin{table}[htbp!]\centering
	\centering
    \footnotesize
    \caption{Frequencies and averages: ISA}
    \label{tab:isa_1_count}
    \begin{tabular}{l|rrrrrrrrr}
\toprule
{} &  N &  Frac. &  Median   & Median    & Mean &     Mean &     Mean & Mean & Mean \\ 
&   &           & earn.  & qual.     & lit.  & num.  & pr.slv. & gender & age \\
\midrule
Over-matched  &      23,065 &       0.32 &  10.59 &     9.0 &  275.48 &  271.30 &  281.33 &      0.54 &  37.94 \\
Under-matched &       9,065 &       0.12 &  13.93 &     5.0 &  262.70 &  259.51 &  274.49 &      0.46 &  42.50 \\
Well-matched  &      38,384 &       0.53 &  13.08 &     7.0 &  274.55 &  271.69 &  280.40 &      0.52 &  39.59 \\
NaN   &       2,049 &       0.03 &   6.61 &     3.0 &  221.91 &  213.39 &  252.34 &      0.45 &  40.35 \\
\bottomrule
\end{tabular}

\end{table}

\begin{tablenotes}
		\footnotesize
    		\item \textit{Notes:} Earnings -- hourly earnings including bonuses (PPP corrected USD). Qualification (qual.) -- ISCED 1997 level.
\end{tablenotes}

\newpage

\begin{table}[htbp!]\centering
	\centering
    \footnotesize
    \caption{Frequencies and averages: DSA}
    \label{tab:dsa_count}
    \begin{tabular}{l | rrrrrrrrr}
\toprule
{} &  N &  Frac. &  Median   & Median    & Mean &     Mean &     Mean & Mean & Mean \\ 
&   &           & earn.  & qual.     & lit.  & num.  & pr.slv. & gender & age \\
\midrule
DP Error &    20,949 &     0.29 &  10.82 &     8.0 &  268.50 &  264.49 &  277.56 &      0.49 &  37.88 \\
Over-matched     &    39,985 &     0.55 &  12.26 &     6.0 &  273.55 &  270.57 &  280.60 &      0.51 &  39.40 \\
Under-matched    &     4,891 &     0.07 &  13.45 &     8.0 &  280.11 &  275 &  285.51 &      0.57 &  40.08 \\
Well-matched     &     6,738 &     0.09 &  13.52 &     6.0 &  266.51 &  262.84 &  275.38 &      0.57 &  44.35 \\
\bottomrule
\end{tabular}

\end{table}

\begin{table}[htbp!]\centering
	\centering
    \footnotesize
    \caption{Frequencies and averages: PF-Literacy}
    \label{tab:pf_lit_005_count}
    \begin{tabular}{l|rrrrrrrrr}
\toprule
{} &  N &  Frac. &  Median   & Median    & Mean &     Mean &     Mean & Mean & Mean \\ 
&   &           & earn.  & qual.     & lit.  & num.  & pr.slv. & gender & age \\
\midrule
Over-matched  &           10,621 &            0.15 &  11.06 &    10.0 &  312.28 &  304.86 &  311.67 &      0.47 &  35.56 \\
Under-matched &            3,896 &            0.05 &  10.55 &     6.0 &  196.14 &  197.09 &  223.60 &      0.51 &  43.22 \\
Well-matched  &           58,004 &            0.80 &  12.29 &     6.0 &  269.57 &  266.51 &  276.95 &      0.53 &  39.94 \\
NaN   &              42 &            0 &  10.10 &     6.0 &  262.92 &  272.97 &  276.31 &      0.38 &  36.31 \\
\bottomrule
\end{tabular}

\end{table}

\begin{table}[htbp!]\centering
	\centering
    \footnotesize
    \caption{Frequencies and averages: PF-Numeracy}
    \label{tab:pf_num_005_count}
    \begin{tabular}{l|rrrrrrrrr}
\toprule
{} &  N &  Frac. &  Median   & Median    & Mean &     Mean &     Mean & Mean & Mean \\ 
&   &           & earn.  & qual.     & lit.  & num.  & pr.slv. & gender & age \\
\midrule
Over-matched  &           10,819 &            0.15 &  10.94 &     9.0 &  303.28 &  310.26 &  306.65 &      0.40 &  36.26 \\
Under-matched &            3,754 &            0.05 &  10.70 &     6.0 &  204.85 &  188.36 &  227.95 &      0.57 &  42.57 \\
Well-matched  &           57,948 &            0.80 &  12.32 &     6.0 &  270.37 &  265.76 &  276.82 &      0.53 &  39.87 \\
NaN   &              42 &            0 &  10.10 &     6.0 &  262.92 &  272.97 &  276.31 &      0.38 &  36.31 \\
\bottomrule 
\end{tabular}

\end{table}

\begin{table}[htbp!]\centering
	\centering
    \footnotesize
    \caption{Frequencies and averages: PF-Problem-Solving}
    \label{tab:pf_psl_005_count}
    \begin{tabular}{l|rrrrrrrrr}
\toprule
{} &  N &  Frac. &  Median   & Median    & Mean &     Mean &     Mean & Mean & Mean \\ 
&   &           & earn.  & qual.     & lit.  & num.  & pr.slv. & gender & age \\
\midrule
Over-matched  &            7,843 &            0.11 &  11.11 &    11.0 &  301.14 &  300.65 &  318.43 &      0.45 &  32.98 \\
Under-matched &            3,797 &            0.05 &   9.94 &     6.0 &  222.45 &  221.10 &  211.39 &      0.52 &  41.67 \\
Well-matched  &           39,118 &            0.54 &  13.99 &     9.0 &  280.60 &  279.76 &  278.74 &      0.54 &  38.50 \\
NaN   &           21,805 &            0.30 &  10.15 &     6.0 &  254.32 &  244.65 &  255.32 &      0.49 &  42.96 \\
\bottomrule 
\end{tabular}

\end{table}

\begin{tablenotes}
		\footnotesize
    		\item \textit{Notes:} Earnings -- hourly earnings including bonuses (PPP corrected USD). Qualification (qual.) -- ISCED 1997 level.
    \end{tablenotes}

\newpage

\begin{table}[htbp!]\centering
	\centering
    \footnotesize
    \caption{Frequencies and averages: ALV-Literacy}
    \label{tab:pf_lit_005_count}
    \begin{tabular}{l|rrrrrrrrr}
\toprule
{} &  N &  Frac. &  Median   & Median    & Mean &     Mean &     Mean & Mean & Mean \\ 
&   &           & earn.  & qual.     & lit.  & num.  & pr.slv. & gender & age \\
\midrule
Over-matched  &            6,254 &            0.09 &   8.77 &     6.0 &  311.45 &  301.87 &  307.56 &      0.54 &  34.82 \\
Under-matched &            6,791 &            0.09 &  11.93 &     6.0 &  215.43 &  215.59 &  239.90 &      0.46 &  41.82 \\
Well-matched  &           59,436 &            0.82 &  12.44 &     7.0 &  274.18 &  270.92 &  281.45 &      0.52 &  39.70 \\
NaN   &              82 &            0 &  11.43 &     8.5 &  264.40 &  258.36 &  285.57 &      0.59 &  40.30 \\
\bottomrule
\end{tabular}

\end{table}

\begin{table}[htbp!]\centering
	\centering
    \footnotesize
    \caption{Frequencies and averages: ALV-Numeracy}
    \label{tab:pf_num_005_count}
    \begin{tabular}{l|rrrrrrrrr}
\toprule
{} &  N &  Frac. &  Median   & Median    & Mean &     Mean &     Mean & Mean & Mean \\ 
&   &           & earn.  & qual.     & lit.  & num.  & pr.slv. & gender & age \\
\midrule
Over-matched  &            6,123 &            0.08 &  13.97 &    11.0 &  315.54 &  320.15 &  309.88 &      0.51 &  37.47 \\
Under-matched &            7,289 &            0.10 &   9.74 &     6.0 &  222.69 &  210.72 &  248.58 &      0.51 &  38.92 \\
Well-matched  &           59,095 &            0.81 &  12.11 &     6.0 &  273.43 &  270.17 &  279.46 &      0.52 &  39.75 \\
NaN   &              56 &            0 &  10.64 &     7.0 &  260.97 &  249.36 &  279.27 &      0.52 &  37.96 \\
\bottomrule
\end{tabular}

\end{table}

\begin{table}[htbp!]\centering
	\centering
    \footnotesize
    \caption{Frequencies and averages: ALV-Problem-Solving}
    \label{tab:pf_psl_005_count}
    \begin{tabular}{l|rrrrrrrrr}
\toprule
{} &  N &  Frac. &  Median   & Median    & Mean &     Mean &     Mean & Mean & Mean \\ 
&   &           & earn.  & qual.     & lit.  & num.  & pr.slv. & gender & age \\
\midrule
Over-matched  &            4,459 &            0.06 &  11.37 &     7.0 &  307.91 &  305.83 &  320.64 &      0.55 &  31.99 \\
Under-matched &            6,481 &            0.09 &  10.97 &     6.0 &  235.84 &  235.31 &  227.36 &      0.51 &  41.59 \\
Well-matched  &           40,216 &            0.55 &  13.72 &     9.0 &  282.89 &  282.12 &  283.52 &      0.53 &  37.92 \\
NaN   &           21,407 &            0.30 &  10.24 &     6.0 &  254.60 &  244.84 &  256.77 &      0.49 &  43.10 \\
\bottomrule
\end{tabular}

\end{table}

\begin{tablenotes}
		\footnotesize
    		\item \textit{Notes:} Earnings -- hourly earnings including bonuses (PPP corrected USD). Qualification (qual.) -- ISCED 1997 level.
    \end{tablenotes}

\clearpage
\section{Mismatch measures output}\label{sec:A_heatmaps}

\begin{figure}[!htbp]
	\centering
	\caption{Country-specific mismatch shares: RM}
	\label{fig:rm_hm}
	{\includegraphics[scale=0.4, trim=0cm 0cm 0cm 0cm, clip]{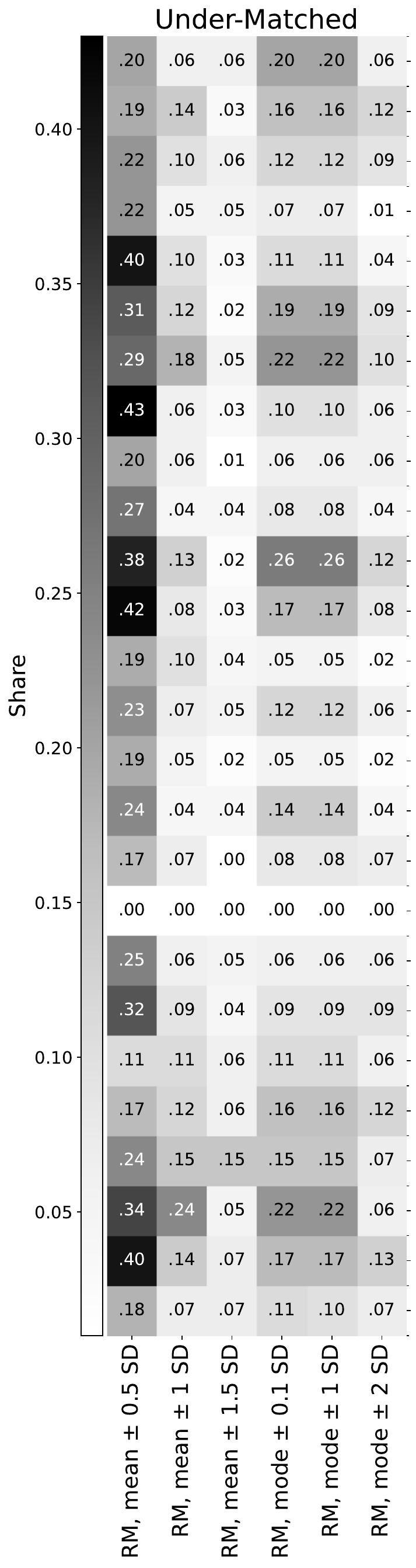}}
	{\includegraphics[scale=0.4, trim=0cm 0cm 0cm 0cm, clip]{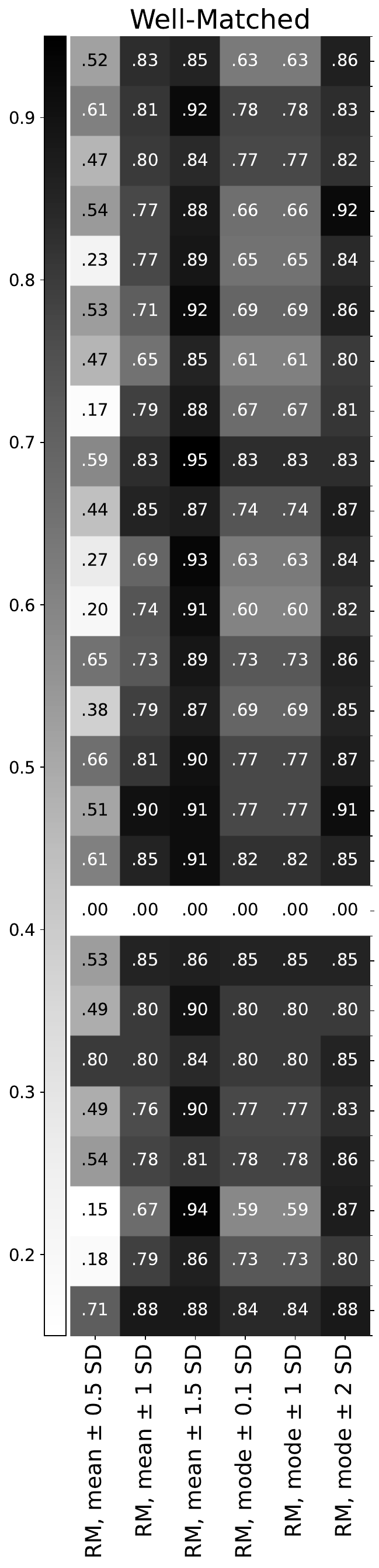}}
	{\includegraphics[scale=0.4, trim=0cm 0cm 0cm 0cm, clip]{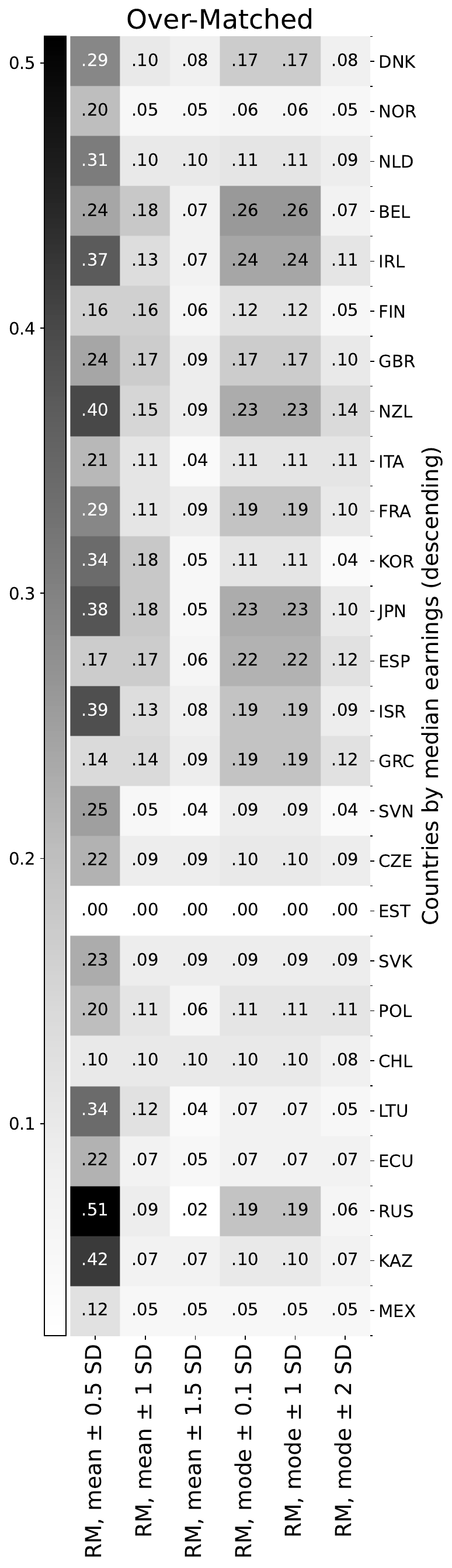}}
	\begin{tablenotes}
		\footnotesize
    		\item \textit{Notes:} Rows are sorted by the country-specific median of hourly earnings including bonuses (PPP corrected USD). Education data is missing for Estonia.
	\end{tablenotes}
\end{figure}

\begin{figure}[!htbp]
	\centering
	\caption{Country-specific mismatch shares: ISA}
	\label{fig:isa_hm}
	{\includegraphics[scale=0.4, trim=0cm 0cm 0cm 0cm, clip]{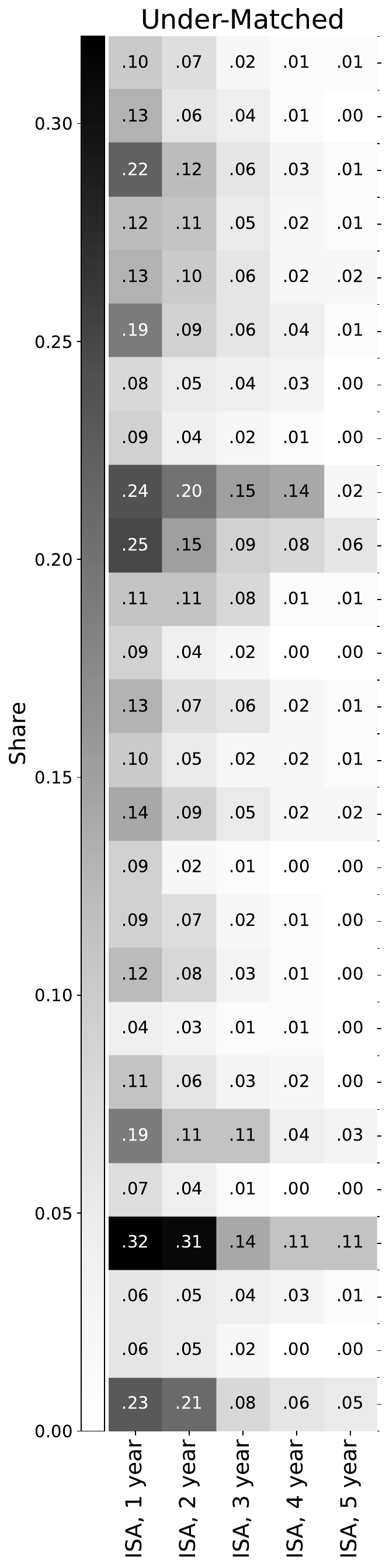}}
	{\includegraphics[scale=0.4, trim=0cm 0cm 0cm 0cm, clip]{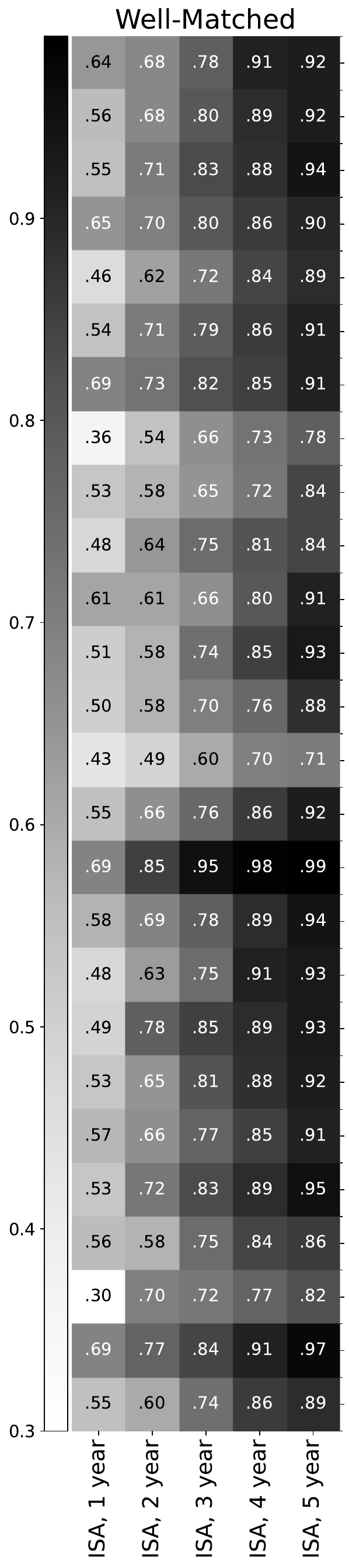}}
	{\includegraphics[scale=0.4, trim=0cm 0cm 0cm 0cm, clip]{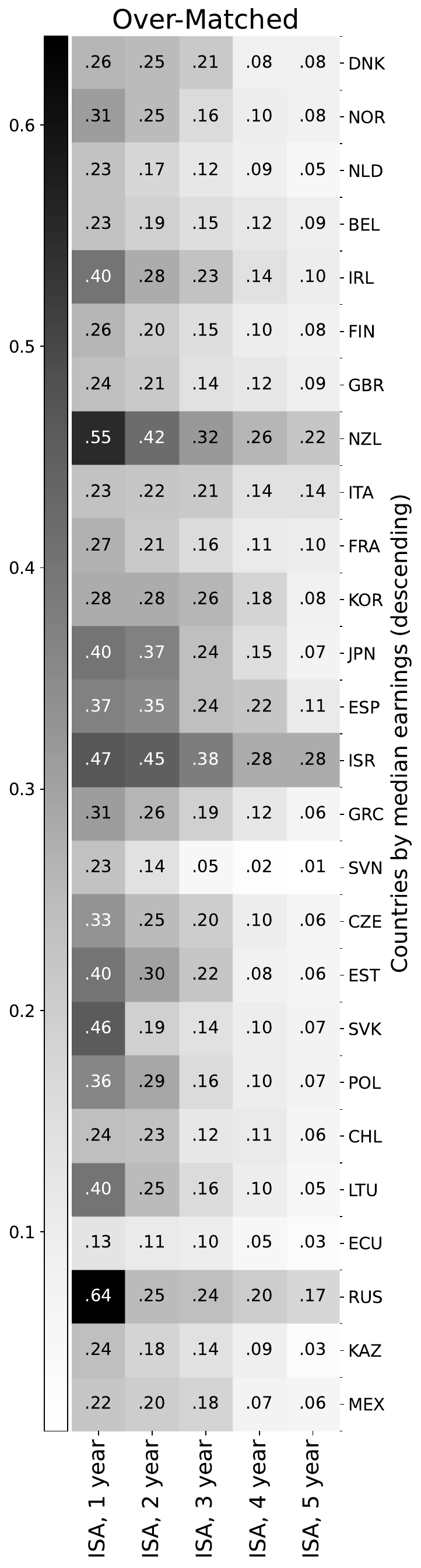}}
	\begin{tablenotes}
		\footnotesize
    		\item \textit{Notes:} Rows are sorted by the country-specific median of hourly earnings including bonuses (PPP corrected USD).
	\end{tablenotes}
\end{figure}

\begin{figure}[!htbp]
	\centering
	\caption{Country-specific mismatch shares: DSA}
	\label{fig:dsa_hm}
	{\includegraphics[scale=0.4, trim=0cm 0cm 0cm 0cm, clip, valign=t]{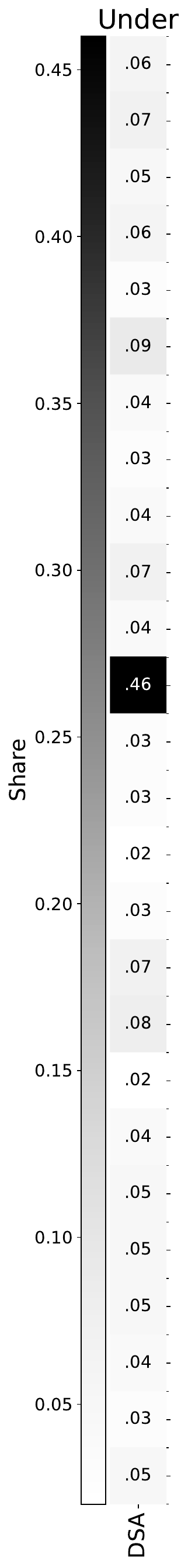}}
	{\includegraphics[scale=0.4, trim=0cm 0cm 0cm 0cm, clip, valign=t]{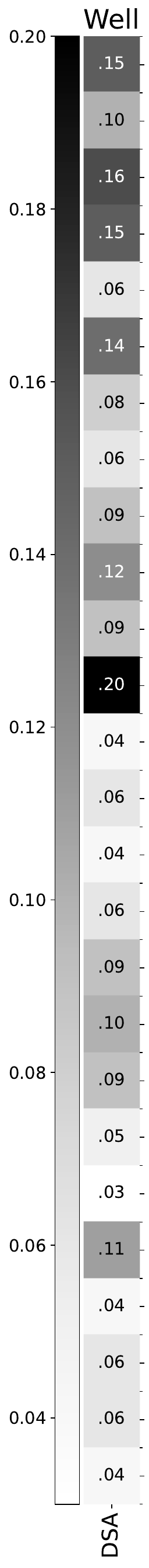}}
	{\includegraphics[scale=0.4, trim=0cm 0cm 0cm 0cm, clip, valign=t]{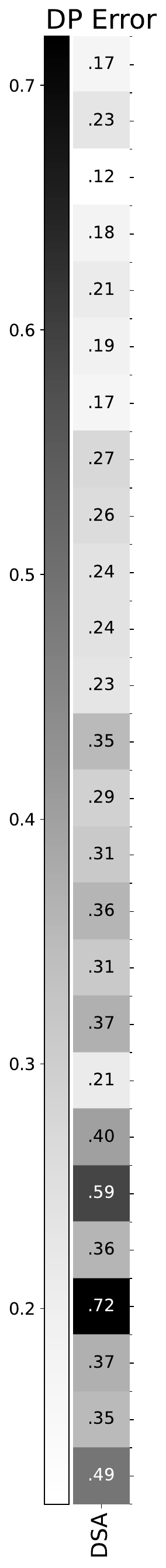}}
	{\includegraphics[scale=0.4, trim=0cm 0cm 0cm 0cm, clip, valign=t]{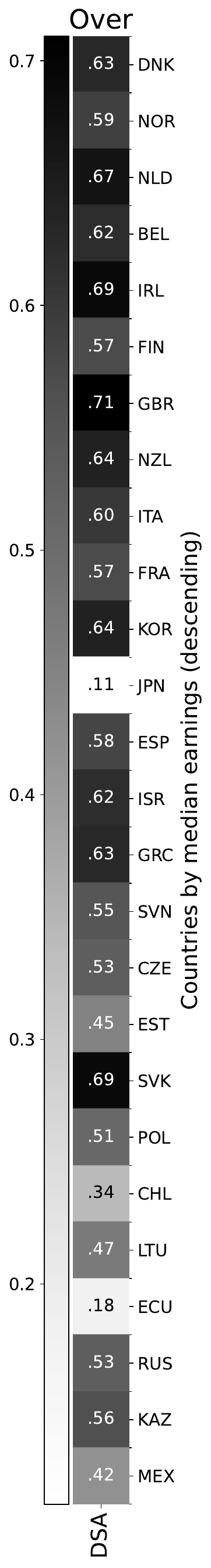}}
	$\; \; \; \;$
	{\includegraphics[scale=0.4, trim=0cm 0cm 0cm 0cm, clip, valign=t]{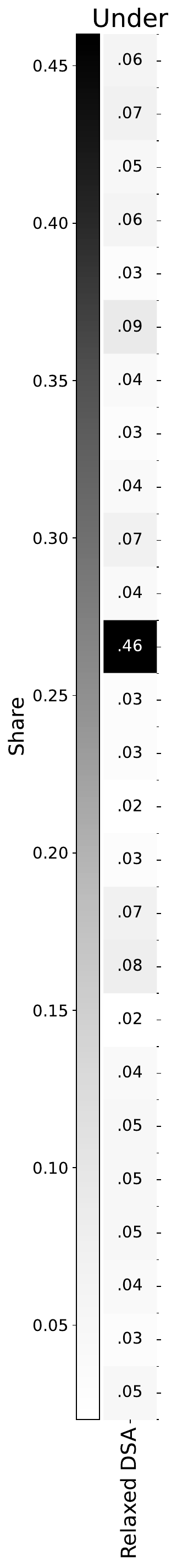}}
	{\includegraphics[scale=0.4, trim=0cm 0cm 0cm 0cm, clip, valign=t]{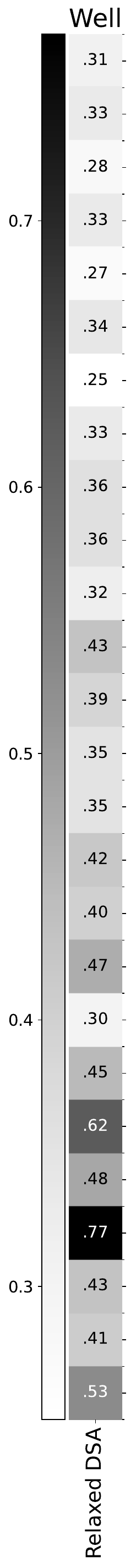}}
	{\includegraphics[scale=0.4, trim=0cm 0cm 0cm 0cm, clip, valign=t]{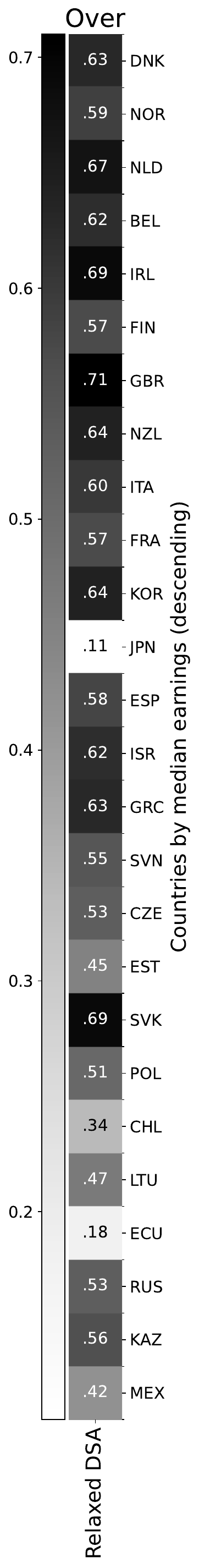}}
	\begin{tablenotes}
		\footnotesize
    		\item \textit{Notes:} Rows are sorted by the country-specific median of hourly earnings including bonuses (PPP corrected USD).
	\end{tablenotes}
\end{figure}

\begin{figure}[!htbp]
	\centering
	\caption{Country-specific mismatch shares (well-skilled): PF and ALV}
	\label{fig:pf_well_hm}
	{\includegraphics[scale=0.4, trim=0cm 0cm 0cm 0cm, clip, valign=t]{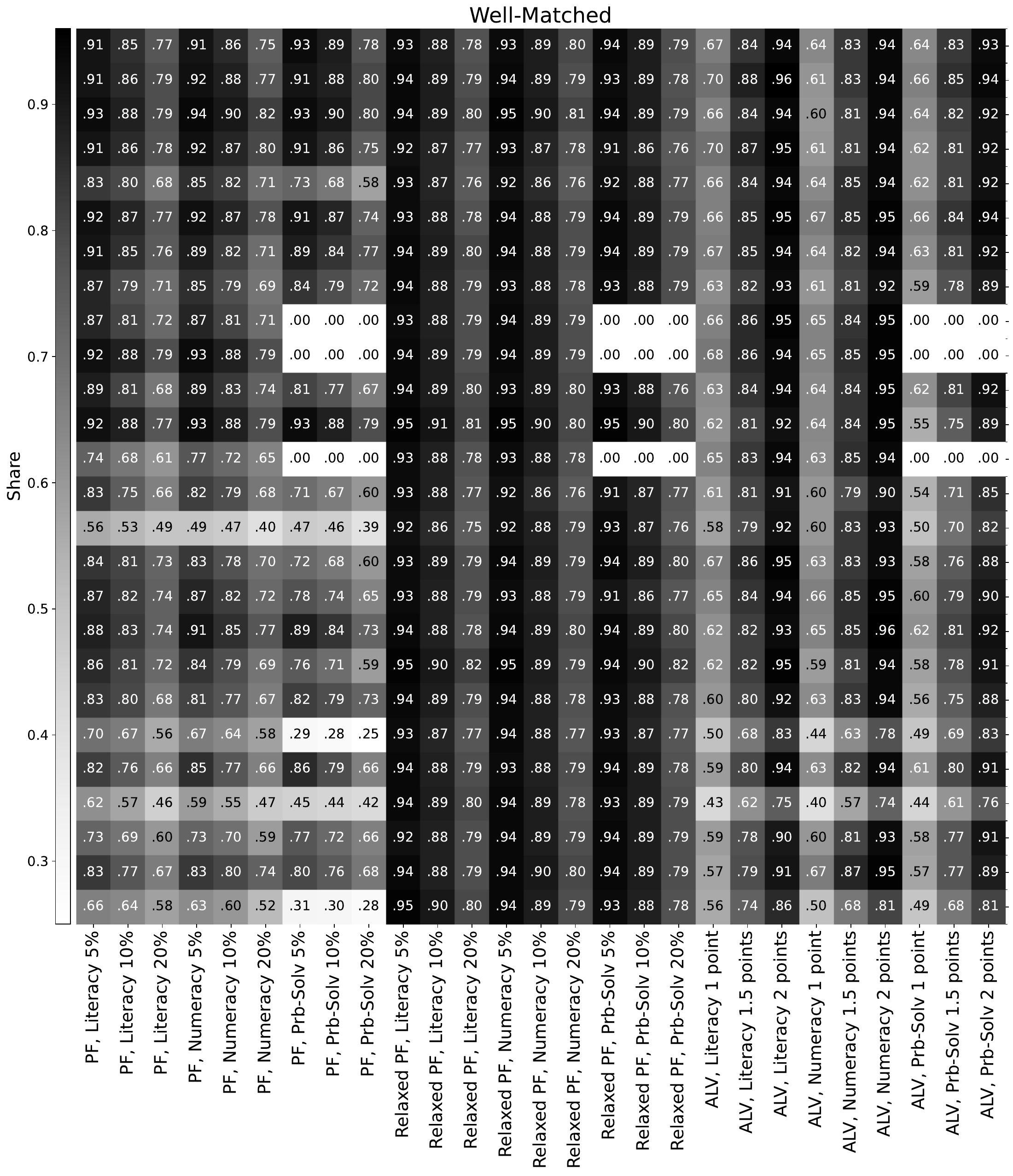}}
	\begin{tablenotes}
		\footnotesize
    		\item \textit{Notes:} Rows are sorted by the country-specific median of hourly earnings including bonuses (PPP corrected USD). Problem-solving data is missing for Italy, France and Spain. 
	\end{tablenotes}
\end{figure}

\begin{figure}[!htbp]
	\centering
	\caption{Country-specific mismatch shares (under-skilled): PF and ALV}
	\label{fig:pf_under_hm}
	{\includegraphics[scale=0.4, trim=0cm 0cm 0cm 0cm, clip, valign=t]{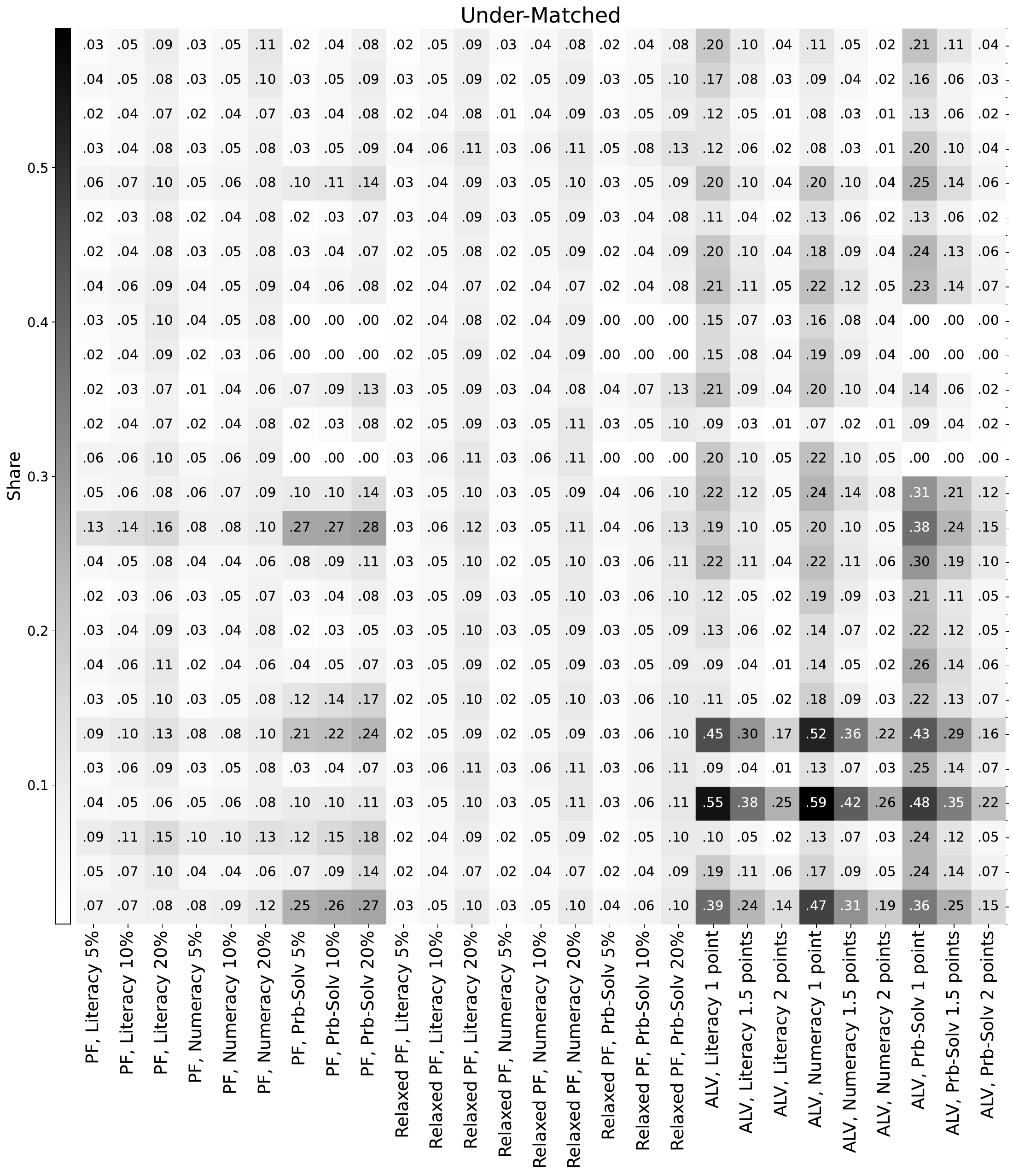}}

	\begin{tablenotes}
		\footnotesize
    		\item \textit{Notes:} Rows are sorted by the country-specific median of hourly earnings including bonuses (PPP corrected USD). Problem-solving data is missing for Italy, France and Spain.
	\end{tablenotes}
\end{figure}

\begin{figure}[!htbp]
	\centering
	\caption{Country-specific mismatch shares (over-skilled): PF and ALV}
	\label{fig:pf_over_hm}
	{\includegraphics[scale=0.4, trim=0cm 0cm 0cm 0cm, clip, valign=t]{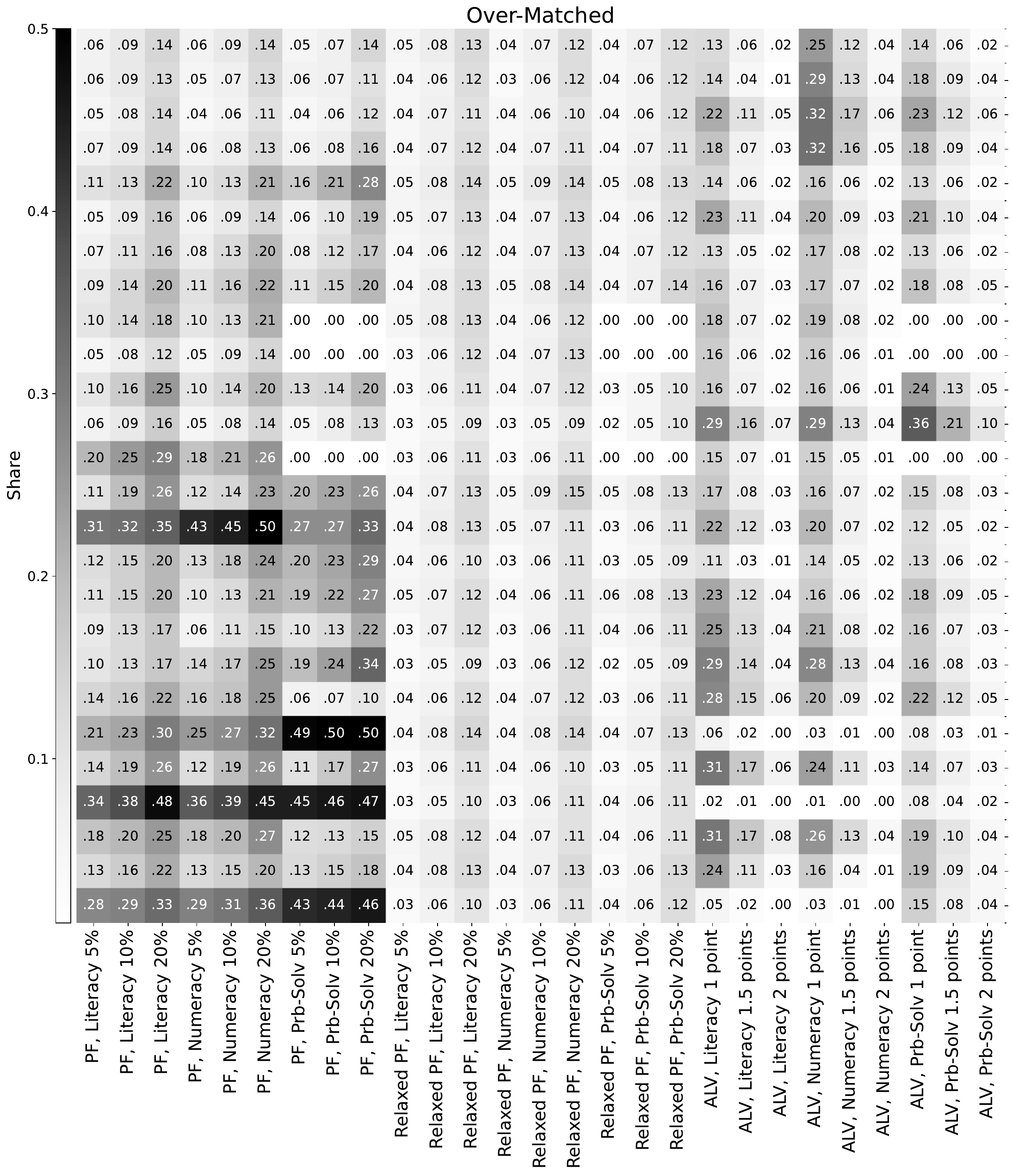}}
	\begin{tablenotes}
		\footnotesize
    		\item \textit{Notes:} Rows are sorted by the country-specific median of hourly earnings including bonuses (PPP corrected USD). Problem-solving data is missing for Italy, France and Spain.
	\end{tablenotes}
\end{figure}

\begin{figure}[!htbp]
	\centering
	\caption{Correlations in the output of mismatch measures}
	\label{fig:corr_hm}
	{\includegraphics[scale=0.4, trim=0cm 4cm 0cm 3cm, clip]{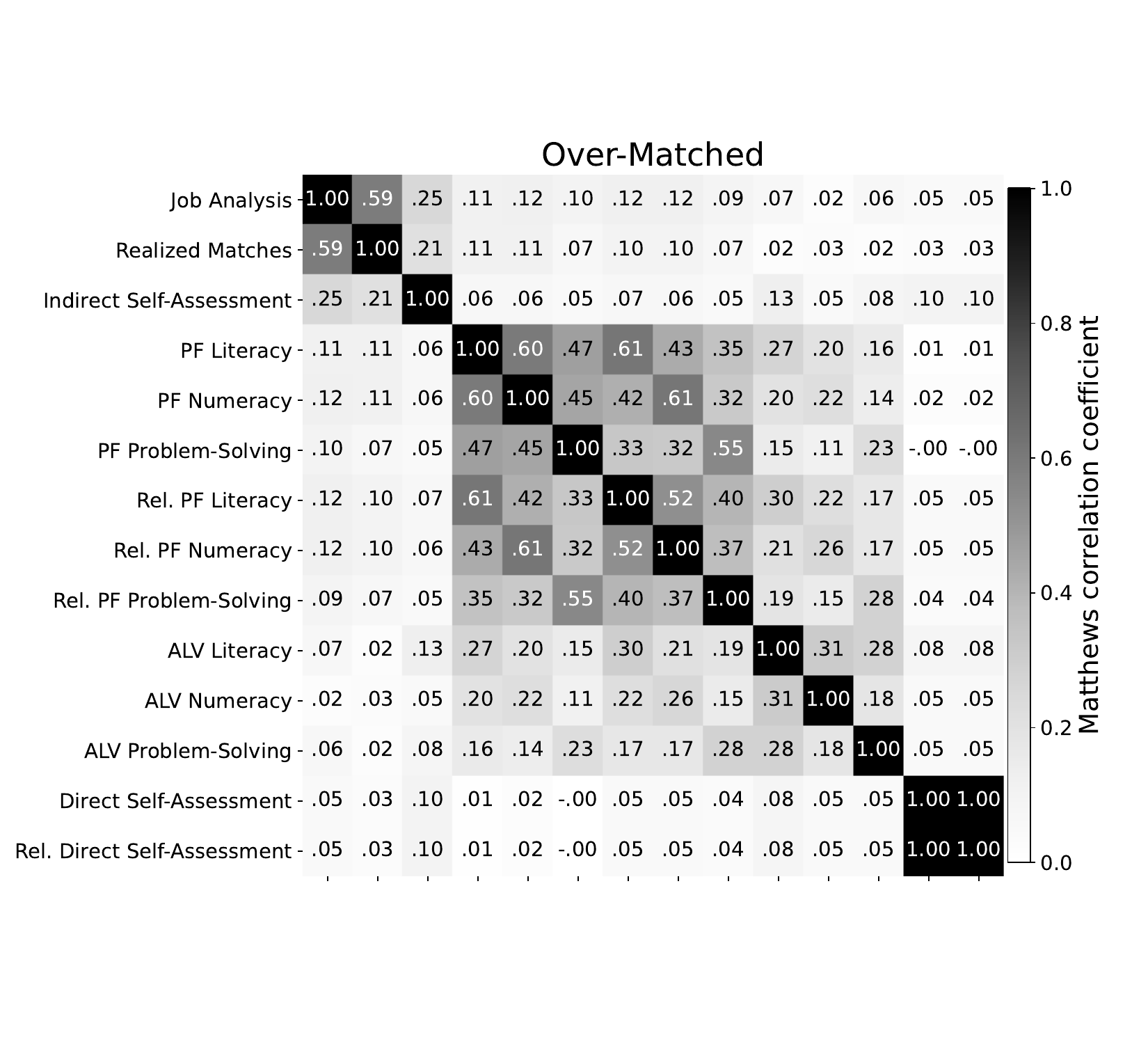}} \\
	{\includegraphics[scale=0.4, trim=0cm 4cm 0cm 3cm, clip]{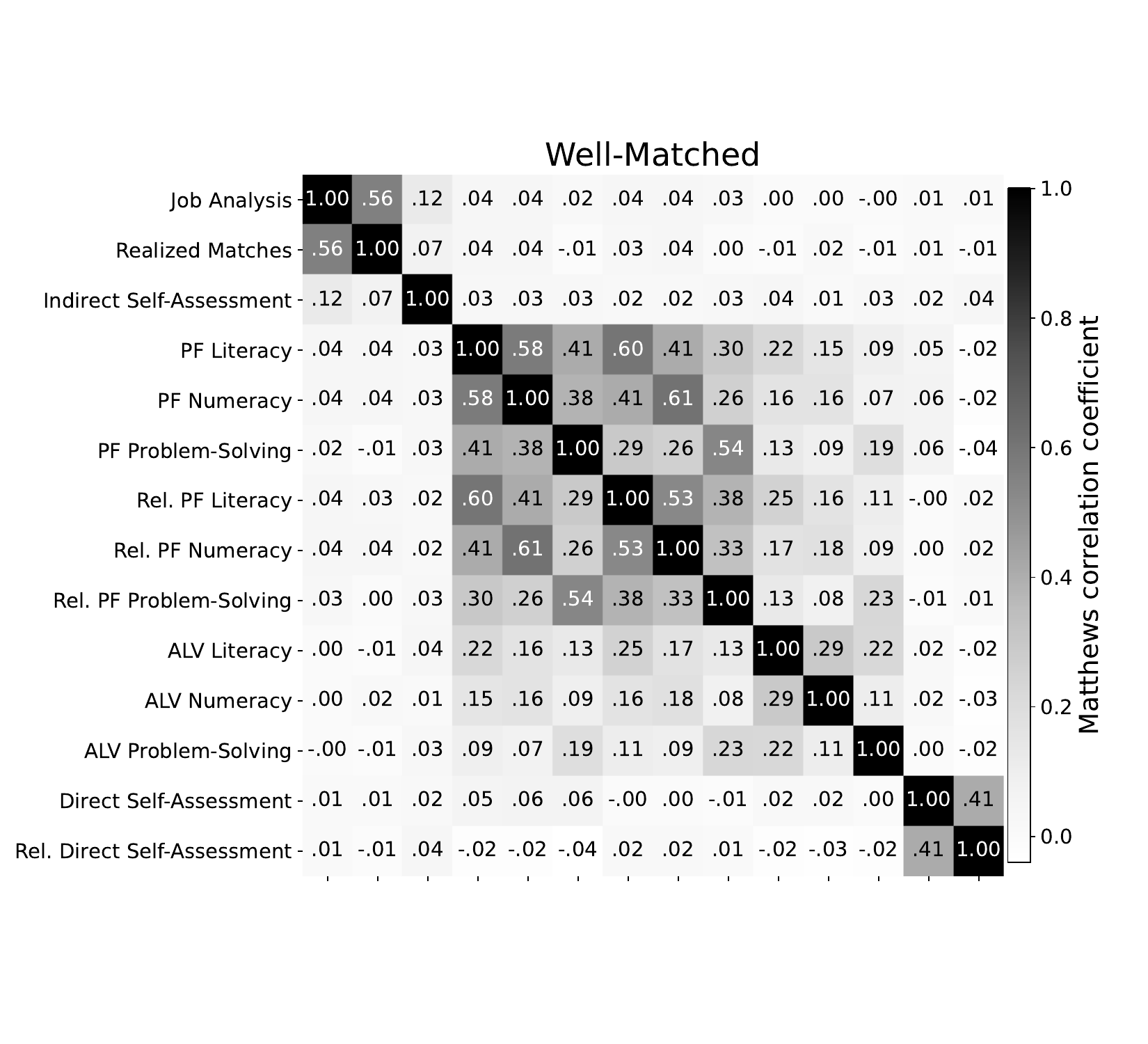}} \\
	{\includegraphics[scale=0.4, trim=0cm 2.5cm 0cm 3cm, clip]{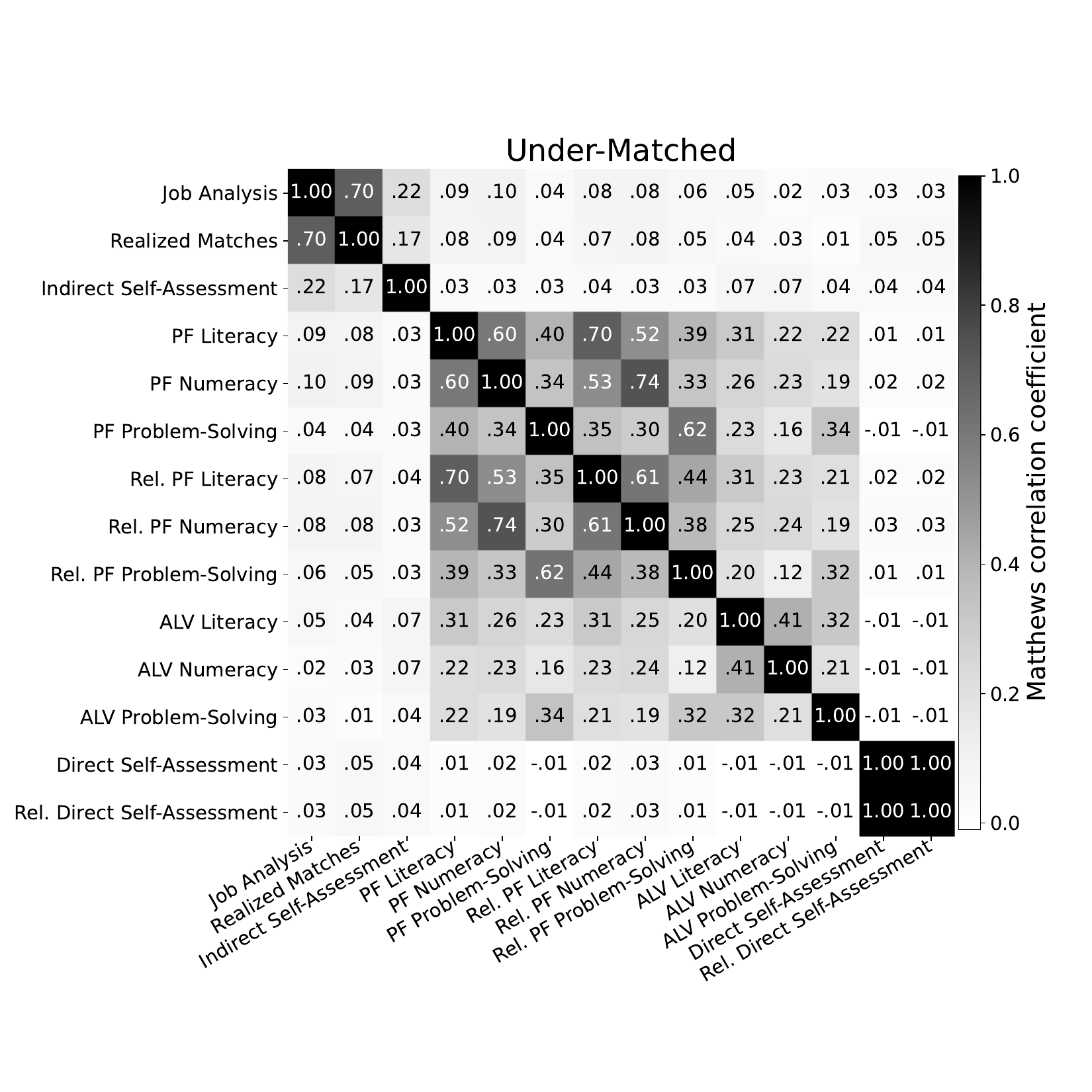}}
\end{figure}

\clearpage
\section{Heterogeneity}\label{sec:A_heterogeneity}

\subsection{Heterogeneity in earnings}
\begin{figure}[!htbp]
    \centering
    \caption{Earnings by gender, age groups, and migration status}
    \includegraphics[scale = 0.6]{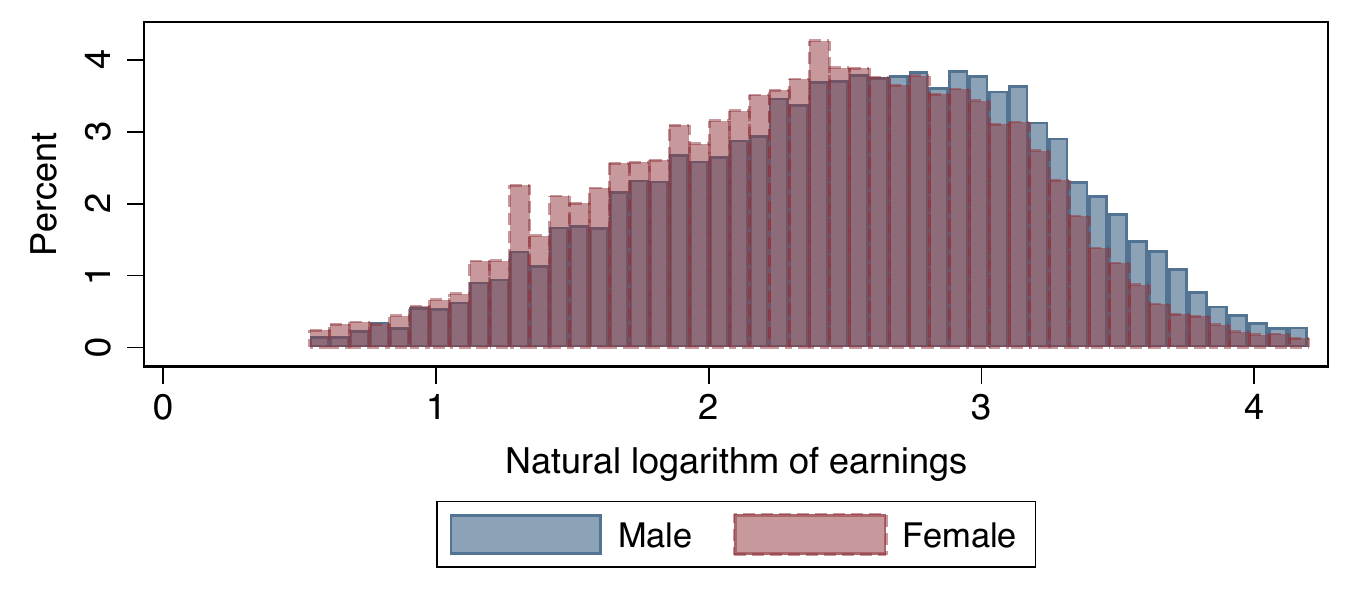} \\
    \includegraphics[scale = 0.6]{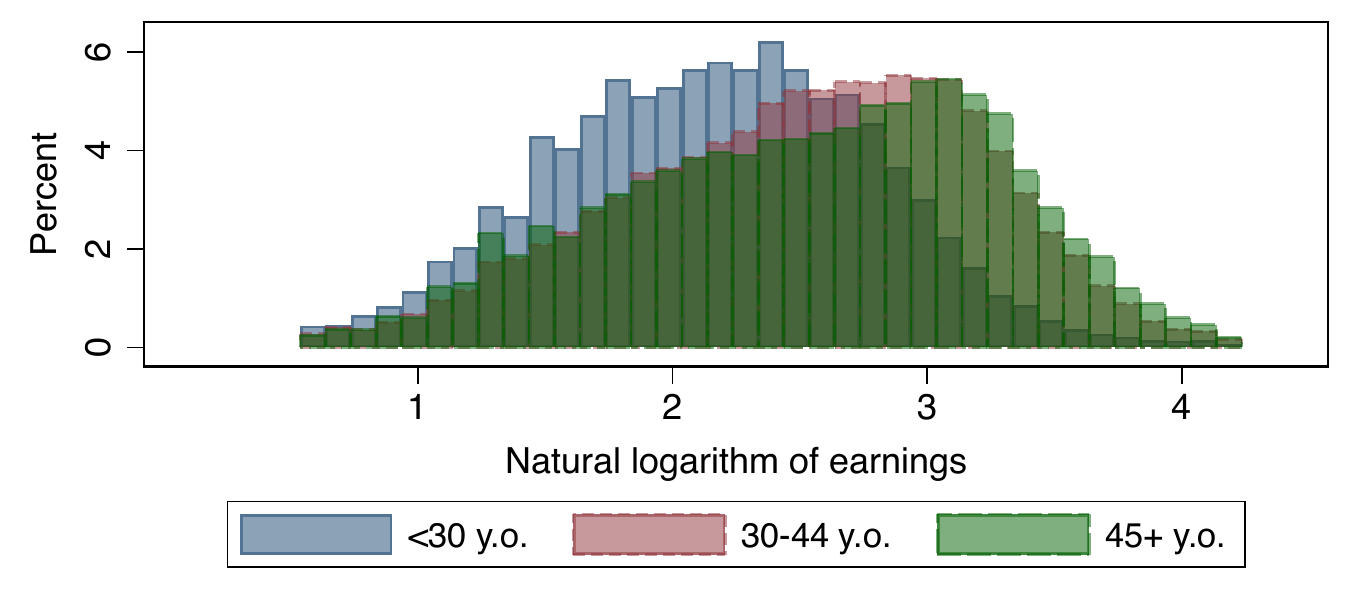} \\
    \includegraphics[scale = 0.6]{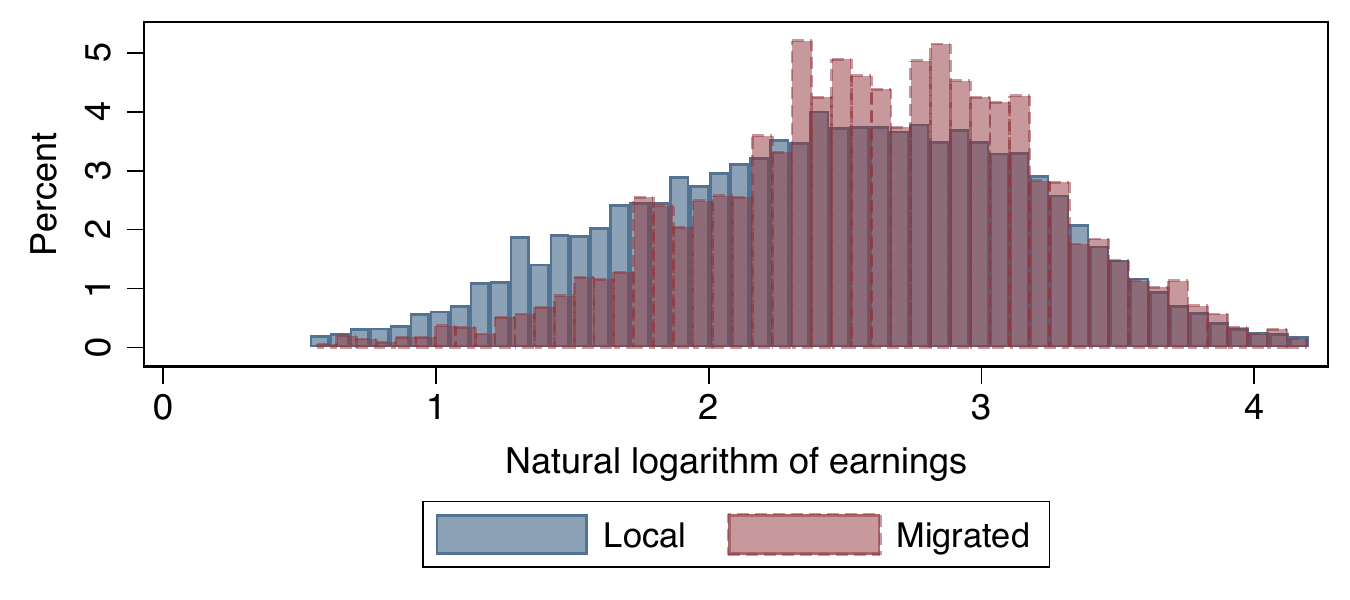}  
    \label{fig:wage_gaps_1}
\end{figure}

\begin{figure}[!htbp]
    \centering
    \caption{Earnings by education, literacy, numeracy, and problem-solving}
    \includegraphics[scale = 0.6]{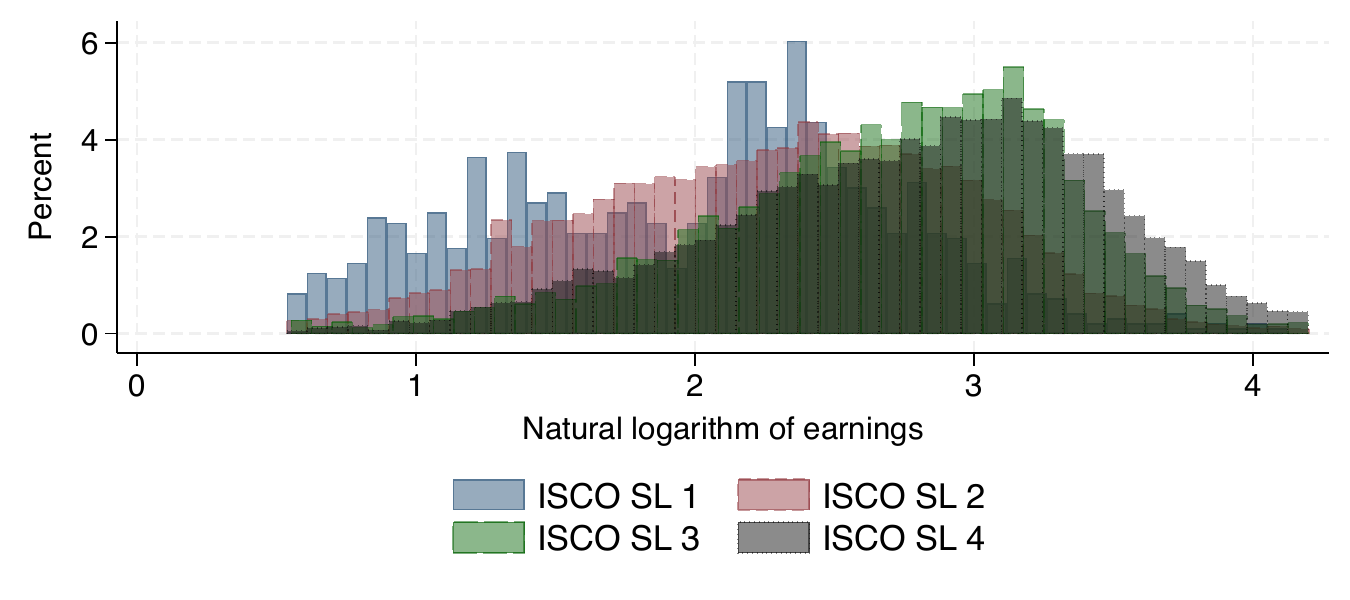} \\
    \includegraphics[scale = 0.6]{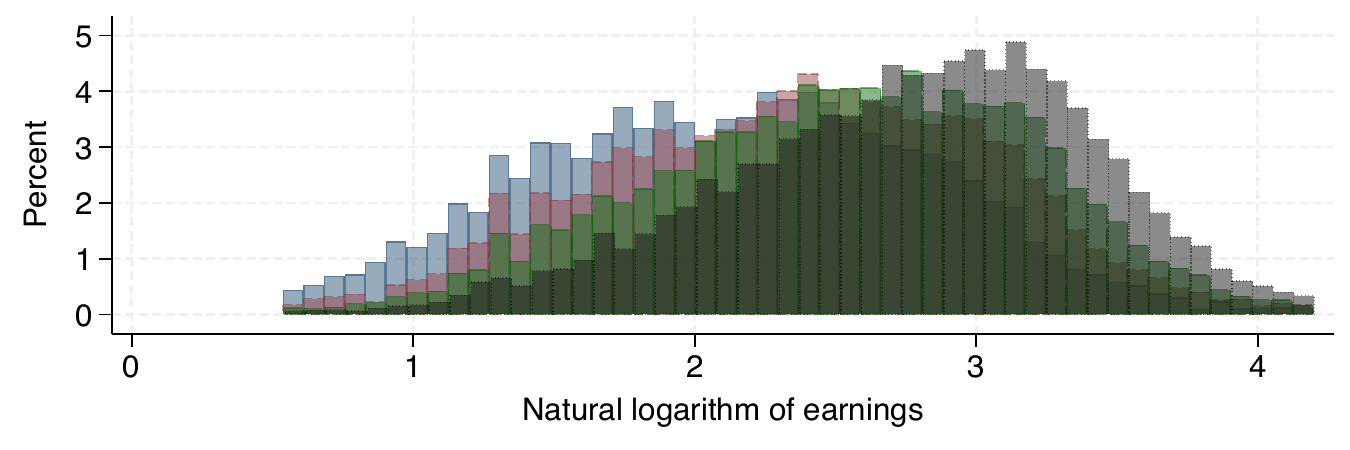} \\
    \includegraphics[scale = 0.6]{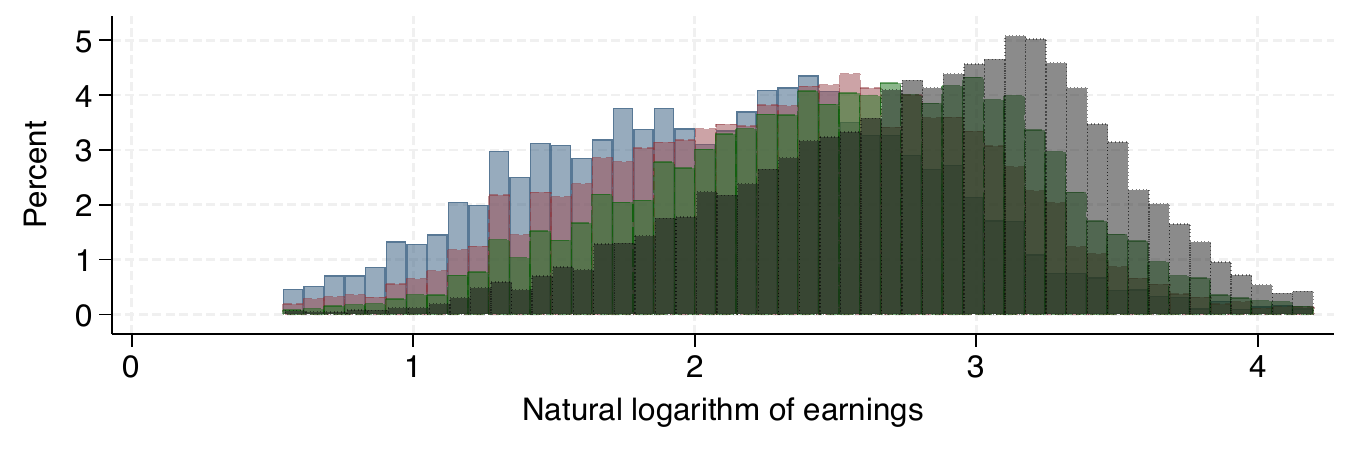} \\
    \includegraphics[scale = 0.6]{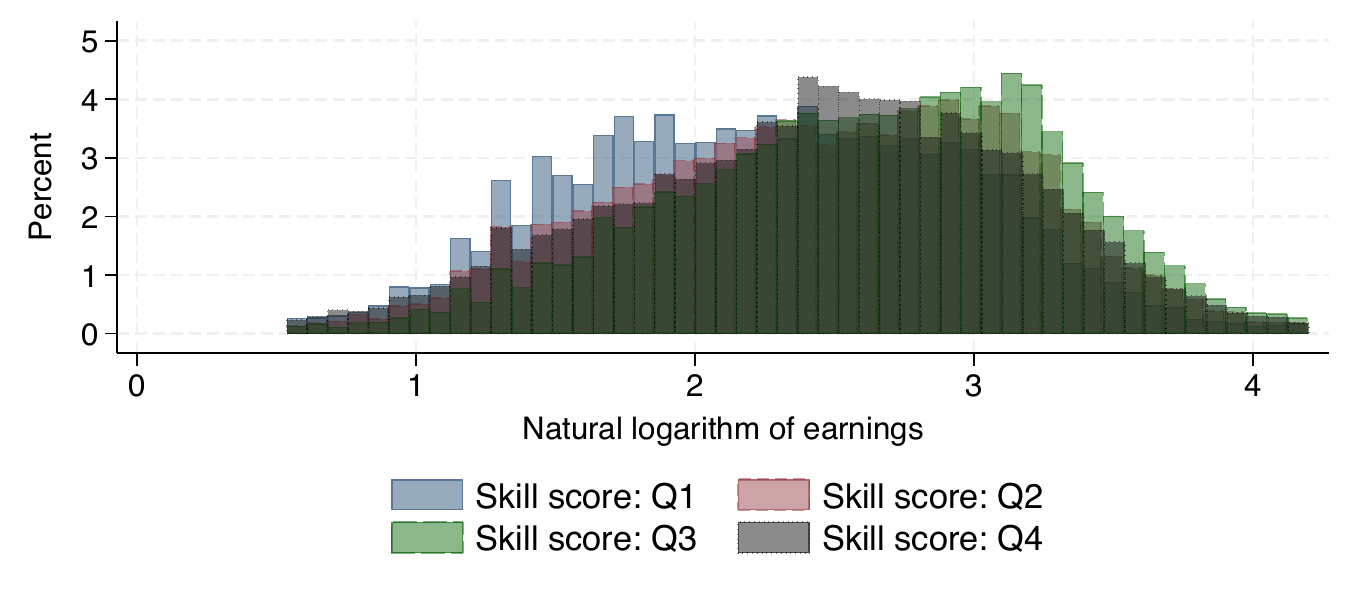}
    \label{fig:wage_gaps_2}
\end{figure}

\clearpage
\subsection{Heterogeneity in labour mismatch}

\begin{figure}[!htbp]
    \centering
    \caption{Shares of well-matched workers by gender}
    \includegraphics[scale = 1.0]{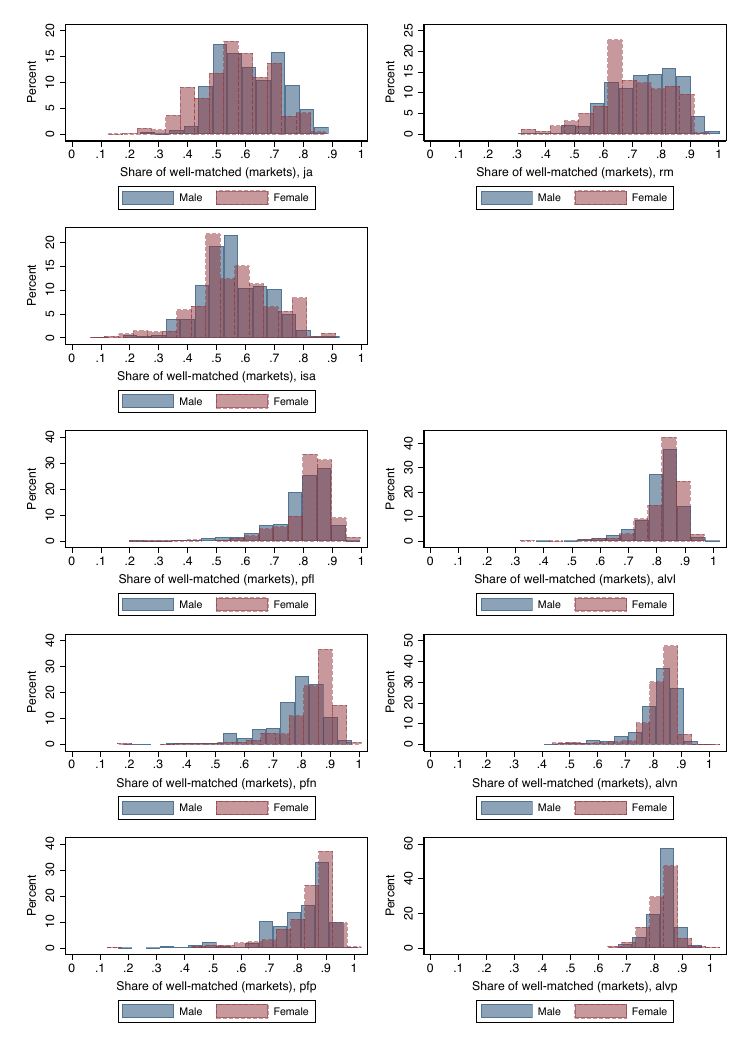}
    \label{fig:hst_w_by_gender}
    \begin{tablenotes}
		\footnotesize
    		\item \textit{Notes:} the shares are calculated at the market level.
    		\item \textit{Notation:} ja – Job Analysis, rm – Realised Matches, isa – Indirect Self Assessment, pfl – Pellizzari-Fichen Literacy, pfn – Pellizzari-Fichen Numeracy, pfp – Pellizzari-Fichen Problem Solving, alvl – Allen-Levels-van-der-Velden Literacy, alvn – Allen-Levels-van-der-Velden Numeracy, alvl – Allen-Levels-van-der-Velden Problem Solving.
	\end{tablenotes}
\end{figure}

\begin{figure}[!htbp]
    \centering
    \caption{Shares of under-matched workers by gender}
    \includegraphics[scale = 1.0]{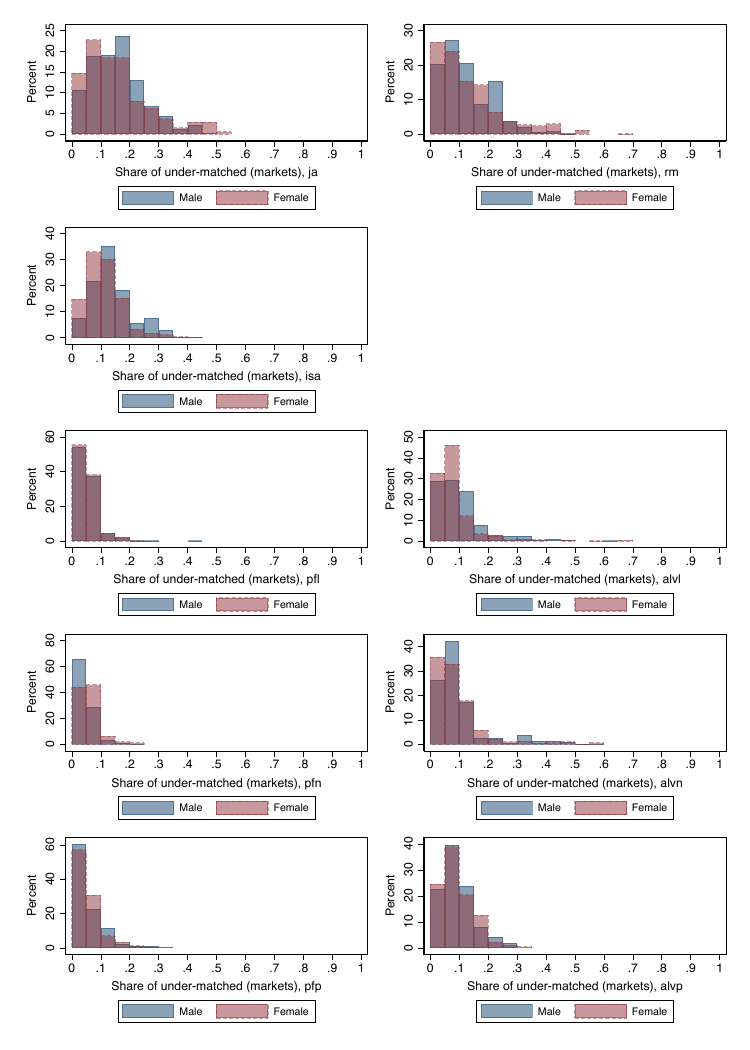}
    \label{fig:hst_u_by_gender}
    \begin{tablenotes}
		\footnotesize
    		\item \textit{Notes:} the shares are calculated at the market level.
    		\item \textit{Notation:} ja – Job Analysis, rm – Realised Matches, isa – Indirect Self Assessment, pfl – Pellizzari-Fichen Literacy, pfn – Pellizzari-Fichen Numeracy, pfp – Pellizzari-Fichen Problem Solving, alvl – Allen-Levels-van-der-Velden Literacy, alvn – Allen-Levels-van-der-Velden Numeracy, alvl – Allen-Levels-van-der-Velden Problem Solving.
	\end{tablenotes}
\end{figure}

\begin{figure}[!htbp]
    \centering
    \caption{Shares of over-matched workers by gender}
    \includegraphics[scale = 1.0]{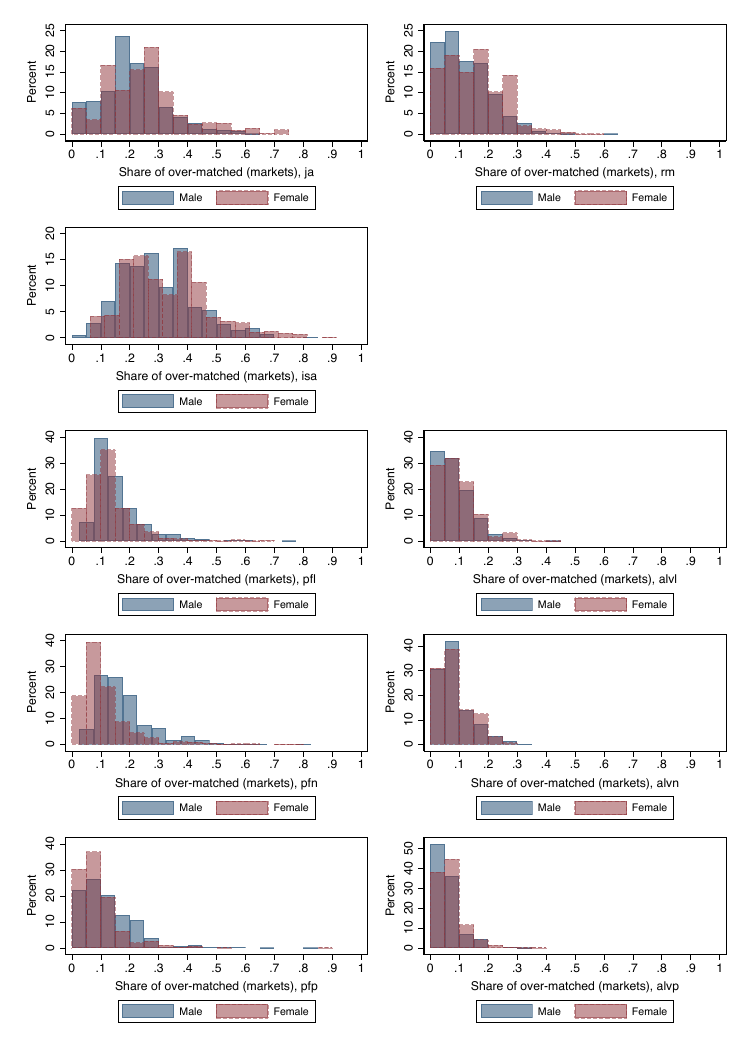}
    \label{fig:hst_o_by_gender}
    \begin{tablenotes}
		\footnotesize
    		\item \textit{Notes:} the shares are calculated at the market level.
    		\item \textit{Notation:} ja – Job Analysis, rm – Realised Matches, isa – Indirect Self Assessment, pfl – Pellizzari-Fichen Literacy, pfn – Pellizzari-Fichen Numeracy, pfp – Pellizzari-Fichen Problem Solving, alvl – Allen-Levels-van-der-Velden Literacy, alvn – Allen-Levels-van-der-Velden Numeracy, alvl – Allen-Levels-van-der-Velden Problem Solving.
	\end{tablenotes}
\end{figure}

\begin{figure}[!htbp]
    \centering
    \caption{Shares of well-matched workers by age groups}
    \includegraphics[scale = 1.0]{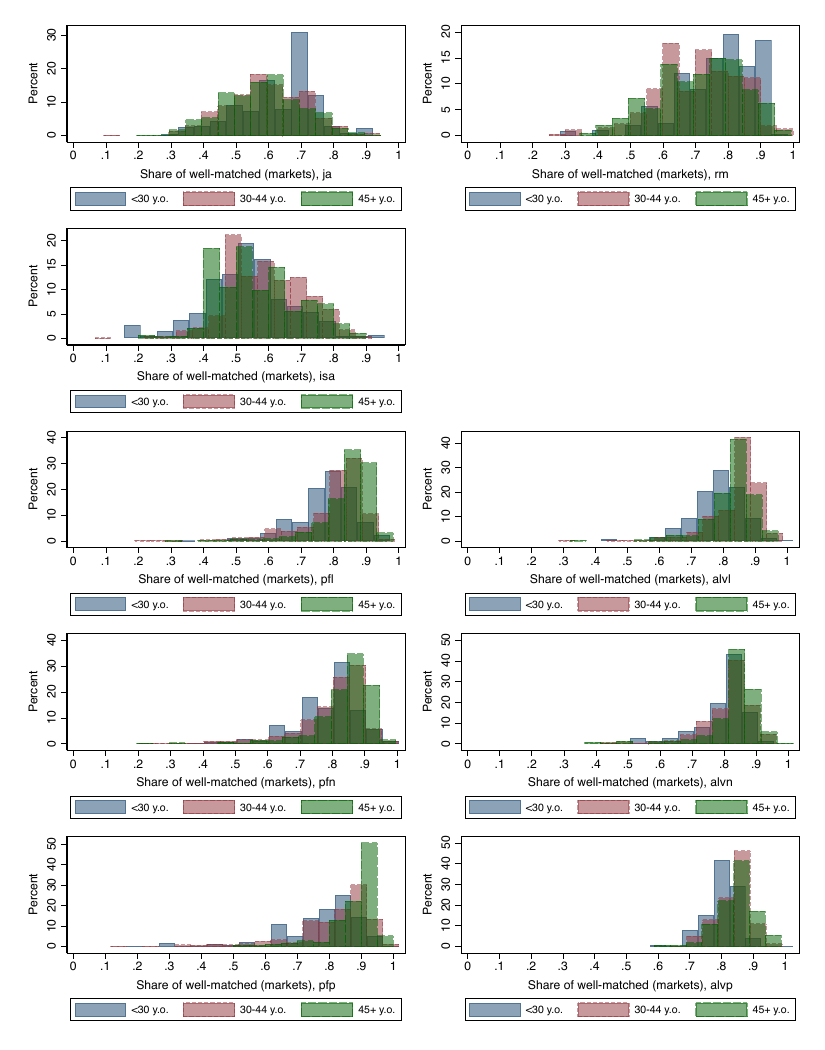}
    \label{fig:hst_w_by_age}
    \begin{tablenotes}
		\footnotesize
    		\item \textit{Notes:} the shares are calculated at the market level.
    		\item \textit{Notation:} ja – Job Analysis, rm – Realised Matches, isa – Indirect Self Assessment, pfl – Pellizzari-Fichen Literacy, pfn – Pellizzari-Fichen Numeracy, pfp – Pellizzari-Fichen Problem Solving, alvl – Allen-Levels-van-der-Velden Literacy, alvn – Allen-Levels-van-der-Velden Numeracy, alvl – Allen-Levels-van-der-Velden Problem Solving.
	\end{tablenotes}
\end{figure}

\begin{figure}[!htbp]
    \centering
    \caption{Shares of under-matched workers by age groups}
    \includegraphics[scale = 1.0]{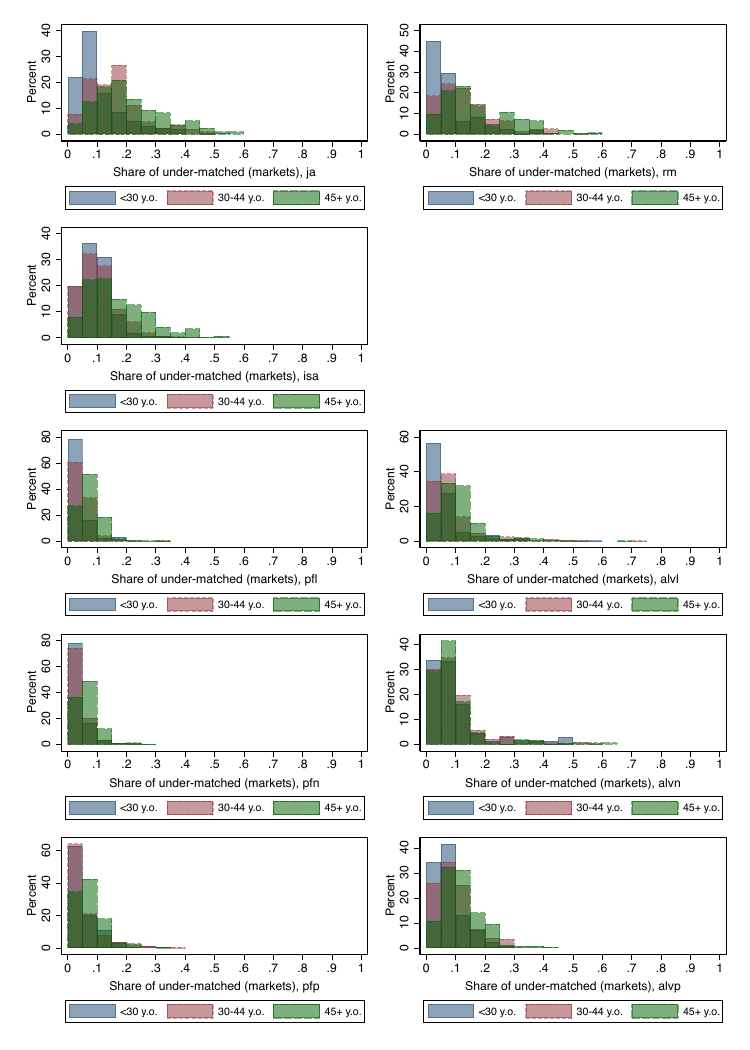}
    \label{fig:hst_u_by_age}
    \begin{tablenotes}
		\footnotesize
    		\item \textit{Notes:} the shares are calculated at the market level.
    		\item \textit{Notation:} ja – Job Analysis, rm – Realised Matches, isa – Indirect Self Assessment, pfl – Pellizzari-Fichen Literacy, pfn – Pellizzari-Fichen Numeracy, pfp – Pellizzari-Fichen Problem Solving, alvl – Allen-Levels-van-der-Velden Literacy, alvn – Allen-Levels-van-der-Velden Numeracy, alvl – Allen-Levels-van-der-Velden Problem Solving.
	\end{tablenotes}
\end{figure}

\begin{figure}[!htbp]
    \centering
    \caption{Shares of over-matched workers by age groups}
    \includegraphics[scale = 1.0]{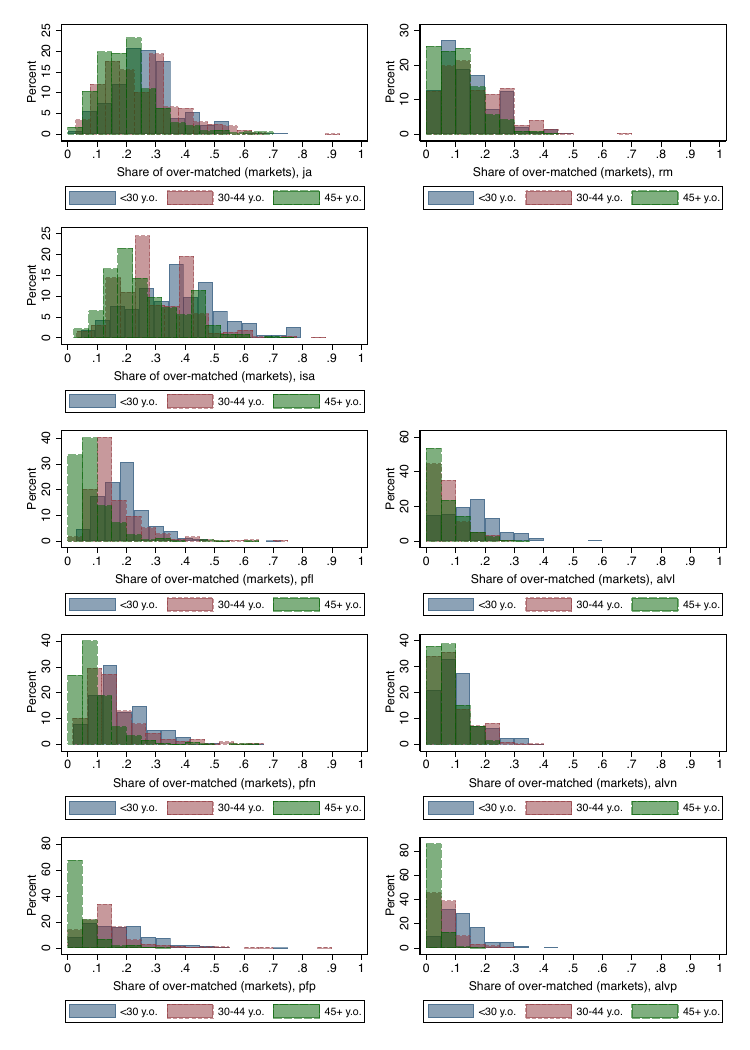}
    \label{fig:hst_o_by_age}
    \begin{tablenotes}
		\footnotesize
    		\item \textit{Notes:} the shares are calculated at the market level.
    		\item \textit{Notation:} ja – Job Analysis, rm – Realised Matches, isa – Indirect Self Assessment, pfl – Pellizzari-Fichen Literacy, pfn – Pellizzari-Fichen Numeracy, pfp – Pellizzari-Fichen Problem Solving, alvl – Allen-Levels-van-der-Velden Literacy, alvn – Allen-Levels-van-der-Velden Numeracy, alvl – Allen-Levels-van-der-Velden Problem Solving.
	\end{tablenotes}
\end{figure}

\begin{figure}[!htbp]
    \centering
    \caption{Shares of well-matched workers by migration status}
    \includegraphics[scale = 1.0]{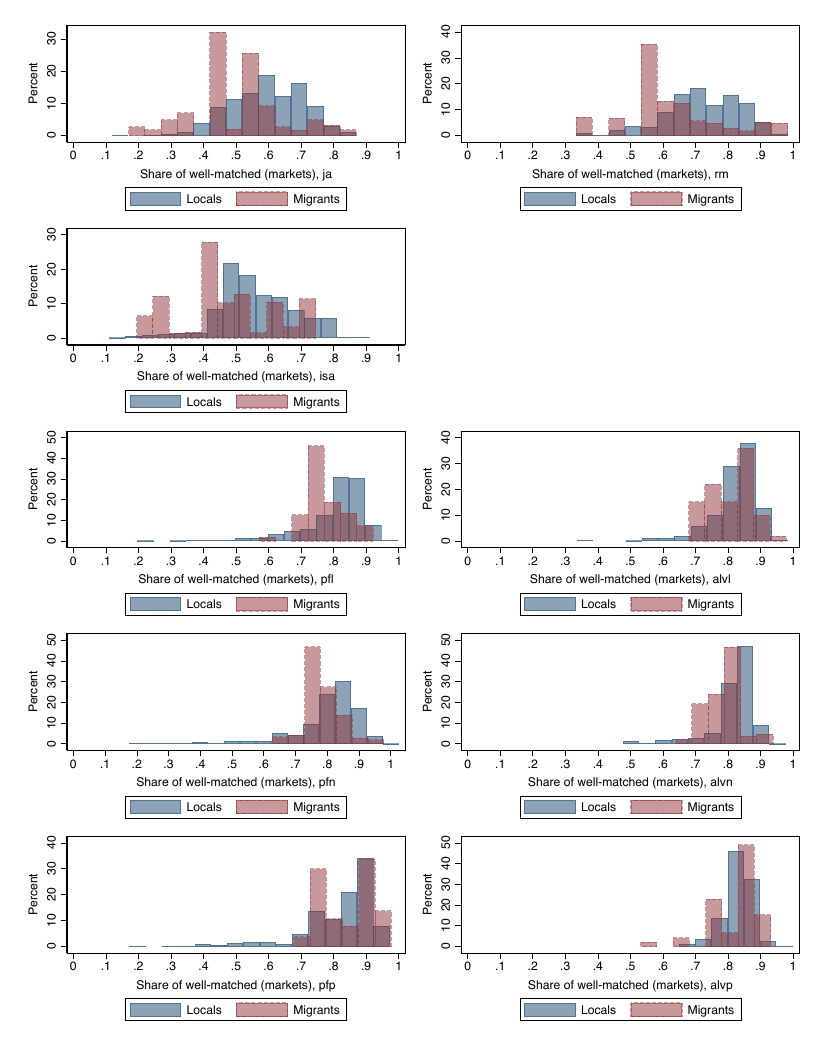}
    \label{fig:hst_w_by_mig}
    \begin{tablenotes}
		\footnotesize
    		\item \textit{Notes:} the shares are calculated at the market level.
    		\item \textit{Notation:} ja – Job Analysis, rm – Realised Matches, isa – Indirect Self Assessment, pfl – Pellizzari-Fichen Literacy, pfn – Pellizzari-Fichen Numeracy, pfp – Pellizzari-Fichen Problem Solving, alvl – Allen-Levels-van-der-Velden Literacy, alvn – Allen-Levels-van-der-Velden Numeracy, alvl – Allen-Levels-van-der-Velden Problem Solving.
	\end{tablenotes}
\end{figure}

\begin{figure}[!htbp]
    \centering
    \caption{Shares of under-matched workers by migration status}
    \includegraphics[scale = 1.0]{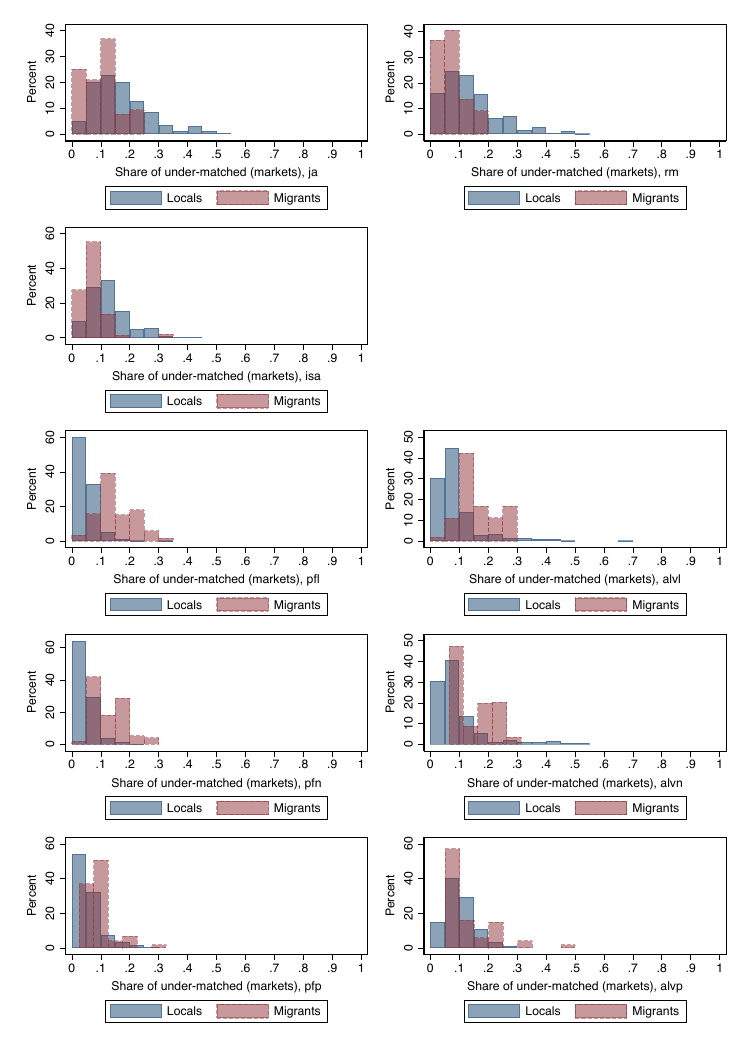}
    \label{fig:hst_u_by_mig}
    \begin{tablenotes}
		\footnotesize
    		\item \textit{Notes:} the shares are calculated at the market level.
    		\item \textit{Notation:} ja – Job Analysis, rm – Realised Matches, isa – Indirect Self Assessment, pfl – Pellizzari-Fichen Literacy, pfn – Pellizzari-Fichen Numeracy, pfp – Pellizzari-Fichen Problem Solving, alvl – Allen-Levels-van-der-Velden Literacy, alvn – Allen-Levels-van-der-Velden Numeracy, alvl – Allen-Levels-van-der-Velden Problem Solving.
	\end{tablenotes}
\end{figure}

\begin{figure}[!htbp]
    \centering
    \caption{Shares of over-matched workers by migration status}
    \includegraphics[scale = 1.0]{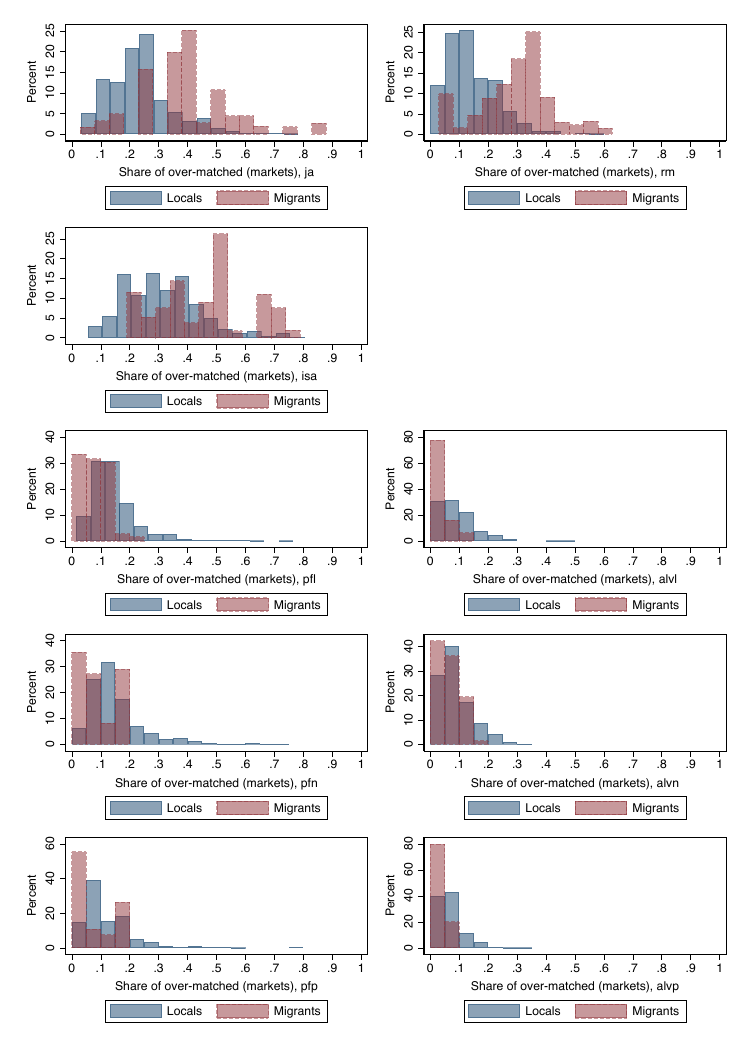}
    \label{fig:hst_o_by_mig}
    \begin{tablenotes}
		\footnotesize
    		\item \textit{Notes:} the shares are calculated at the market level.
    		\item \textit{Notation:} ja – Job Analysis, rm – Realised Matches, isa – Indirect Self Assessment, pfl – Pellizzari-Fichen Literacy, pfn – Pellizzari-Fichen Numeracy, pfp – Pellizzari-Fichen Problem Solving, alvl – Allen-Levels-van-der-Velden Literacy, alvn – Allen-Levels-van-der-Velden Numeracy, alvl – Allen-Levels-van-der-Velden Problem Solving.
	\end{tablenotes}
\end{figure}

\begin{figure}[!htbp]
    \centering
    \caption{Shares of well-educated workers by literacy, numeracy, and problem-solving scores}
    \includegraphics[scale = 1.0]{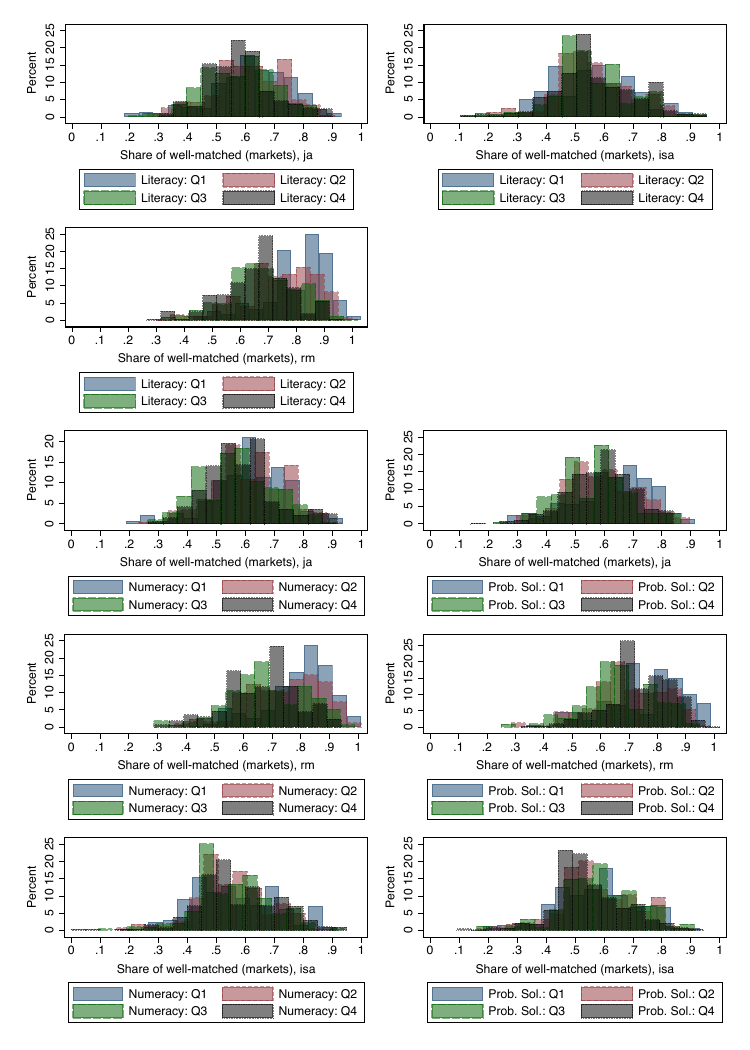}
    \label{fig:hst_w_by_skill}
    \begin{tablenotes}
		\footnotesize
    		\item \textit{Notes:} the shares are calculated at the market level.
    		\item \textit{Notation:} ja – Job Analysis, rm – Realised Matches, isa – Indirect Self Assessment, pfl – Pellizzari-Fichen Literacy, pfn – Pellizzari-Fichen Numeracy, pfp – Pellizzari-Fichen Problem Solving, alvl – Allen-Levels-van-der-Velden Literacy, alvn – Allen-Levels-van-der-Velden Numeracy, alvl – Allen-Levels-van-der-Velden Problem Solving.
	\end{tablenotes}
\end{figure}

\begin{figure}[!htbp]
    \centering
    \caption{Shares of under-educated workers by literacy, numeracy, and problem-solving scores}
    \includegraphics[scale = 1.0]{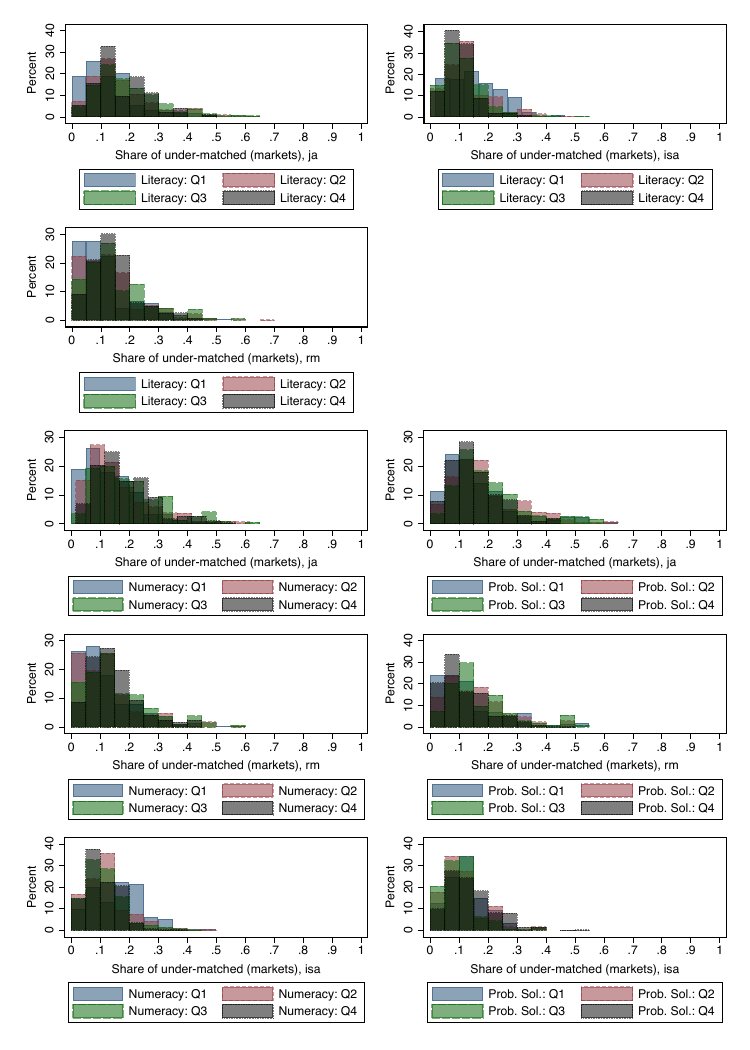}
    \label{fig:hst_u_by_skill}
    \begin{tablenotes}
		\footnotesize
    		\item \textit{Notes:} the shares are calculated at the market level.
    		\item \textit{Notation:} ja – Job Analysis, rm – Realised Matches, isa – Indirect Self Assessment, pfl – Pellizzari-Fichen Literacy, pfn – Pellizzari-Fichen Numeracy, pfp – Pellizzari-Fichen Problem Solving, alvl – Allen-Levels-van-der-Velden Literacy, alvn – Allen-Levels-van-der-Velden Numeracy, alvl – Allen-Levels-van-der-Velden Problem Solving.
	\end{tablenotes}
\end{figure}

\begin{figure}[!htbp]
    \centering
    \caption{Shares of over-educated workers by literacy, numeracy, and problem-solving scores}
    \includegraphics[scale = 1.0]{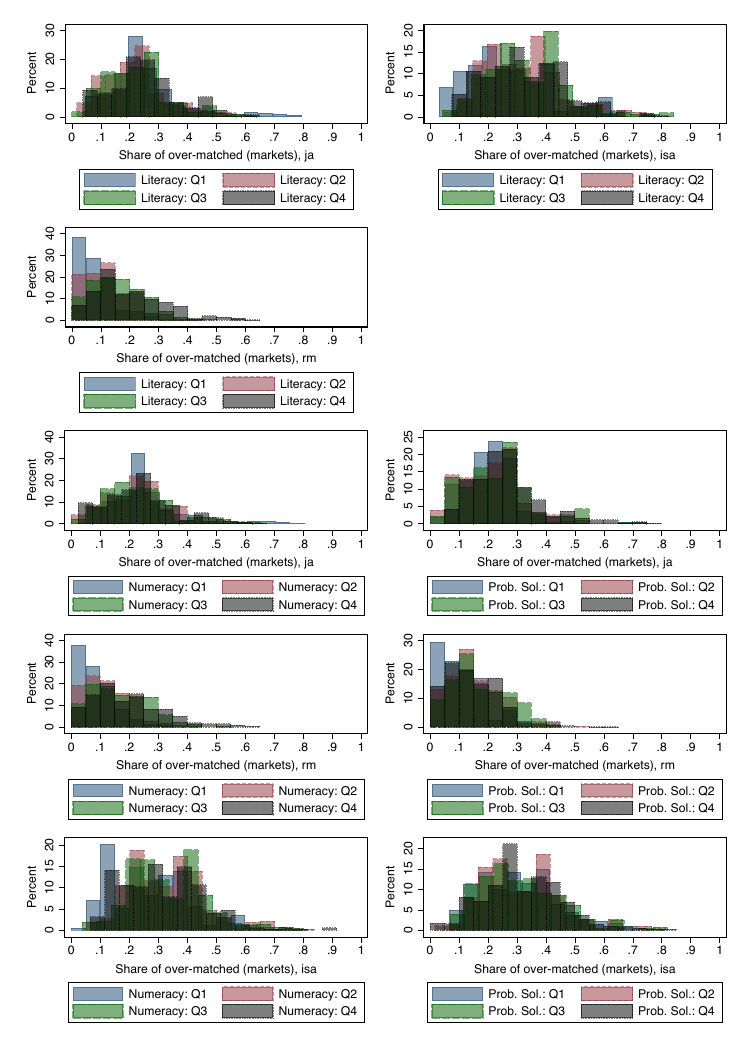}
    \label{fig:hst_o_by_skill}
    \begin{tablenotes}
		\footnotesize
    		\item \textit{Notes:} the shares are calculated at the market level.
    		\item \textit{Notation:} ja – Job Analysis, rm – Realised Matches, isa – Indirect Self Assessment, pfl – Pellizzari-Fichen Literacy, pfn – Pellizzari-Fichen Numeracy, pfp – Pellizzari-Fichen Problem Solving, alvl – Allen-Levels-van-der-Velden Literacy, alvn – Allen-Levels-van-der-Velden Numeracy, alvl – Allen-Levels-van-der-Velden Problem Solving.
	\end{tablenotes}
\end{figure}

\begin{figure}[!htbp]
    \centering
    \caption{Shares of well-skilled workers by education}
    \includegraphics[scale = 1]{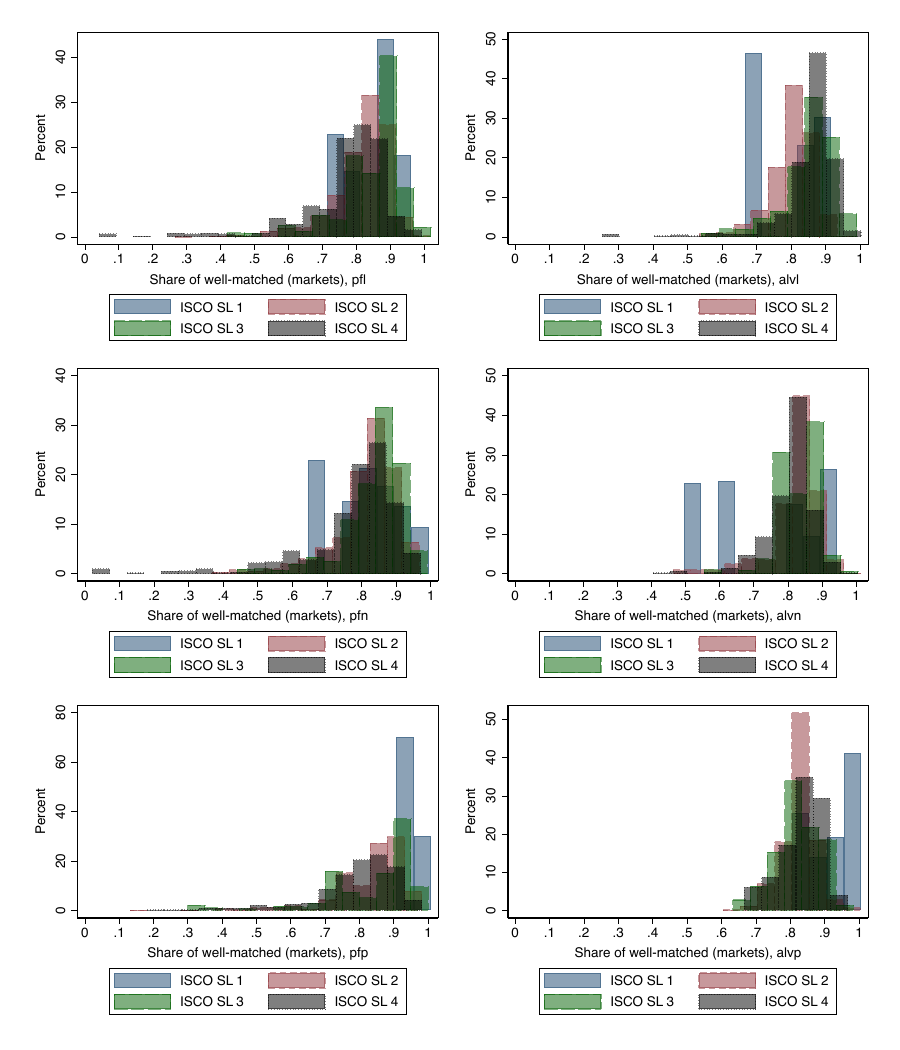}
    \label{fig:hst_w_by_sl}
    \begin{tablenotes}
		\footnotesize
    		\item \textit{Notes:} the shares are calculated at the market level.
    		\item \textit{Notation:} ja – Job Analysis, rm – Realised Matches, isa – Indirect Self Assessment, pfl – Pellizzari-Fichen Literacy, pfn – Pellizzari-Fichen Numeracy, pfp – Pellizzari-Fichen Problem Solving, alvl – Allen-Levels-van-der-Velden Literacy, alvn – Allen-Levels-van-der-Velden Numeracy, alvl – Allen-Levels-van-der-Velden Problem Solving.
	\end{tablenotes}
\end{figure}

\begin{figure}[!htbp]
    \centering
    \caption{Shares of under-skilled workers by education}
    \includegraphics[scale = 1]{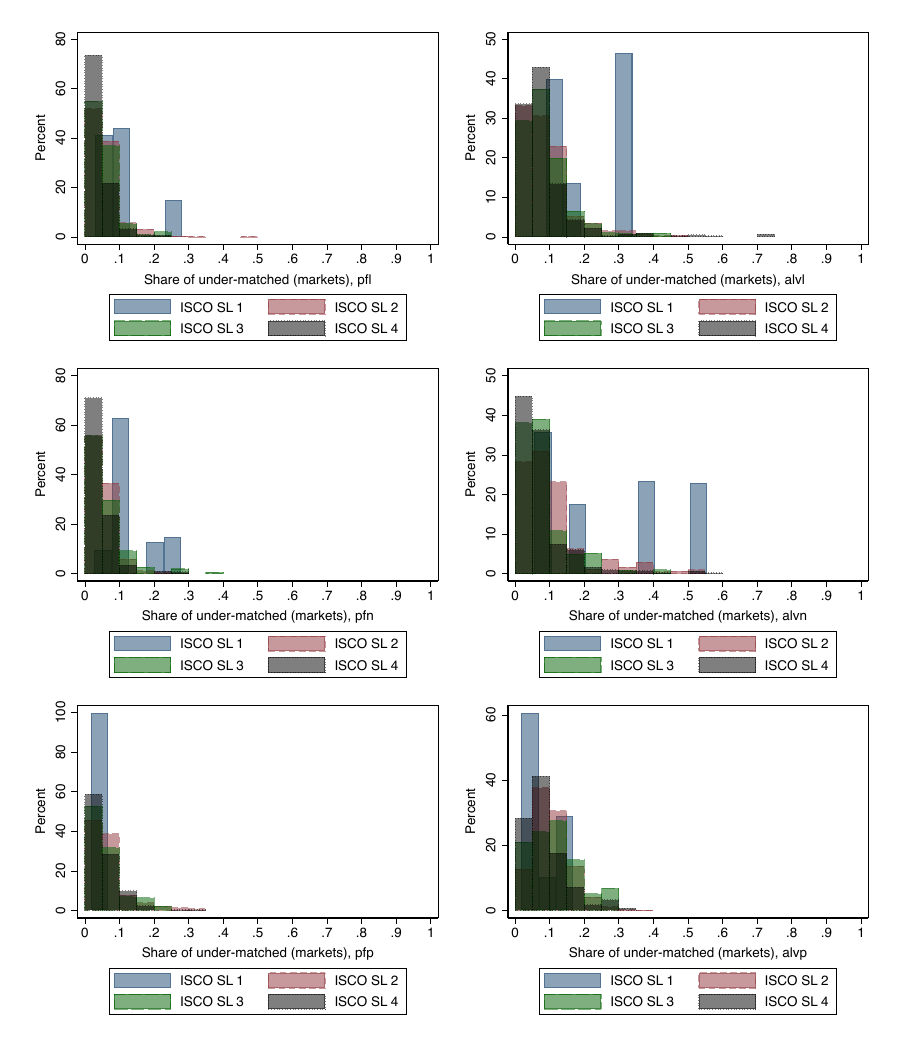}
    \label{fig:hst_u_by_sl}
    \begin{tablenotes}
		\footnotesize
    		\item \textit{Notes:} the shares are calculated at the market level.
    		\item \textit{Notation:} ja – Job Analysis, rm – Realised Matches, isa – Indirect Self Assessment, pfl – Pellizzari-Fichen Literacy, pfn – Pellizzari-Fichen Numeracy, pfp – Pellizzari-Fichen Problem Solving, alvl – Allen-Levels-van-der-Velden Literacy, alvn – Allen-Levels-van-der-Velden Numeracy, alvl – Allen-Levels-van-der-Velden Problem Solving.
	\end{tablenotes}
\end{figure}

\begin{figure}[!htbp]
    \centering
    \caption{Shares of over-skilled workers by education}
    \includegraphics[scale = 1]{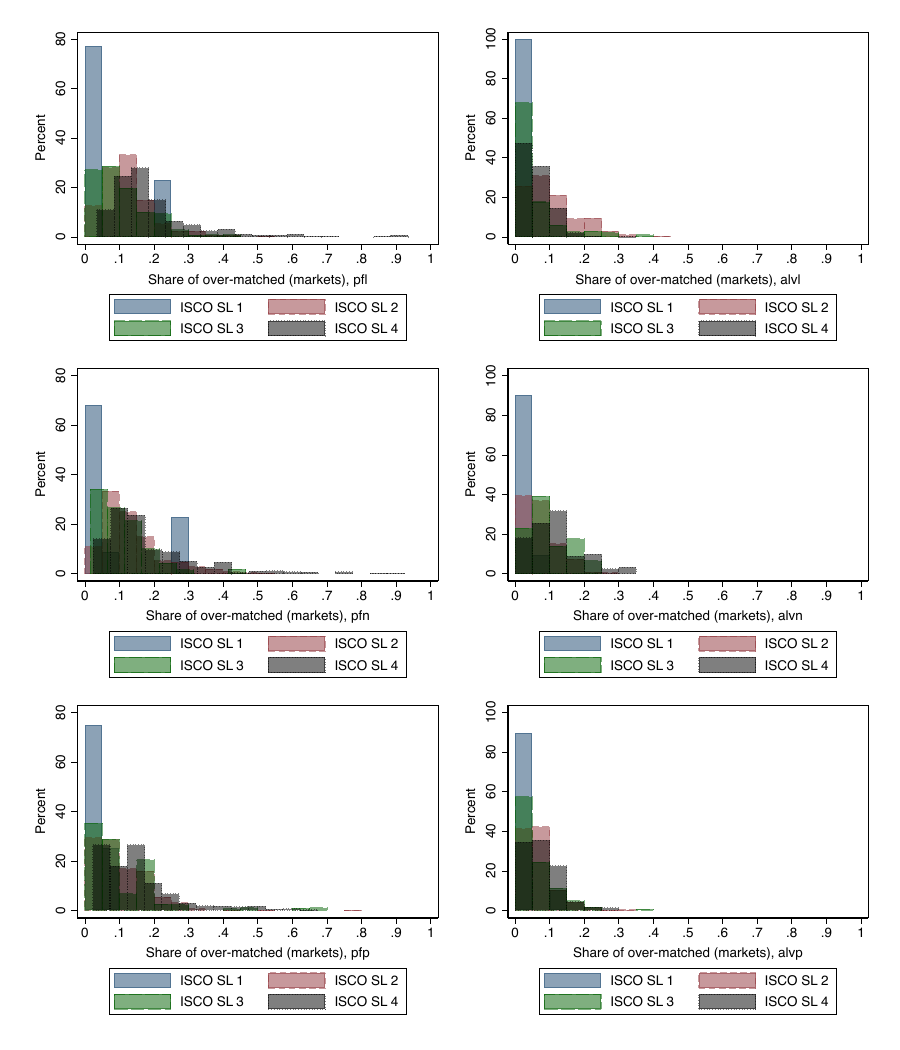}
    \label{fig:hst_o_by_sl}
    \begin{tablenotes}
		\footnotesize
    		\item \textit{Notes:} the shares are calculated at the market level.
    		\item \textit{Notation:} ja – Job Analysis, rm – Realised Matches, isa – Indirect Self Assessment, pfl – Pellizzari-Fichen Literacy, pfn – Pellizzari-Fichen Numeracy, pfp – Pellizzari-Fichen Problem Solving, alvl – Allen-Levels-van-der-Velden Literacy, alvn – Allen-Levels-van-der-Velden Numeracy, alvl – Allen-Levels-van-der-Velden Problem Solving.
	\end{tablenotes}
\end{figure}

\clearpage
\section{Estimation results}
\subsection{Lasso model selection}\label{sec:A_cvlasso}

{
\footnotesize
\def\sym#1{\ifmmode^{#1}\else\(^{#1}\)\fi}
\begin{longtable}{l*{3}{c}}
\caption{Lasso post-estimation results} \label{tab:cvlasso}\\
\toprule\endfirsthead\midrule\endhead\midrule\endfoot\endlastfoot
                &\multicolumn{1}{c}{Adaptive Lasso}         &\multicolumn{1}{c}{Lasso, $\lambda_{lopt}$}         &\multicolumn{1}{c}{Lasso, $\lambda_{lse}$}         \\
\midrule
Female          							&          -0.128         &          -0.127         &          -0.124         \\
Age             							&           0.032         &           0.029         &           0.015         \\
Age $\times$ Age							&          -0.000         &          -0.000         &          -0.000         \\
Tenure          							&           0.007         &           0.007         &           0.007         \\
Migrated after 16							&          -0.065         &          -0.053         &          -0.019         \\
Years in country							&           0.001         &           0.001         &                         \\
CHL             							&          -0.649         &          -0.626         &          -0.529         \\
CZE             							&          -0.818         &          -0.799         &          -0.719         \\
DNK             							&           0.130         &           0.143         &           0.195         \\
ECU             							&          -0.982         &          -0.956         &          -0.851         \\
FIN             							&          -0.143         &          -0.106         &          -0.024         \\
GBR             							&          -0.198         &          -0.177         &          -0.096         \\
GRC             							&          -0.671         &          -0.645         &          -0.545         \\
IRL             							&          -0.022         &           0.010         &           0.072         \\
ISR             							&          -0.474         &          -0.452         &          -0.368         \\
JPN             							&          -0.379         &          -0.360         &          -0.277         \\
KAZ             							&          -1.258         &          -1.236         &          -1.137         \\
KOR             							&          -0.334         &          -0.313         &          -0.223         \\
LTU             							&          -1.045         &          -1.026         &          -0.942         \\
MEX             							&          -1.033         &          -1.007         &          -0.910         \\
NLD             							&          -0.054         &          -0.036         &           0.018         \\
NOR             							&           0.081         &           0.099         &           0.161         \\
POL             							&          -0.857         &          -0.838         &          -0.764         \\
RUS             							&          -1.347         &          -1.325         &          -1.230         \\
SVK             							&          -0.918         &          -0.900         &          -0.819         \\
SVN             							&          -0.744         &          -0.724         &          -0.631         \\
ISIC section B               				&           0.487         &           0.431         &           0.372         \\
ISIC section C               				&           0.144         &           0.093         &           0.063         \\
ISIC section D               				&           0.209         &           0.151         &           0.084         \\
ISIC section E               				&           0.088         &           0.027         &                         \\
ISIC section F               				&           0.159         &           0.108         &           0.074         \\
ISIC section G               				&           0.043         &          -0.008         &          -0.037         \\
ISIC section H               				&           0.132         &           0.081         &           0.048         \\
ISIC section I               				&           0.032         &          -0.020         &          -0.055         \\
ISIC section J               				&           0.161         &           0.108         &           0.073         \\
ISIC section K               				&           0.217         &           0.164         &           0.129         \\
ISIC section L               				&           0.166         &           0.107         &           0.041         \\
ISIC section M               				&           0.096         &           0.043         &           0.006         \\
ISIC section N               				&           0.057         &           0.003         &          -0.010         \\
ISIC section O               				&           0.114         &           0.062         &           0.031         \\
ISIC section P               				&           0.010         &          -0.040         &          -0.062         \\
ISIC section Q               				&           0.048         &          -0.001         &          -0.021         \\
ISIC section R               				&           0.035         &          -0.016         &          -0.040         \\
ISIC section S               				&           0.031         &          -0.018         &          -0.036         \\
ISIC section T               				&           0.128         &           0.066         &                         \\
ISIC missing         						&           0.095         &           0.029         &                         \\
ISCO SL attained=2							&          -0.118         &          -0.043         &          -0.012         \\
ISCO SL attained=3							&          -0.079         &                         &                         \\
ISCO SL attained=4							&          -0.038         &           0.055         &           0.075         \\
ISCO SL required=2   						&           0.027         &                         &          -0.005         \\
ISCO SL required=3   						&           0.128         &           0.090         &           0.075         \\
ISCO SL required=4   						&           0.238         &           0.185         &           0.170         \\
Years of schooling							&           0.016         &           0.015         &           0.013         \\
Years to get job							&           0.014         &           0.015         &           0.018         \\
Not challenged       						&          -0.006         &          -0.004         &                         \\
Need training     							&          -0.007         &          -0.008         &          -0.007         \\
Literacy        							&           0.000         &           0.000         &           0.000         \\
Numeracy        							&           0.001         &           0.001         &           0.001         \\
Problem-Solving 							&           0.001         &           0.001         &           0.000         \\
Literacy use     							&           0.054         &           0.056         &           0.062         \\
Numeracy use     							&          -0.001         &           0.001         &           0.002         \\
Problem-Solving use     					&           0.001         &           0.002         &           0.002         \\
JA under-matched     						&          -0.046         &          -0.029         &                         \\
JA over-matched     						&           0.017         &          -0.003         &          -0.007         \\
DSA under-matched            				&           0.006         &                         &                         \\
DSA over-matched            				&          -0.007         &                         &                         \\
DSA (relaxed) under-matched    				&          -0.006         &                         &                         \\
DSA (relaxed) over-matched    				&           0.009         &                         &                         \\
ISA (1 year) over-matched          			&          -0.027         &          -0.026         &          -0.021         \\
ISA (2 years) under-matched          		&           0.024         &           0.017         &           0.003         \\
ISA (2 years) over-matched          		&          -0.003         &          -0.001         &                         \\
ISA (3 years) under-matched          		&          -0.010         &          -0.001         &                         \\
ISA (3 years) over-matched          		&           0.001         &                         &                         \\
ISA (4 years) under-matched          		&           0.081         &           0.068         &           0.029         \\
ISA (4 years) over-matched         			&          -0.018         &          -0.012         &                         \\
ISA (5 years) under-matched       			&          -0.058         &          -0.047         &                         \\
ISA (5 years) over-matched          		&           0.019         &           0.013         &                         \\
RM (mean $\pm 0.5\sigma$) under-matched     &          -0.005         &          -0.005         &          -0.010         \\
RM (mean $\pm 0.5\sigma$) over-matched      &          -0.009         &          -0.000         &                         \\
RM (mean $\pm1\sigma$) under-matched      	&           0.029         &           0.023         &                         \\
RM (mean $\pm1\sigma$) over-matched      	&           0.006         &           0.001         &                         \\
RM (mean $\pm 1.5\sigma$) under-matched     &          -0.034         &          -0.016         &                         \\
RM (mean $\pm1.5\sigma$) over-matched     	&          -0.002         &          -0.004         &                         \\
RM (mode $\pm0.1\sigma$) under-matched     	&          -0.154         &          -0.023         &          -0.000         \\
RM (mode $\pm 0.1\sigma$) over-matched     	&          -0.082         &          -0.010         &                         \\
RM (mode $\pm1\sigma$) under-matched      	&           0.121         &                         &                         \\
RM (mode $\pm1\sigma$) over-matched      	&           0.029         &          -0.030         &                         \\
RM (mode $\pm2\sigma$) under-matched      	&           0.035         &           0.025         &                         \\
RM (mode $\pm2\sigma$) over-matched      	&          -0.005         &          -0.013         &          -0.038         \\
PF-L (2.5\%) under-matched   			 	&          -0.006         &                         &                         \\
PF-L (2.5\%) over-matched   			 	&           0.025         &           0.022         &           0.011         \\
PF-L (5\%) under-matched     			 	&           0.044         &           0.033         &           0.001         \\
PF-L (5\%) over-matched     				&           0.022         &           0.018         &           0.002         \\
PF-L (10\%) under-matched      				&          -0.035         &          -0.025         &                         \\
PF-L (10\%) over-matched      				&          -0.006         &          -0.004         &                         \\
PF-N PF-N (2.5\%) under-matched    			&           0.026         &           0.014         &           0.001         \\
PF-N (2.5\%) over-matched    				&           0.001         &                         &                         \\
PF-N (5\%) under-matched     				&          -0.018         &          -0.005         &                         \\
PF-N (5\%) over-matched     				&          -0.007         &          -0.005         &          -0.003         \\
PF-N (10\%) under-matched      				&           0.005         &                         &                         \\
PF-N (10\%) over-matched      				&           0.001         &          -0.001         &          -0.005         \\
PF-P (2.5\%) under-matched    				&           0.009         &                         &                         \\
PF-P (2.5\%) over-matched    				&          -0.003         &          -0.000         &          -0.001         \\
PF-P (5\%) under-matched     				&          -0.014         &          -0.007         &                         \\
PF-P (5\%) over-matched     				&           0.004         &                         &                         \\
PF-P (10\%) under-matched      				&          -0.008         &          -0.006         &                         \\
PF-P (10\%) over-matched      				&          -0.020         &          -0.018         &          -0.010         \\
PF-L (2.5\%, relaxed) under-matched 		&          -0.052         &          -0.041         &                         \\
PF-L (2.5\%, relaxed) over-matched 			&          -0.016         &          -0.008         &                         \\
PF-L (5\%, relaxed) under-matched 			&           0.007         &           0.001         &                         \\
PF-L (5\%, relaxed) over-matched 			&           0.014         &           0.007         &                         \\
PF-L (10\%, relaxed) under-matched 			&           0.024         &           0.021         &           0.010         \\
PF-L (10\%, relaxed) over-matched 			&          -0.009         &          -0.004         &                         \\
PF-N (2.5\%, relaxed) under-matched 		&           0.023         &           0.016         &           0.006         \\
PF-N (2.5\%, relaxed) over-matched 			&          -0.016         &          -0.011         &                         \\
PF-N (5\%, relaxed) under-matched 			&          -0.009         &                         &                         \\
PF-N (5\%, relaxed) over-matched 			&           0.011         &           0.002         &                         \\
PF-N(10\%, relaxed) under-matched 			&           0.014         &           0.014         &           0.015         \\
PF-N (10\%, relaxed) over-matched 			&          -0.003         &          -0.000         &                         \\
PF-P (2.5\%, relaxed) under-matched 		&           0.004         &                         &                         \\
PF-P (2.5\%, relaxed) over-matched 			&          -0.042         &          -0.035         &          -0.012         \\
PF-P (5\%, relaxed) under-matched 			&          -0.006         &                         &                         \\
PF-P (5\%, relaxed) over-matched 			&           0.025         &           0.016         &                         \\
PF-P (10\%, relaxed) under-matched 			&           0.039         &           0.035         &           0.018         \\
PF-P (10\%, relaxed) over-matched 			&          -0.006         &          -0.002         &                         \\
ALV-L\_1 under-matched      				&           0.002         &                         &                         \\
ALV-L (1 z-score) over-matched      		&          -0.018         &          -0.017         &          -0.014         \\
ALV-L (1.5 z-score) over-matched     		&          -0.004         &          -0.002         &                         \\
ALV-L (2 z-score) under-matched      		&           0.008         &           0.004         &                         \\
ALV-L (2 z-score) over-matched      		&           0.022         &           0.017         &                         \\
ALV-N (1 z-score) under-matched      		&           0.013         &           0.009         &                         \\
ALV-N (1 z-score) over-matched      		&          -0.006         &          -0.003         &                         \\
ALV-N (1.5 z-score) under-matched     		&          -0.018         &          -0.013         &                         \\
ALV-N (1.5 z-score) over-matched     		&          -0.000         &                         &                         \\
ALV-N (2 z-score) under-matched      		&           0.033         &           0.027         &           0.003         \\
ALV-N (2 z-score) over-matched      		&          -0.019         &          -0.015         &                         \\
ALV-P (1 z-score) under-matched      		&          -0.006         &          -0.004         &                         \\
ALV-P (1 z-score) over-matched      		&          -0.034         &          -0.033         &          -0.027         \\
ALV-P (1.5 z-score) under-matched     		&           0.002         &                         &                         \\
ALV-P (1.5 z-score) over-matched     		&          -0.002         &                         &                         \\
ALV-P (2 z-score) under-matched      		&           0.010         &           0.008         &                         \\
ALV-P (2 z-score) over-matched      		&           0.020         &           0.015         &                         \\
Constant        &           1.135         	&           1.157         &           1.286         \\\midrule
Observations    &           42922         	&           42922         &           42922         \\
\bottomrule
\end{longtable}
	\begin{tablenotes}
		\footnotesize
    		\item \textit{Notes:} ISCO SL --- International Standard Classification of Occupations Skill Level; ISIC --- International Standard Industrial Classification of All Economic Activities. The mismatch measures and their specifications: 
    		\begin{itemize}[itemsep=0cm,parsep=0cm]
    			\item[] JA --- Job Analysis
    			\item[] RM --- Realised Matches: mean-based with 0.5, 1 or 1.5 SDs thresholds or mode-based with 0.1, 1 or 2 SDs thresholds 
    			\item[] DSA --- Direct Self Assessment: regular or relaxed
    			\item[] ISA --- Indirect Self Assessment: 1-5 year gaps
    			\item[] PF --- Pellizzari-Fichen: regular or relaxed, literacy (-L); numeracy (-N) or problem-solving (-P) based; 0.025, 0.05 or 0.1 quantile thresholds 
    			\item[] ALV --- Allen-Levels-van-der-Velden: literacy (-L); numeracy (-N) or problem-solving (-P) based; 1, 1.5 or 2 z-score gaps
    		\end{itemize} 
	\end{tablenotes}}

\clearpage
\subsection{Error components model}\label{sec:A_error_comp}

{\footnotesize
{
\def\sym#1{\ifmmode^{#1}\else\(^{#1}\)\fi}
\begin{longtable}{l*{4}{c}}
\caption{Wage penalties: JA \label{tab:ja}}\\
\toprule\endfirsthead\midrule\endhead\midrule\endfoot\endlastfoot
                & \multicolumn{1}{c}{POLS} &\multicolumn{1}{c}{FE}&\multicolumn{1}{c}{BE}&\multicolumn{1}{c}{RE}\\
\midrule
Over-matched    &          -0.103\sym{***}&          -0.126\sym{***}&          -0.185         &          -0.126\sym{***}\\
                &   [-0.14,-0.07]         &   [-0.14,-0.11]         &    [-0.56,0.19]         &   [-0.14,-0.11]         \\
Under-matched   &           0.106\sym{***}&           0.074\sym{***}&           0.413         &           0.076\sym{***}\\
                &     [0.07,0.15]         &     [0.06,0.09]         &    [-0.04,0.86]         &     [0.06,0.10]         \\
Over-matched (mean)&                         &          -0.059         &                         &                         \\
                &                         &    [-0.43,0.31]         &                         &                         \\
Under-matched (mean)&                         &           0.338         &                         &                         \\
\end{longtable}
}
{
\def\sym#1{\ifmmode^{#1}\else\(^{#1}\)\fi}
\begin{longtable}{l*{4}{c}}
\caption{Wage penalties: RM \label{tab:rm}}\\
\toprule\endfirsthead\midrule\endhead\midrule\endfoot\endlastfoot
                & \multicolumn{1}{c}{POLS} &\multicolumn{1}{c}{FE}&\multicolumn{1}{c}{BE}&\multicolumn{1}{c}{RE}\\
\midrule
Over-matched    &          -0.063\sym{*}  &          -0.164\sym{***}&           0.306         &          -0.162\sym{***}\\
                &   [-0.12,-0.01]         &   [-0.18,-0.15]         &    [-0.15,0.77]         &   [-0.18,-0.14]         \\
Under-matched   &           0.103\sym{***}&           0.071\sym{***}&           0.773\sym{***}&           0.073\sym{***}\\
                &     [0.06,0.15]         &     [0.05,0.09]         &     [0.32,1.22]         &     [0.05,0.09]         \\
Over-matched (mean)&                         &           0.470\sym{*}  &                         &                         \\
                &                         &     [0.01,0.93]         &                         &                         \\
Under-matched (mean)&                         &           0.702\sym{**} &                         &                         \\
\end{longtable}
}
{
\def\sym#1{\ifmmode^{#1}\else\(^{#1}\)\fi}
\begin{longtable}{l*{4}{c}}
\caption{Wage penalties: ISA \label{tab:isa}}\\
\toprule\endfirsthead\midrule\endhead\midrule\endfoot\endlastfoot
                & \multicolumn{1}{c}{POLS} &\multicolumn{1}{c}{FE}&\multicolumn{1}{c}{BE}&\multicolumn{1}{c}{RE}\\
\midrule
Over-matched    &          -0.188\sym{***}&          -0.130\sym{***}&          -0.783\sym{***}&          -0.131\sym{***}\\
                &   [-0.23,-0.15]         &   [-0.14,-0.12]         &   [-1.08,-0.49]         &   [-0.14,-0.12]         \\
Under-matched   &           0.127\sym{***}&           0.098\sym{***}&           1.587\sym{***}&           0.098\sym{***}\\
                &     [0.09,0.16]         &     [0.08,0.12]         &     [1.09,2.09]         &     [0.08,0.12]         \\
Over-matched (mean)&                         &          -0.653\sym{***}&                         &                         \\
                &                         &   [-0.95,-0.36]         &                         &                         \\
Under-matched (mean)&                         &           1.490\sym{***}&                         &                         \\
\end{longtable}
}

{
\def\sym#1{\ifmmode^{#1}\else\(^{#1}\)\fi}
\begin{longtable}{l*{4}{c}}
\caption{Wage penalties: PF-Literacy \label{tab:pfl}}\\
\toprule\endfirsthead\midrule\endhead\midrule\endfoot\endlastfoot
                & \multicolumn{1}{c}{POLS} &\multicolumn{1}{c}{FE}&\multicolumn{1}{c}{BE}&\multicolumn{1}{c}{RE}\\
\midrule
Over-matched    &          -0.289\sym{***}&          -0.051\sym{***}&           0.089         &          -0.058\sym{***}\\
                &   [-0.33,-0.25]         &   [-0.08,-0.02]         &    [-0.31,0.48]         &   [-0.08,-0.03]         \\
Under-matched   &           0.240\sym{***}&           0.140\sym{***}&          -1.031\sym{*}  &           0.144\sym{***}\\
                &     [0.19,0.29]         &     [0.11,0.17]         &   [-2.04,-0.02]         &     [0.12,0.17]         \\
Over-matched (mean)&                         &           0.140         &                         &                         \\
                &                         &    [-0.25,0.53]         &                         &                         \\
Under-matched (mean)&                         &          -1.171\sym{*}  &                         &                         \\
\end{longtable}
}

{
\def\sym#1{\ifmmode^{#1}\else\(^{#1}\)\fi}
\begin{longtable}{l*{4}{c}}
\caption{Wage penalties: PF-Numeracy \label{tab:pfn}}\\
\toprule\endfirsthead\midrule\endhead\midrule\endfoot\endlastfoot
                & \multicolumn{1}{c}{POLS} &\multicolumn{1}{c}{FE}&\multicolumn{1}{c}{BE}&\multicolumn{1}{c}{RE}\\
\midrule
Over-matched    &          -0.329\sym{***}&          -0.079\sym{***}&          -0.051         &          -0.086\sym{***}\\
                &   [-0.37,-0.29]         &   [-0.10,-0.06]         &    [-0.40,0.30]         &   [-0.11,-0.06]         \\
Under-matched   &           0.254\sym{***}&           0.146\sym{***}&          -0.250         &           0.150\sym{***}\\
                &     [0.19,0.32]         &     [0.12,0.17]         &    [-1.36,0.86]         &     [0.12,0.18]         \\
Over-matched (mean)&                         &           0.028         &                         &                         \\
                &                         &    [-0.32,0.37]         &                         &                         \\
Under-matched (mean)&                         &          -0.396         &                         &                         \\
\end{longtable}
}

{
\def\sym#1{\ifmmode^{#1}\else\(^{#1}\)\fi}
\begin{longtable}{l*{4}{c}}
\caption{Wage penalties: PF-Problem-Solving \label{tab:pfp}}\\
\toprule\endfirsthead\midrule\endhead\midrule\endfoot\endlastfoot
                & \multicolumn{1}{c}{POLS} &\multicolumn{1}{c}{FE}&\multicolumn{1}{c}{BE}&\multicolumn{1}{c}{RE}\\
\midrule
Over-matched    &          -0.283\sym{***}&          -0.057\sym{***}&           0.047         &          -0.064\sym{***}\\
                &   [-0.34,-0.23]         &   [-0.08,-0.03]         &    [-0.27,0.36]         &   [-0.09,-0.04]         \\
Under-matched   &           0.132\sym{***}&           0.090\sym{***}&          -0.570\sym{*}  &           0.091\sym{***}\\
                &     [0.07,0.19]         &     [0.07,0.11]         &   [-1.03,-0.11]         &     [0.07,0.11]         \\
Over-matched (mean)&                         &           0.104         &                         &                         \\
                &                         &    [-0.22,0.42]         &                         &                         \\
Under-matched (mean)&                         &          -0.660\sym{**} &                         &                         \\
\end{longtable}
}

{
\def\sym#1{\ifmmode^{#1}\else\(^{#1}\)\fi}
\begin{longtable}{l*{4}{c}}
\caption{Wage penalties: ALV-Literacy \label{tab:alvl}}\\
\toprule\endfirsthead\midrule\endhead\midrule\endfoot\endlastfoot
                & \multicolumn{1}{c}{POLS} &\multicolumn{1}{c}{FE}&\multicolumn{1}{c}{BE}&\multicolumn{1}{c}{RE}\\
\midrule
Over-matched    &          -0.389\sym{***}&          -0.161\sym{***}&          -1.731\sym{***}&          -0.171\sym{***}\\
                &   [-0.43,-0.34]         &   [-0.18,-0.14]         &   [-2.23,-1.23]         &   [-0.19,-0.15]         \\
Under-matched   &           0.226\sym{***}&           0.137\sym{***}&           1.938\sym{***}&           0.141\sym{***}\\
                &     [0.19,0.26]         &     [0.12,0.15]         &     [1.43,2.45]         &     [0.12,0.16]         \\
Over-matched (mean)&                         &          -1.570\sym{***}&                         &                         \\
                &                         &   [-2.07,-1.07]         &                         &                         \\
Under-matched (mean)&                         &           1.801\sym{***}&                         &                         \\
\end{longtable}
}

{
\def\sym#1{\ifmmode^{#1}\else\(^{#1}\)\fi}
\begin{longtable}{l*{4}{c}}
\caption{Wage penalties: ALV-Numeracy \label{tab:alvn}}\\
\toprule\endfirsthead\midrule\endhead\midrule\endfoot\endlastfoot
                & \multicolumn{1}{c}{POLS} &\multicolumn{1}{c}{FE}&\multicolumn{1}{c}{BE}&\multicolumn{1}{c}{RE}\\
\midrule
Over-matched    &          -0.082\sym{**} &          -0.075\sym{***}&           0.675         &          -0.077\sym{***}\\
                &   [-0.13,-0.03]         &   [-0.09,-0.06]         &    [-0.05,1.40]         &   [-0.09,-0.06]         \\
Under-matched   &           0.057\sym{*}  &           0.100\sym{***}&           1.317\sym{***}&           0.101\sym{***}\\
                &     [0.01,0.10]         &     [0.08,0.12]         &     [0.75,1.88]         &     [0.08,0.12]         \\
Over-matched (mean)&                         &           0.750\sym{*}  &                         &                         \\
                &                         &     [0.02,1.48]         &                         &                         \\
Under-matched (mean)&                         &           1.216\sym{***}&                         &                         \\
\end{longtable}
}
{
\def\sym#1{\ifmmode^{#1}\else\(^{#1}\)\fi}
\begin{longtable}{l*{4}{c}}
\caption{Wage penalties: ALV-Problem-Solving \label{tab:alvp}}\\
\toprule\endfirsthead\midrule\endhead\midrule\endfoot\endlastfoot
                & \multicolumn{1}{c}{POLS} &\multicolumn{1}{c}{FE}&\multicolumn{1}{c}{BE}&\multicolumn{1}{c}{RE}\\
\midrule
Over-matched    &          -0.198\sym{***}&          -0.092\sym{***}&          -1.356\sym{***}&          -0.096\sym{***}\\
                &   [-0.23,-0.17]         &   [-0.11,-0.08]         &   [-2.12,-0.59]         &   [-0.11,-0.08]         \\
Under-matched   &           0.087\sym{***}&           0.068\sym{***}&          -0.587         &           0.069\sym{***}\\
                &     [0.06,0.12]         &     [0.05,0.08]         &    [-1.46,0.29]         &     [0.06,0.08]         \\
Over-matched (mean)&                         &          -1.264\sym{**} &                         &                         \\
                &                         &   [-2.03,-0.50]         &                         &                         \\
Under-matched (mean)&                         &          -0.656         &                         &                         \\
\end{longtable}
}

}

\end{document}